\documentclass[12 pt, preprint]{aastex}

\shorttitle{RSGs in M31}
\shortauthors{Massey et al.}

\begin{document}

\title{Red Supergiants in the Andromeda Galaxy (M31)\altaffilmark{1}}

\author{
Philip Massey\altaffilmark{2}}
\affil{Lowell Observatory, 1400 West Mars Hill Rd., Flagstaff, AZ 86001; phil.massey@lowell.edu}

\author{David R. Silva\altaffilmark{2}}
\affil{National Optical Astronomy Observatory, 950 North Cherry Avenue,
Tucson, AZ 85748; dsilva@noao.edu}

\author{Emily M. Levesque\altaffilmark{3}}
\affil{Institute for Astronomy, University of Hawaii, 2680 Woodlawn Drive, Honolulu, HI 96822;
emsque@ifa.hawaii.edu}

\author{Bertrand Plez\altaffilmark{4}}
\affil{GRAAL, Universit\'{e} Montpellier II, CNRS, 34095 Montpellier, France; bertrand.plez@graal.univ-montp2.fr}

\author{Knut A. G. Olsen}
\affil{National Optical Astronomy Observatory, 950 North Cherry Avenue,
Tucson, AZ 85748; kolsen@noao.edu}

\author{Geoffrey C. Clayton\altaffilmark{2}}
\affil{Department of Physics and Astronomy, Louisiana State University, Baton Rouge, LA 70803; gclayton@fenway.phys.lsu.edu}

\author{Georges Meynet and Andre Maeder}
\affil{Geneva University, Geneva Observatory, CH-1290 Versoix, Switzerland; georges.meynet@unige.ch, andre.maeder@unige.ch}

\altaffiltext{1}{Observations reported here were obtained at the MMT Observatory, a joint facility of the University of Arizona and the Smithsonian Institution.}
\altaffiltext{2}{Visiting Astronomer, Kitt Peak National Observatory,
National Optical Astronomy Observatory (NOAO) , which is operated by the
Association of Universities for Research in Astronomy, Inc. (AURA)
under cooperative agreement with the National Science Foundation (NSF).}
\altaffiltext{3}{Smithsonian Astrophysical Observatory Predoctoral Fellow.  Current
address: Center for Astrophysics, 60 Garden Street, Cambridge,  MA 02138.}
\altaffiltext{4}{Also Department of Physics and Astronomy, Uppsala University, SE-75120 Uppsala, Sweden.}

\begin{abstract}
Red supergiants (RSGs) are a  short-lived stage in the evolution of moderately massive
stars (10-25$M_\odot$), and as such their location in the H-R diagram provides
an exacting test of stellar evolutionary models.  Since massive star evolution is
strongly affected by the amount of mass-loss a star suffers, and since the mass-loss
rates depend upon metallicity, it is highly desirable to study the physical properties
of these stars in galaxies of various metallicities.  Here 
we identify a sample of red supergiants in M31, the most metal-rich of the Local Group
galaxies.  We determine the physical
properties of these stars using both moderate resolution spectroscopy and broad-band $V-K$
photometry.  We find that on average the RSGs are our sample  are variable in $V$
by 0.5~mag,  smaller but comparable to the 0.9~mag found for Magellanic Cloud (MC) RSGs.
No such variability is seen at $K$, also in accord with what we know of Galactic and
MC RSGs.   We find that there is a saturation effect in the model TiO band strengths
with metallicities higher than solar.
The physical properties we derive for the RSGs from our analysis with stellar atmosphere
models  agree well with the current evolutionary tracks, a truly remarkable achievement given
the complex physics involved in each.
We do not confirm an earlier result that
the upper luminosities of RSGs depends upon metallicity; instead, the most luminous
RSGs have $\log L/L_\odot \sim$5.2-5.3, broadly consistent but slightly larger than 
that recently observed by Smartt et al.\ as the upper
luminosity limit to Type II-P supernovae, believed to have come from RSGs.  We find that, on average,
the RSGs are considerably more reddened than O and B stars, suggesting that
circumstellar dust is adding a significant amount of extra
extinction, $\sim$0.5~mag, on average.  This is in accord with our earlier findings
on Milky Way and Magellanic Cloud stars.  Finally, we call attention to  a peculiar star  whose
spectrum appears to be heavily veiled, possibly due to scattering by an expanding dust
shell.
\end{abstract}


\section{Introduction}

Red supergiants (RSGs) are the evolved descendants of moderately massive (10-25 $M_\odot$)
stars. Until recently, the physical properties of RSGs have been at variance with stellar evolutionary theory
(Massey 2003; Massey \& Olsen 2003), with the ``observed" location of these stars in the H-R
diagram (HRD)
being cooler and more luminous than theory allows.  In some ways this disagreement was hardly surprising.
After all, how far the evolutionary tracks extend to cooler temperatures depends on how the mixing length
is treated (see Figure 9 of Maeder \& Meynet 1987), and for that matter RSGs themselves are known to violate
just about all of the simplifying assumptions made in stellar atmospheres and evolution models: their atmospheres
are highly extended, while most models assume plane-parallel geometry, and the velocities of the
convective layers are nearly sonic, or even supersonic in the atmosphere layers, leading to shocks (Freytag et al.\ 2002),
which violates mixing-length assumptions, etc.

However, it turned out that the problem was not with
the location of the evolutionary tracks but rather with the ``observed" location of the stars in the
HRD.  The effective
temperature scale commonly used for RSGs was largely derived from lunar occultation measurements of red {\it giants}
(and not supergiants), as discussed in Massey \& Olsen (2003).  Levesque et al.\ (2005, hereafter Paper I) measured
new effective temperatures and bolometric luminosities
of Milky Way RSGs by fitting moderate-resolution spectrophotometry with the new generation of MARCS stellar
atmospheres (Gustafsson et al.\ 2008), finding an effective temperature scale that was about 5-10\% 
warmer.  With the corresponding change in bolometric luminosities, this was sufficient
to bring the location of these RSGs into excellent agreement with that of the evolutionary tracks.  Levesque et al.\ (2006,
hereafter Paper II) then extended this work to RSGs found in the lower metallicity Magellanic Clouds.  As might
be expected, due to the lower elemental abundances, stars needed to be cooler in order to produce the same
band strength of TiO, which forms the basis for assigning spectral types.   Thus M supergiants in the LMC and SMC
are cooler by $\sim$50 and $\sim$150 K, respectively, than their Galactic counterparts on average;  see Table 4 in Paper II.
Furthermore, Paper II again found excellent agreement between the evolutionary tracks and the location of LMC stars in the HRD.
The agreement for the SMC stars was more mixed: although the situation was drastically improved since the work of Massey \& Olsen (2003),
there were still a number of SMC stars that were cooler than the evolutionary 
tracks allow.    Papers I and II also established that the MARCS
models predicted broad-band $V-K$ colors that were systematically warmer by about 100~K for Milky Way and LMC RSGs, and by
about 170~K for the SMC RSGs.  The reason for this remains unclear; it may be an intrinsic limitation of one-dimensional static models,
as radiative-hydrodynamic three-dimensional models of RSGs show a pattern of large warm and cool patches on the surface (Freytag et al.\ 2002),
which may explain this wavelength-dependent result for the effective temperatures.

We were naturally keen to extend this work both to lower and higher metallicities.  The range covered by the SMC to the Milky Way
is only a factor of 4 in metallicity (as measured by the oxygen abundances; see Russell \& Dopita 1990, and Westerlund 1997).
By contrast, the star-forming galaxies of the Local Group span a factor of 15 in metallicity, from WLM
whose HII regions yield an oxygen abundance of log O/H+12=7.8 (Hodge \& Miller 1995) to the extra-solar metallicity galaxy M31 with log O/H+12=9.0 (Zaritsky et al.\ 1994)\footnote{Zaritsky et al.\ (1994)
measure the oxygen abundance as log O/H+12=9.0-9.1 at several radii, and determine 
a very shallow gradient, 0.02~dex/kpc.  Given this, we use the short-hand here of referring to
characterizing the metallicity of M31 with a single value, where we are of course referring to the
disk metallicity of the youngest population, corresponding to  the HII regions.  Similarly,  what we refer to as ``solar metallicity" is that of the HII regions
measured in the solar neighborhood, which yield a log O/H+12=8.7 (Esteban \& Piembert 1995).}; 
see Table 1 of Massey (2003).  Since massive stars lose copious amounts of mass on the main-sequence, and this mass-loss scales
with metallicity $Z$ as roughly $(Z/Z_\odot)^{0.7}$ (Vink et al.\ 2001),  a study of RSGs provides an important additional test
of stellar evolutionary theory in addition to the changes we expect in the effective temperature
scale due to the different metallicities (really controlled by the abundance of Ti; see discussion in Plez 2003).
Of course, such a study would be incomplete without some unanticipated surprises.
For instance, in the SMC and LMC we discovered several stars that were variable by hundreds of degrees in effective temperature on the time-scale of months
(Massey et al.\ 2007a; Levesque et al.\ 2007), a phenomenon not known in Milky Way RSGs.

In this study we turn our attention to M31, the most metal-rich of any of the galaxies in the
Local Group.  In this, we are aided by the Local Group Galaxies Survey (LGGS) photometry of Massey et al.\ (2006b, 2007b),
which allow us to select candidates.  We use these broad-band photometry to first identify 
new RSG candidates, and then determine if their
radial velocities are consistent with their locations within M31 (\S~\ref{Sec-sample}).  We obtain spectroscopy and new photometry
of these objects (\S~\ref{Sec-data}) that will allow us to determine their physical properties (\S~\ref{Sec-analysis}).  We then
use these data to compare the physical properties of these stars to those predicted by stellar evolution theory (\S~\ref{Sec-hrd}).
In \S~\ref{Sec-wacky} we comment on one of the intriguing new objects we found, J004047.84+405602.6, which appears to be
heavily reddened and veiled.  We close by summarizing our results,
and laying out our future plans (\S~\ref{Sec-summary}).

\section{Sample Selection and Membership Confirmation}
\label{Sec-sample}

\subsection{Sample Selection from the LGGS}

One of the basic problems confronting studies of RSGs in nearby
galaxies is distinguishing these extragalactic RSGs from foreground stars,
primarily Milky Way disk dwarfs but also some Milky Way halo giants.
This is not a problem for the blue supergiants, where there is essentially no
foreground contamination, but it {\it is} significant for RSGs.  We
first became aware of the magnitude of this issue by comparing 
the distributions of blue and red stars seen towards M33.  This problem is
nicely illustrated by comparing Figures 21 and 22
of Humphreys \& Sandage (1980).  While the distribution of
blue stars is clumpy (as might be expected), the distribution of red stars
is quite a bit more uniform.
Given that RSGs are descended from the blue supergiants, but are not much
different in age, we expect the two to have similar spatial distributions.  The
fact that the distribution of red stars is smooth suggests that their numbers are 
not dominated
by the RSGs, but rather by foreground stars.  Massey et al.\ (2007b) use the 
Bahcall \& Soneira (1980) model to estimate that
roughly 80\% of the red stars ($1.2 < B-V < 1.8$) seen towards M31 with $16<V<20$
will be foreground stars.

\begin{figure}
\epsscale{0.6}
\plotone{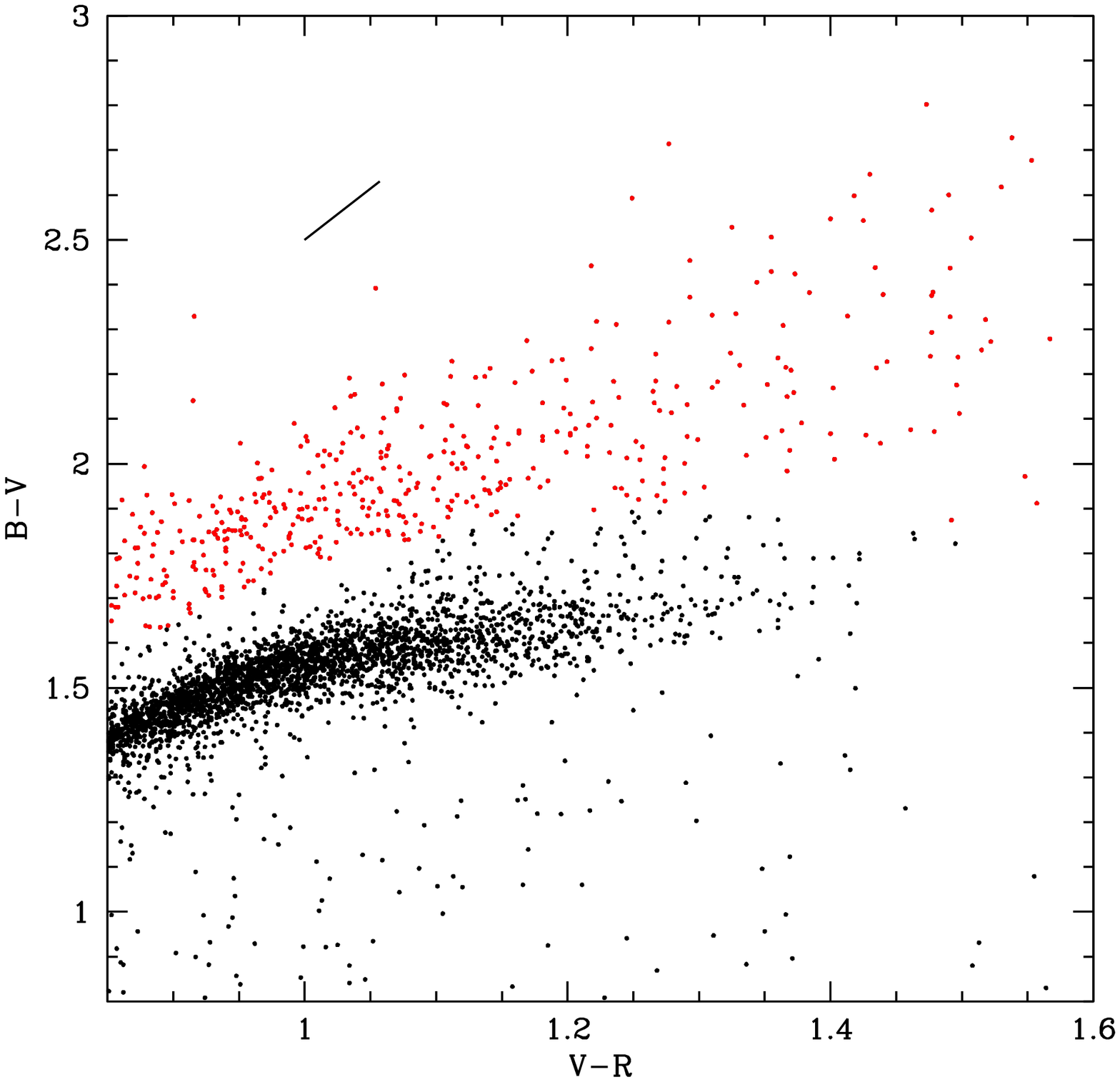}
\caption{\label{fig:2color1}Two-color diagram showing separation of candidate RSGs (red) from foreground stars (black).
Stars with lower surface gravities (i.e., supergiants) will have larger $B-V$ values than stars with higher surface gravities
(foreground dwarfs) for a given $V-R$.  A reddening line corresponding to a ``typical" $E(B-V) = 0.13$ for early-type stars in M31
(Massey et al.\ 2007b) is shown in the upper left.)}
\end{figure}

However, Massey (1998) showed that a  $V-R$, $B-V$ two-color diagram
can be used to readily distinguish between the two groups, as $V-R$ is sensitive 
primarily to effective temperature, while $B-V$ is sensitive both to effective temperature
and surface 
gravity\footnote{This is well shown by inspection of photometric indices derived
by Buser \& Kurucz (1978).  The MARCS model atmosphere colors agree with these
over the limited temperature regime we consider here, although there may be a problem
with the spectral energy distributions of the MARCS models at cooler ($<$3600 K) effective
temperatures.  We are investigating this, but it does not affect the present study.}.  
At a given $V-R$ stars with larger $B-V$ values are expected to be
supergiants.  Massey et al.\ (2006b) used the LGGS photometry to show a fairly
clean separation in $B-V$ for red stars ($V-R>0.6$).  We show a new version of this
two-color plot in Fig.~\ref{fig:2color1}.  We have color coded the RSG candidates
in red, adopting a dividing line at
\begin{equation}
B-V =  -1.599 (V-R)^2 + 4.18 (V-R) -0.83,\end{equation}
similar to the relationship used by
Massey (1998).  Note that the reddening vector (corresponding to a ``typical" $E(B-V) = 0.13$ and shown in the
upper left corner) is nearly parallel to the two sequences; i.e., stars with a little more
or less reddening should not be confused between the two sequences.

We have picked $V=20$ as a  magnitude cut-off for two reasons, one practical, and one
theoretical.  First, it represents a reasonable limit to what can be 
studied with moderate-resolution optical spectroscopy with sufficient signal-to-noise
for model fitting, using 4 to 6.5-m class telescopes.    Second, it is sufficiently bright
that it helps avoid confusion with lower mass
asymptotic giant branch (AGB) stars.  As first emphasized by Brunish et al.\ (1986),
there is overlap in the luminosities of AGB and RSG stars.  Following
Brunish et al.\ (1986), Massey \& Olsen (2003) adopted a luminosity cut-off of $\log L/L_\odot=4.9$
 in order to avoid these.  However, we note that there is considerable uncertainty
in the luminosities of the most luminous AGBs, including, for instance, whether or not ``super" AGBs even exist.
(For some recent studies, see, for example, Siess 2006, 2007; Eldridge et al.\ 2007; and Poelarends et al.\ 2008.)     
We adopt a distance to M31 of 0.76~Mpc (van den Bergh 2000),
and a value for the average reddening of $E(B-V)=0.13$ (Massey et al.\ (2007b).  A cut-off of $V=20$ then includes stars
down to a luminosity of $\log L/L_\odot \sim 4.2$ for early K-types 
(BC$\sim -0.9$~mag), and $\log L/L_\odot \sim 4.6$ for mid M types  (BC$\sim -2.0$~mag).  Thus our sample could
include some AGBs at the lower luminosity end.  These might reveal themselves as stars with cooler effective temperatures
at the lower luminosities than the tracks predict (see Eldridge et al.\ 2007) but otherwise have little effect on our study.  This may be significant, however, for future comparisons of the relative number of RSGs
either as a function of luminosity or compared to other populations (i.e., the number of Wolf-Rayet stars).

In addition to the $V$ requirement, we also
applied the requirement that $V-R\ge 0.85$ (and correspondingly to $B-V\ge 1.5$)
to restrict the sample to K-type stars and later.
This color roughly
corresponds to a $(V-R)_0$=0.81, which, according to the MARCS models
(described below) is characteristic of a star with $\log g=0.0$ and $T_{\rm eff}=4000$ K,
roughly that of a K2-K3 I at Galactic metallicities (Paper I).  

There were 437 stars
in the LGGS catalog that met these photometric criteria.  We list these stars in Table~\ref{tab:all},
where we also include data discussed later in the paper.  The number of stars meeting these photometric criteria is not very
sensitive to the adopted $V-R$ cutoff: if instead we had chosen to count probable
RSGs from  K1~I  and later, corresponding to 4100~K at Galactic metallicity,
 $(V-R)_0=0.77$ (or $V-R>0.81$) we would find 471 stars.  If we were to count on stars from M0~I
and later, corresponding to  3800 K,  $(V-R)_0=0.90$ (or $V-R=0.94$), we would find 395.
Going a magnitude deeper would drastically increase these numbers, to 1750$\pm$500,
so care is needed to strictly understand the luminosity limits in comparing any sample
of RSGs to that of any other population.

\subsection{Membership Confirmation Using Radial Velocities from WIYN}

We consider the candidate RSGs selected photometrically to be just that,
{\it candidate} RSGs.  To confirm their membership in M31 we measured 
radial velocities using the Ca II triplet ($\lambda \lambda 8498, 8542, 8662$)
observed with the Hydra multi-object fiber spectrograph on the WIYN 3.5-m telescope.  
The observations were obtained on 2005 Sep 27, 28, and 30. An 860 line mm$^{-1}$
grating with a blaze angle of 30.9$^\circ$ was used in first order red with the
Bench Spectrograph Camera.  An RG-610 filter was used to block second-order
blue light.   The detector was T2KA, a 2048$\times$2048
Tektronix CCD with 24$\mu$m pixels.  We used the red fiber bundle, consisting
of 90 fibers, each of which subtend a diameter of 2.0 arcsec on the sky. The configuration provided wavelength
coverage from 7565-9665 \AA\ with a spectral resolution of 1.6 \AA, with a dispersion of 0.95 \AA\ pixel$^{-1}$.

We observed four configurations on M31, each consisting of 10-17 sky observations
and 46-72 RSG candidates.  The exposure times varied from 1-2 hrs, in 
1200-1800 sec pieces.  Comparison arcs were run before and after each sequence,
as well as in the middle for the longer exposures.  Flat-fielding was provided by
observations of the illuminated dome spot.  In addition, 9
observations were made of 3 late-type radial velocity standards, $\alpha$ Cet,
$\alpha$ Tau, and $\mu$ Psc, selected as close matches in spectral types
from the list of radial velocity standards in the
US Naval Observatory's Astronomical Almanac. Those exposures were 0.5 sec.

After basic reductions and spectral extractions with IRAF\footnote{IRAF is distributed by NOAO,
which is operated by AURA under cooperative
agreement with the NSF.  We appreciate the on-going support of IRAF by the
volunteers at the IRAF help ``desk", http://www.iraf.net.}  (see Massey 1997), we measured
radial velocities by Fourier cross-correlation of the region surrounding the Ca II triplet, using the ``fxcor" routine,
also in IRAF.   The spectrum of each
red supergiant candidate was cross-correlated against that of each of the 9 radial
velocity standards, and the results averaged.  The standard deviations
of the means were typically 0.4 km s$^{-1}$.    We obtained good radial velocity
data for 126 stars, or about 16\% of the original photometrically-selected sample.
We list these stars in Table~\ref{tab:rvs}.   

\begin{figure}
\epsscale{0.4}
\plotone{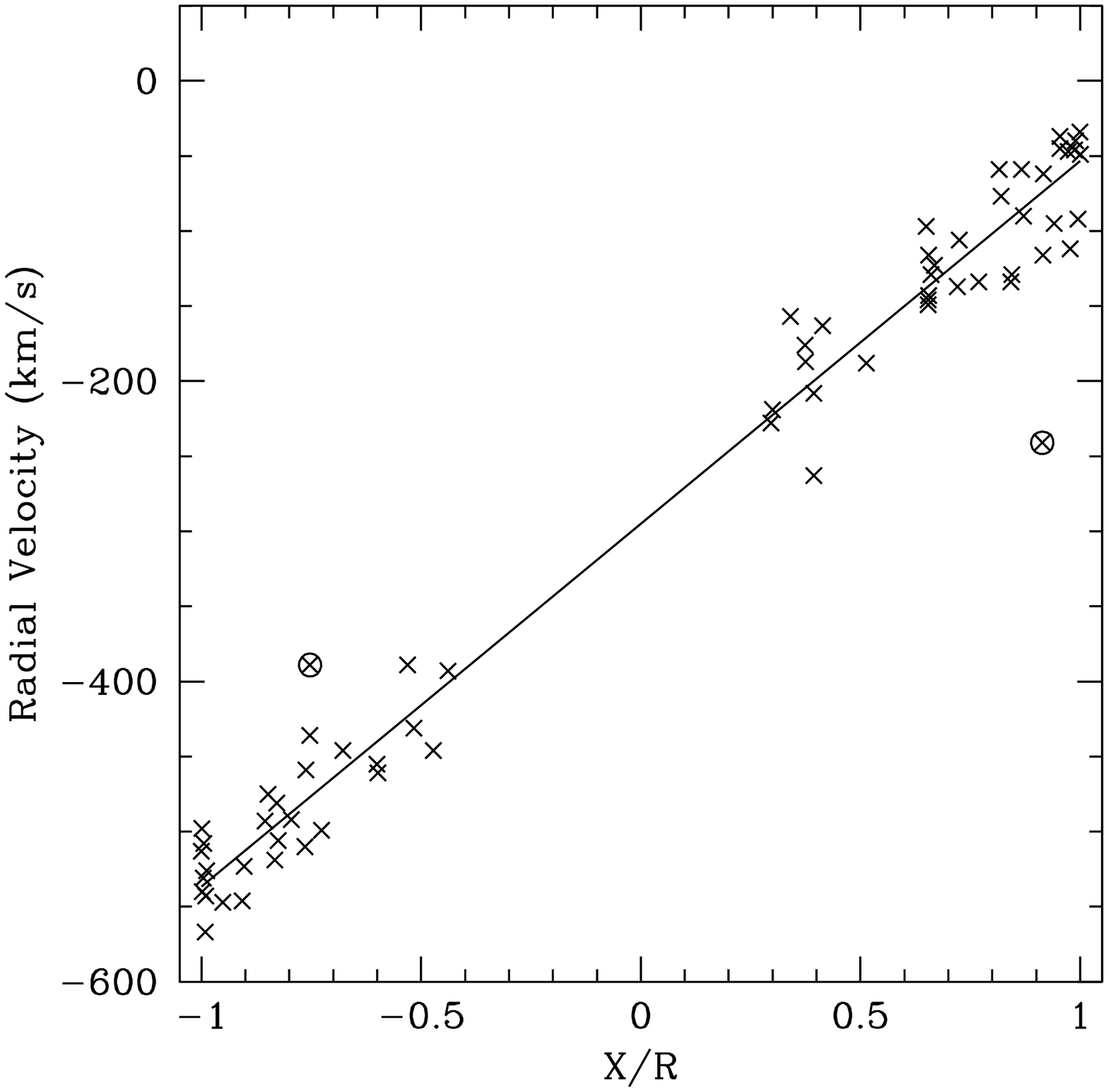}
\plotone{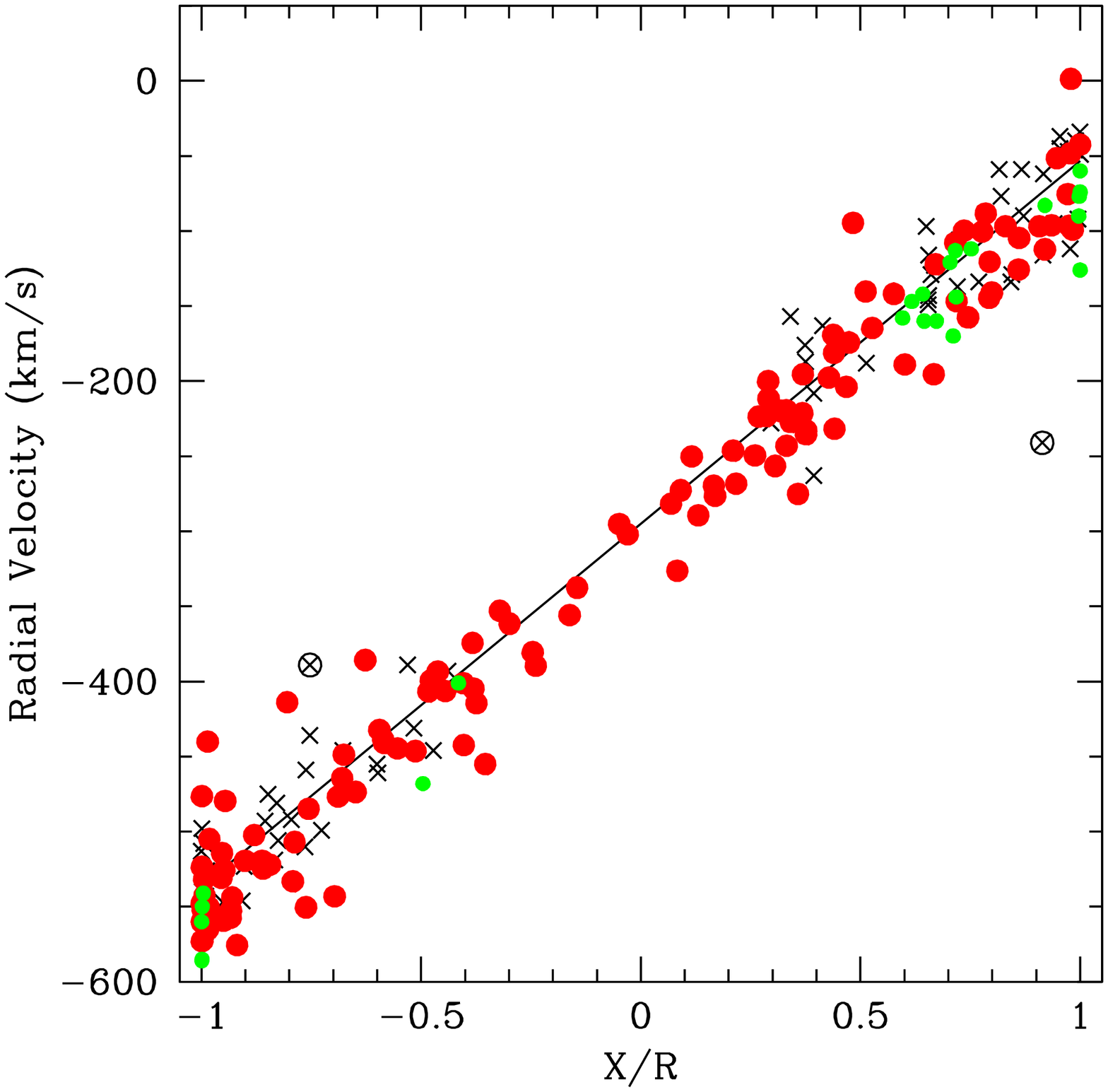}
\caption{\label{fig:rubin} Radial velocities as a function of position $X/R$ in M31.  {\it Left} The radial velocities of HII regions
from Rubin \& Ford (1970) are shown plotted against the quantity $X/R$, where $X$ is the distance along the semi-major axis
of M31, and $R$ is the radial distance of the object within the plane of M31, where we have adopted the geometric parameters
and radius given by Rubin \& Ford (1970) and references therein.  We have fit their data with a straight line, ignoring two outliers
denoted by circled points.  {\it Right} We show the radial velocities of the RSG candidates (red)
from Table~\ref{tab:rvs} superimposed on
the HII region data and fit.  The agreement is excellent.  Green denotes the M31 RSG stars spectroscopically confirmed by Massey (1998). }

\end{figure}

So which stars are members?  To evaluate this, we turn to the seminal study by
Rubin \& Ford (1970), which demonstrated that  M31's rotation curve is essentially flat, 
requiring the mass distribution to be dominated by dark matter. 
They measured the radial velocities of HII regions throughout M31's disk.  In Fig.~\ref{fig:rubin} {\it (left)}
we show their data, where we have plotted their radial velocity against the quantity $X/R$, where $X$ is the distance along
the semi-major axis, and $R$ is the radial distance within the plane of M31.   In general, the radial velocity RV will
be ${\rm RV}_0+  V(R) \sin{\xi} \cos{\theta}$ for a circular velocity $V(R)$, where RV$_0$ is the systemic radial velocity of
M31 with respect to the sun, $\xi$ is the angle between the line of sight and the perpendicular to the plane of M31 (taken to be $77^\circ$, following Rubin \& Ford 1970), and $\theta$ is the azimuthal angle within the plane of M31; i.e., $\cos{\theta}=(X/R)$.
  We have fit their HII region data using a straight line, and find
${\rm RV}=-295+241.5(X/R)$ (after two outliers are removed) with a dispersion of 25 km s$^{-1}$.  The fact that this is well
approximated by a straight-line is equivalent to stating that $V(R)$ is a constant, emphasizing the flatness of the rotation curve.
The simple approximation is in good agreement 
with the more complex two dimension velocity field (Sofue \& Kato 1981), and with other recent
approximations (Hurley-Keller et al.\  2004).   

In Fig.~\ref{fig:rubin} ({\it right}) we now superimpose the radial velocities of our RSG candidates from Table~\ref{tab:rvs}.  
The agreement is striking: {\it every} candidate RSG with a measured radial velocity has the velocity expected from 
its location within M31!    One caveat needs to be clearly stated, and that is that for stars with RVs $>-200$ this agreement
doesn't {\it prove} membership.  Of the $\sim -300$ km s$^{-1}$ systemic motion of M31, about two-thirds of
it is the reflex motion of the sun according to equation 4 of
Coutreau \& van den Bergh (1999), $\sim$-180 km s$^ {-1}$.  So a halo red giant {\it could} be confused
with a red supergiant were it to be located in the NE quadrant of M31 where $X/R>0.5$.  However, 
halo giants will be quite rare compared to the foreground disk dwarfs that dominate
the contamination, and we note that {\it all} of
the other stars that have been photometrically selected do have radial velocities which necessitate their being M31 members.
We conclude that the photometric selection is in fact quite good at selecting bona-fide RSGs.

\begin{figure}
\epsscale{0.6}
\plotone{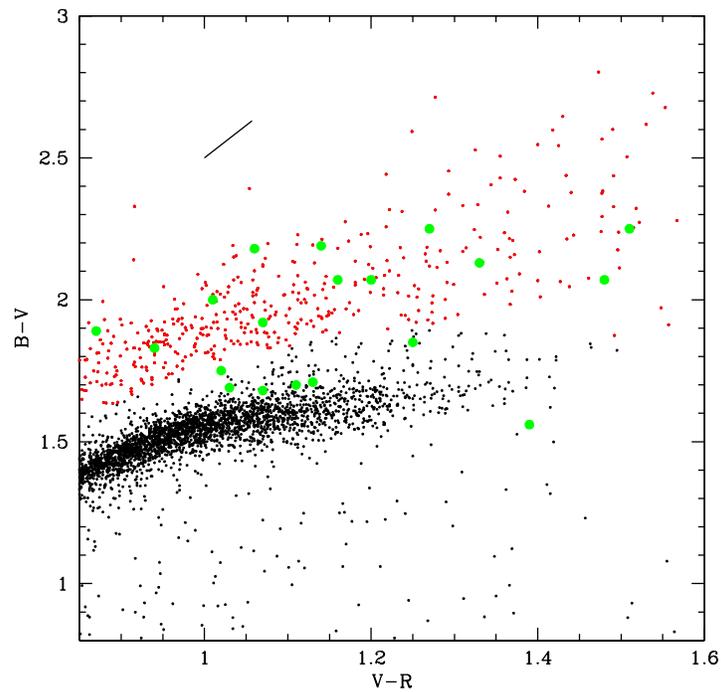}
\caption{\label{fig:2color2} Same as Fig.~\ref{fig:2color1} with the addition of previously confirmed RSGs.
The RSGs confirmed by Massey (1998) are shown in green dots superimposed on the same data
from Fig.~\ref{fig:2color1}.  Note that
several of these stars fall below the cut-off in B-V, and would not be considered RSG candidates here.
Reasons are discussed in the text.}
\end{figure}

Before we leave this subject we must note an apparent inconsistency with the previous RSG list of Massey (1988).  That sample
had been selected primarily on the same photometric basis as the sample presented here, using imaging over a few small regions of M31, and targeted to include some of the more interesting
(crowded) OB associations.  However, with the LGGS photometry not all of the spectroscopically ``confirmed" RSGs from that
paper  would make our photometric cut.  This problem was alluded to by Massey et al.\ (2006b) Section 4.2, and we illustrate
it in Fig.~\ref{fig:2color2}.  The green dots are the LGGS photometry of this older sample.  In fact, a significant fraction
(5 out of 19) would have been marginally below 
the dividing line between what we consider RSGs and foreground stars, and yet their radial velocities are consistent with membership
(green dots in Fig.~\ref{fig:rubin} {\it right}).  

{\it All} of these discrepant stars are in OB48, and in fact 3 of these 5 stars were also not ``photometric" RSG candidates in
the list of Massey (1998).  Instead, their designation of RSG came about via the follow-up spectroscopy which showed radial
velocities consistent with that expected for M31.  OB48 is located at a $X/R\sim0.72$, and we expect a rotational velocity of about
$-120$ km s$^{-1}$; as we argue above, a foreground halo giant could have this velocity.  An additional criterion used by
Massey (1998) was the
strength of the Ca II triplet lines, which should be a luminosity indicator (Jaschek \& Jaschek 1990), but which could
be confused between supergiants and (foreground) giants (see Figure 4 of Mallik 1994).  Still, it would be astonishing to
find three such halo stars concentrated in such a small area of the sky. 

So, here we prefer to retain only
those stars which have met our photometric cut-off.  Others {\it might} be RSGs, but for now the evidence clearly supports that
the photometric candidates {\it are} all RSGs.  

We did impose one further criterion before gathering additional data (and arguably should have
applied this before obtaining radial velocities),  and
that was the additional requirement that the star be ``isolated" in some sense, so that our spectrophotometry would be of a single  object---at least, as far as we could tell.  Using the LGGS photometry, we considered the flux contribution of all other stars within
10" of the position of each of our RSGs with radial velocities.  We did this for a variety of reasonable seeing conditions
(0.8-1.5 arcsec) and slit widths (1.0-1.5 arcsec wide).   We indicate the degree of crowding in
Table~\ref{tab:all} by the amount of contamination for a 1.2 arcsec slit in 1.2 arcsec
seeing.  We note any stars in Table~\ref{tab:rvs} for which the crowding might affect our
results.  In choosing stars for further spectroscopy we restricted the sample to those with
no crowding issues.

\section{New Spectroscopy and Photometry}

Our study of RSGs in M31 differs from our earlier work in two important ways.  The first
of these concerns the sample size.   In Papers I and II were  able to analyze a large
number of RSGs, 74
in the Milky Way, 36 in the LMC, and 37 in the SMC.
However, the faintness of M31's RSGs  requires
a large aperture telescope for the spectroscopy, and this precludes studying a similar
size sample.  We therefore
decided to concentrate on stars whose colors
suggested they were ``moderate" M stars, i.e., M2-3 in type.  

The second point represents a shift in technique, as discussed below, with a greater
emphasis on $K$-band photometry to determine the bolometric luminosity of the star.
  We have previously proceeded by obtaining
optical spectrophotometry of our stars, and using the MARCS models to determine effective
temperature and $A_V$.  We then applied these to the $V$ in order to determine $M_V$, and
then used the models and the effective temperature to determine the bolometric correction to
determine the total luminosity of the star.  While this yielded good results, we also wanted to
further explore using  $K$-band to determine the luminosity in conjunction with the
effective temperature and reddening determined from the optical spectroscopy.
 In addition, we can use $V-K$ to get a nearly independent determination of the effective
 temperatures of these stars.

\label{Sec-data}
\subsection{Optical Spectroscopy at the MMT for Model Fitting}

Our spectroscopy was carried out with the 6.5-m MMT telescope on 2.5  nights of 2006 October 26-28 using the Blue Channel 
spectrograph.   The third of these nights was split between this project and another.
 The detector was a 2688$\times$512 15$\mu$m CCD made as part of a University of Arizona Imaging Technology Laboratory
 foundry run, and has excellent cosmetics,
low read-noise (2.5 e$^-$), and high full-well (129,000 e$^{-1}$).  We used the 300 line mm$^{-1}$ grating in first order to achieve 
wavelength coverage from 3800-7500\AA\ with 7.2\AA\ spectral resolution with a 1.25 arcsec wide slit.  A L-42 blocking filter 
cut any light with a wavelength $<3800$ that could have interfered with the red end of the spectrum; this obviously was not of much
concern for the RSGs, but was a potential issue for the spectrophotometric standards.  The length of the slit
was 180 arcsec, and the slit was kept oriented near the parallactic angle for each observation.  

A typical observing sequence consisted of 3$\times$900 s consecutive exposures, with the fainter stars requiring more (3$\times$1200 s)
and the brighter stars requiring less (3$\times$600 s).  We typically observed three spectrophotometric standards at the beginning
of the night, two in the middle of the night, and three at the end of the night.  Conditions were generally good, although there were some
light cirrus on the second night.  Comparison arcs were obtained at the beginning of each night, and exposures of an internal
quartz lamp were used for flat-fielding.  Observations of twilight sky allowed for corrections of the slit functions.

The reductions were carried out with IRAF, using the standard optimal extraction and cleaning algorithms.  Flux calibration was conducted separately for each night, and (after a grey shift) the residuals were $\sim 2$\%, in accord with our expectations.
The signal-to-noise ratio was low (10-15 per spectral resolution element) in the blue (4500\AA) 
but high ($>$300) in the red.  We obtained good data on 16 M31 M-type RSGs when all was said and done, plus one peculiar star
that will be discussed separately below (\S~\ref{Sec-wacky}).   These are the stars with new spectral
types listed in Table~\ref{tab:rvs}, based upon our analysis in \S~\ref{Sec-analysis}.

\subsection{New Photometry}

As alluded to above, our analysis in this paper explores using $K$-band photometry
to determine the bolometric luminosity more precisely than we have done in the past
using $V$-band photometry.  In addition to this improvement, $V-K$ colors give us 
a chance to explore
 the differences between the physical
properties determined from the MARCS models by fitting the spectral
features versus those determined from the models using the $V-K$ colors.  
In our previous work,  we found 
a metallicity-dependent difference between the two, with larger differences
seen at smaller metallicities (Paper II).  One problem with the interpretation of these values
is that the $V$-band and $K$-band observations were seldom obtained contemporaneously, either with each other or with the spectral data.  Although
$K$ appears to be relatively constant for RSGs, $V$ photometry is known to
differ significantly, by as much as a magnitude, or (in some cases) even more;
see discussion in Josselin et al.\ (2000) and Levesque et al.\ (2007).    In some cases this
$V$-band variability is due (at least in part) to variable dust extinction (Massey et al.\ 2005; 
Massey et al.\ 2007a).

All of the stars in our sample had broad band optical photometry in the LGGS,
obtained circa 2000-2001, and many of
the stars have $K$-band (actually $K_s$; see discussion in \S~\ref{Sec-analysis})
from 2MASS, obtained 1997-2000.   These are {\it relatively} contemporary with
each other, although variations in the $V$ band magnitude take place on the
order of several months, not years.   Therefore, we decided to obtain new $V$- and
$K$-band data, as we wanted photometry that was taken at nearly the
same time as our optical spectrophotometry, and because many of the
2MASS observations were near the limit of the catalog and have large errors.
We also wanted $K$-band observations obtained with a better spatial scale than the
2MASS 2 arcsec pixels given potential crowding problems within M31.  Although we have
selected stars for this study that are relatively isolated in the optical, we feared that crowding
would be substantially more of an issue in the near IR.

\subsubsection{Flamingos $K_s$ Photometry at the KPNO 4-m}

The $K_s$ photometry data described here were all obtained on 2006
October 4 UTC at the Kitt Peak National Observatory Mayall 4m
telescope using the Florida Multi-object Imaging Near-IR Grism
Observational Spectrometer (FLAMINGOS).   This instrument uses a Hawaii II
2048x2048 HgCdTe science grade array, divided into four quadrants. One
quadrant of this array was significantly noiser than the other three
quadrants. In the raw data, a reflection from the 
warm MOS exchange
unit above the detector dewar is clearly visible. Fortunately, this
reflection was removable simply by subtracting successive pairs of
frames (see below).

At the KPNO 4m, the FLAMINGOS plate scale was 0.316 arcsec
pixel$^{-1}$, corresponding to a total field-of-view of 10.8 arcmin x 10.8 arcmin in size. During these observations, the delivered image quality was
typically 2.5 -- 3.5 pixels (0.79 -- 1.1 arcsec) FWHM. Care was taken
to keep the integrated peak intensity level in our target objects plus
background below 25,000 ADU, i.e. so that all pixels were linear to
better than 0.5\%.

During our observations, the mean background level (after dark
correction) was circa 780 ADU pixel$^{-1}$ sec$^{-1}$ (or 3820 e
pixel$^{-1}$ sec$^{-1}$ for an assumed gain of 4.9 ADU
electron$^{-1}$) corresponding to an effective $K_s$ background of
12.4 mag arcsec$^{-2}$. Judging from the stability of the background
and the frame-by-frame zeropoints (see below), the night was
photometric until the end of the night. Data taken under
non-photometric conditions have been discarded.

Each field of interest was observed at least twice, with a small (less
than 30 arcsec in magnitude) telescope offset between each
repetition. Each repetition consisted of 25 five second exposures,
executed in a 5x5 square grid pattern with 30 arcsec spaces between
grid positions.

Each set of 25 frames was processed independently. Before processing
began, individual frames corrupted by detector readout errors were
rejected. Most sets contained no corrupted frames, and a few sets had one
or two bad frames.

Each uncorrupted frame was corrected for background, dark current, and
internal reflection contributions by subtracting the next frame in
the sequence. Pixel-to-pixel sensitivity variations were corrected by
dividing by a normalized dome flat. The normalized dome flat was
constructed from observations of the telescope enclosure interior with
incandescent lights on and then again with lights off. The average
lamp-off frame was subtracted from the average lamp-on frame to remove
dark current and enclosure thermal background effects.  The result was
then normalized to a median value of 1.

Each flat-field corrected frame was then geometrically aligned to the
first frame in the sequence. After alignment, an astrometric solution
for each frame was established using stars listed in the 2MASS Point
Source Catalog with $K_s$ uncertainties less than 0.03 mags while
being faint enough that the peak pixel intensity did not exceed the
linearity range of the detector. Typically, 10 - 15 such 2MASS stars
were visible in each frame.

Those same stars were used to determine a frame-specific $K_s$
photometric zeropoint. For each star, the instrumental magnitude was
measured within a circular aperture with radius 2.1 arcsecs (6.6
pixels), after measuring and subtracting the local background. Such a
relatively large radius (equivalent to approximately 2.5 $\times$ FWHM for
the typical frame) is a compromise between maximizing signal-to-noise
and minimizing the effect of frame-to-frame image quality
variations. It was chosen by testing a range of aperture sizes and
selecting a size that minimized frame-to-frame instrumental magnitude
scatter.

The frame-specific zeropoint was computed by subtracting each
instrumental magnitude from the appropriate 2MASS catalog value and
then averaging all differences. Averaging was done in an iterative
manner -- during each iteration, values with more than a 2.5$\sigma$
difference from the mean were rejected and the mean was computed
again. This process was repeated until no more values were rejected.
Typically, only one or two measurements per frame were rejected during
this process and the final standard deviation within a given frame was
less than 0.04 mag. This scatter is dominated by a combination of
2MASS measurement uncertainties and flat-fielding inaccuracies.

These frame-specific zeropoint values were then averaged together to
create a repetition-specific zeropoint value. Again, the average was
determined iteratively with 1 or 2 frame-specific zeropoints being
rejected overall. The scatter associated with the mean field-specific
zeropoints results from a combination of atmospheric transparency
("photometric") and image quality variations during the 25 frame
sequence. The repetition-specific zeropoint standard deviation for the
majority of our fields (42 out of 51) was less than 0.02 mag, another
six fields have zeropoint scatter between 0.02 and 0.06 mag, and the
last six fields have scatter larger than 0.10 as the quality of the night
began to deteriorate.

In parallel to this zeropoint determination process, the $K_s$
magnitudes of the RSG candidates were measured. Recall that two sets
of 25 dithered, five second integrations were obtained for each M31
field. Within each set, the instrumental magnitude of each RSG candidate
was measured on each frame that was free of detector readout
corruption. As above, a 2.1 arcsec radius circular aperture was
used. These instrumental magnitudes were then transformed to $K_s$ in
the 2MASS system using the appropriate repetition-specific zeropoint
value. A final mean $K_s$ value was determined from these individual
$K_s$ values using the iterative method discussed above.\footnote{The
alternative approach would have been to transform instrumental
magnitudes to $K_s$ values on a frame-by-frame basis using the
frame-specific zeropoints and then averaging those
values. Arithmetically, both approaches are essentially identical.}
Both standard deviation ($\sigma$) and the uncertainty in the mean
(i.e. $\sigma/\sqrt{N}$ where N is the number of measurements) were
computed. These uncertainities arise from frame-to-frame atmospheric
transparency variations and image quality variations combined with
photon noise (all the RSG candidates are significantly fainter than
the background).

Our final $K_s$ measurements are presented in Table~\ref{tab:dave}. For each star,
the following information is provided: LGGS identification number, the
new $K_s$ measurement with the standard deviation of the mean ($\sigma_\mu$),
and the number of frames $N$ used to measure $K_s$.  For comparison,
we include the 2MASS PSC $K_s$ values and their uncertainties, when
available, and compute the difference between our $K_s$
values and the 2MASS values, along with the total uncertainty (the sum
of the two errors added in quadrature, which is always dominated by the
2MASS uncertainty).  The final column ($N\sigma$) shows how many sigma different
our results were from the 2MASS values.  We can see from this that the 2MASS 
values always agree with ours within a few sigma.  Of course, we
believe that our new measurements are more accurate due to lower
photon noise errors (due to large telescope aperture and longer
effective exposure times) and better spatial resolution.

\subsubsection{PRISM $V$-band Photometry at the Perkins 1.8-m}

For the $V$-band photometry we availed ourselves of two nights
on the Lowell Observatory Perkins 1.8-m telescope, 2008 November 21 and 22.
We used the ``PRISM  camera", which has a 13.6 arcmin by 13.6 arcmin field
of view, and a scale of 0.39 arcsec pixel$^{-1}$.   The detector was a
Fairchild 2048x2048 device with 15$\mu$m pixels. The delivered image quality as
measured on the frames was 1.6-1.9 arcseconds.  The nights were mostly clear, although there
were signs of cirrus detected on satellite images on the second night.

We set on the same fields containing multiple RSGs as we had with the $K$-band
observations, as well as some that we had not gotten to during the $K$-band
run.  We tried to observe each field multiple times during the two nights.
Some RSGs are also found on more than one field, so typically
there are multiple observations.  In total, we obtained 277 observations
of 66 M31 RSGs from 86 $V$-band CCD frames.

Both the basic CCD reductions and photometry were done using 
IRAF.  We removed the overscan and bias structure, 
and flat-fielded the data using exposures
of the twilight sky. Photometry was then performed using a 5-pixel (2 arcsec) radius
aperture for
the M31 RSGs and calibration
stars.  The calibration stars were drawn from the LGGS catalog,
restricting the sample to isolated, bright ($V<20$), 
stars.  (An ``isolated" star in this case is taken to mean that it has no
star within 10 arcsec any brighter than two magnitudes fainter than itself,
and that there are no other stars within 2 arcsec, regardless of how bright they are,
at least as judged by the LGGS catalog.)   In general there were $\sim$100 calibration stars
on each frame that were used to determine the photometric zero-point.
The averages and the medians always agreed
to within the standard deviation of the mean ($<$0.0-0.02~mag), 
and we adopted the median as a
more robust estimator.    We list our final $V$-band photometry in Table~\ref{tab:Vphot}.

We found above that our M31 RSGs were constant at $K$ (to within the photometric errors) 
in agreement with our expectations.  But if M31 RSGs behave like Galactic and Magellanic Cloud
RSGs, we expect to find $V$-band variability on the order of several tenths of a magnitude.  We therefore compare our
new $V$ photometry to that from the LGGS in Table~\ref{tab:Vphot}. 
 Indeed, we find that in
general the photometry from these two epochs differ by several tenths of a magnitude, and,
in three cases out of 66, by 0.7-1.1~mag.

Since there is a significant difference in the delivered image quality of the Perkins
data and the LGGS, and since we measured the Perkins data with a relatively large
(2 \arcsec radius) aperture compared to the PSF-fitting from the LGGS data, we 
checked whether or not there were neighboring stars listed in the LGGS catalog
which could have
contaminated our Perkins results.  We found 7 such instances where the 
contamination in 2\arcsec seeing would increase the measured flux by 0.02-0.05~mag
(none showed greater contamination than 0.05~mag).
 We indicate those in Table~\ref{tab:Vphot}, where we have retained the Perkins measurements but increased the
listed errors by the amount of the contaminations;  we do not consider stars so flagged 
to have their variability established.

Using data from the All Sky Automated Survey (ASAS) project (Pojmanski 2002)
and their own photometry, Levesque et al.\ (2007) investigate the V-band
variability a sample of ``normal" RSGs
in the Magellanic Clouds, for comparison to several peculiar RSGs.  They found that in
general the normal RSGs in the SMC and LMC varied by (on average) about 0.9~mag in $V$.
That number presents the full range (bright-to-faint) in the well-sampled photometry.  Here
we have only two observations, the current data and the LGGS.
 Obviously from simple sampling considerations, we expect
that the full range of variability will be larger than the difference of just two observations.
If the actual variations were sinusoidal we would expect the full range to be 2.6 times
greater (on average) than the differences from our observations.  If the light variations
were purely random then our two observations would underestimate the true range
by a factor of 3.2.  The average (absolute) difference we find in Table~\ref{tab:Vphot} is
0.16~mag, excluding the ones where crowding might affect the results. Thus we expect that
on average the full range of variability is $3\times$ this value, or 0.5~mag.  This is smaller,
but comparable, to the photometric variability see in the Magellanic Cloud RSGs.

\section{Analysis}
\label{Sec-analysis}

\subsection{Physical Properties from the Spectra}
\label{Sec-props}

We began by assigning spectral types to the 17 stars with spectroscopy.  For this, we used our previous 
spectra (Papers I and II), which were of comparable dispersion and wavelength coverage.  
For one star, J004047.84+405602.6, we failed utterly in assigning even
a spectral type; this star is discussed separately below (\S~\ref{Sec-wacky}).

For fitting the spectra, we used MARCS stellar atmosphere models computed for a metallicity that was
2$\times$ solar, in accordance with the analysis of the oxygen abundances of 
HII region data by Blair et al.\ (1982) and Zaritsky et al.\ (1994). We note that there have been some recent 
suggestions that the metallicity in M31 is nearly solar instead: abundance studies of several A- and B-type supergiants by Venn
et al.\ (2000) and Smartt et al.\ (2001) found oxygen abundances that were essentially solar, not twice that.  However,
using the improved photometry of these stars provided by the LGGS, N. Przybilla (2007, private communication)
now derives significantly higher
abundances for some of these stars (see comments following Massey et al.\ 2008, and discussion below in \S~\ref{Sec-hrd}).

The MARCS models have been described in some detail by Gustafsson et al.\ (1975, 2003), Plez et al\ (1992), 
Plez (2003), and most recently by Gustafsson et al.\ (2008).  These are static, LTE, opacity-sampled models that have
up-to-date opacities (including TiO, VO, and H$_2$O) and include sphericity.  The grid of models we used covered
3300-4600 K, and had been interpolated to 25 K using models computed at 100 K intervals.  The surface gravities 
included $\log g=-0.5$, 0.0, and +0.5.  The model spectra had been smoothed to match the 7 \AA\ resolution of our spectra.

 In performing the fit, we followed the same
procedures as in Papers I and II.  The TiO band strengths are sensitive to the effective temperature, and once we have adopted
an effective temperature, then the amount of reddening needed to match the continuum shape gives the value of $A_V$.
 For each star, we began with a $\log g=0.0$ model, and attempted to match the depths
of the molecular bands (primarily TiO) by trying various effective temperatures.  (The TiO band strengths have little sensitivity
to the adopted $\log g$.) At the same time, we reddened the model spectra
by various amounts of $A_V$ using the Cardelli et al.\ (1989) reddening law, with $R_V=3.1$\footnote{Although a much larger
value, such as 4.2, is needed as the {\it effective} $R_V$ for broad-band colors of RSGs, a value of $R_V=3.1$ is appropriate
for spectrophotometry under average Galactic conditions; see Massey et al.\ (2005).  Although we do not have any knowledge
of the appropriate $R_V$ for the dust in M31, we do note that the UV extinction law curve is similar to that in the 
Milky Way (Bianchi et al.\ 1996), and so this is as good a choice as any for now.}.  Once we were satisfied with the fit (as judged
by eye using an interactive IDL program) we then computed the star's other physical properties.  The bolometric magnitude
$M_{\rm bol}$ followed from: $$V_0=V-A_V$$ $$M_V=V_0-24.40$$ $$M_{\rm bol}=M_V+{\it BC}_V,$$
where we adopted a distance of 0.76 Mpc (distance modulus of 24.40), following van den Bergh (2000) and references therein.
For the bolometric corrections BC$_V$ we adopted a simple fit to the output of the MARCS models:
\begin{equation}
{\rm BC}_V=-330.228+239.964(\frac{T_{\rm eff}}{{\rm 1000\ K}})-58.51434(\frac{T_{\rm eff}}{{\rm 1000\ K}})^2+4.77590(\frac{T_{\rm eff}}{{\rm 1000\ K}})^3, \label{equ-BCV}
\end{equation}
 very similar to the relations found in Papers I and II.   The fits should be good to a few hundredths of a magnitude.
The stellar radius then follows, by definition, from the bolometric luminosity: $$R/R_\odot =(L/L_\odot)^{0.5} (5770/T_{\rm eff})^2,$$
where we adopt $M_{\rm bol}=4.75$ and $T_{\rm eff}$=5770 K for the sun.

At this point in the process we asked ourselves if $\log g=0.0$ was reasonable, or if $+0.5$ or $-0.5$ would have been more
appropriate.  To answer this question we used the bolometric luminosity to obtain a {\it very} crude estimate of 
the mass $m$ based upon a heuristic examination of evolutionary tracks: $\log (m/m_\odot) \sim 0.50 - 0.10 M_{\rm mbol}$.
The corresponding surface gravity then follows from $\log g=4.438+ \log (m/m_\odot) - 2\log (R/R_\odot)$.  If the computed
$\log g$ was closer to 0.5 or -0.5 than to 0.0, we refit.  Only one star were affected, J004424.94+412322.3, which required $\log g=-0.5$.
The new fit did not change the effective temperature, but did change the required $A_V$ by 0.5~mag.  As a result, we 
rechecked the new parameters to see if they were consistent with the lower $\log g$, and they were.

\begin{figure}
\epsscale{0.28}
\plotone{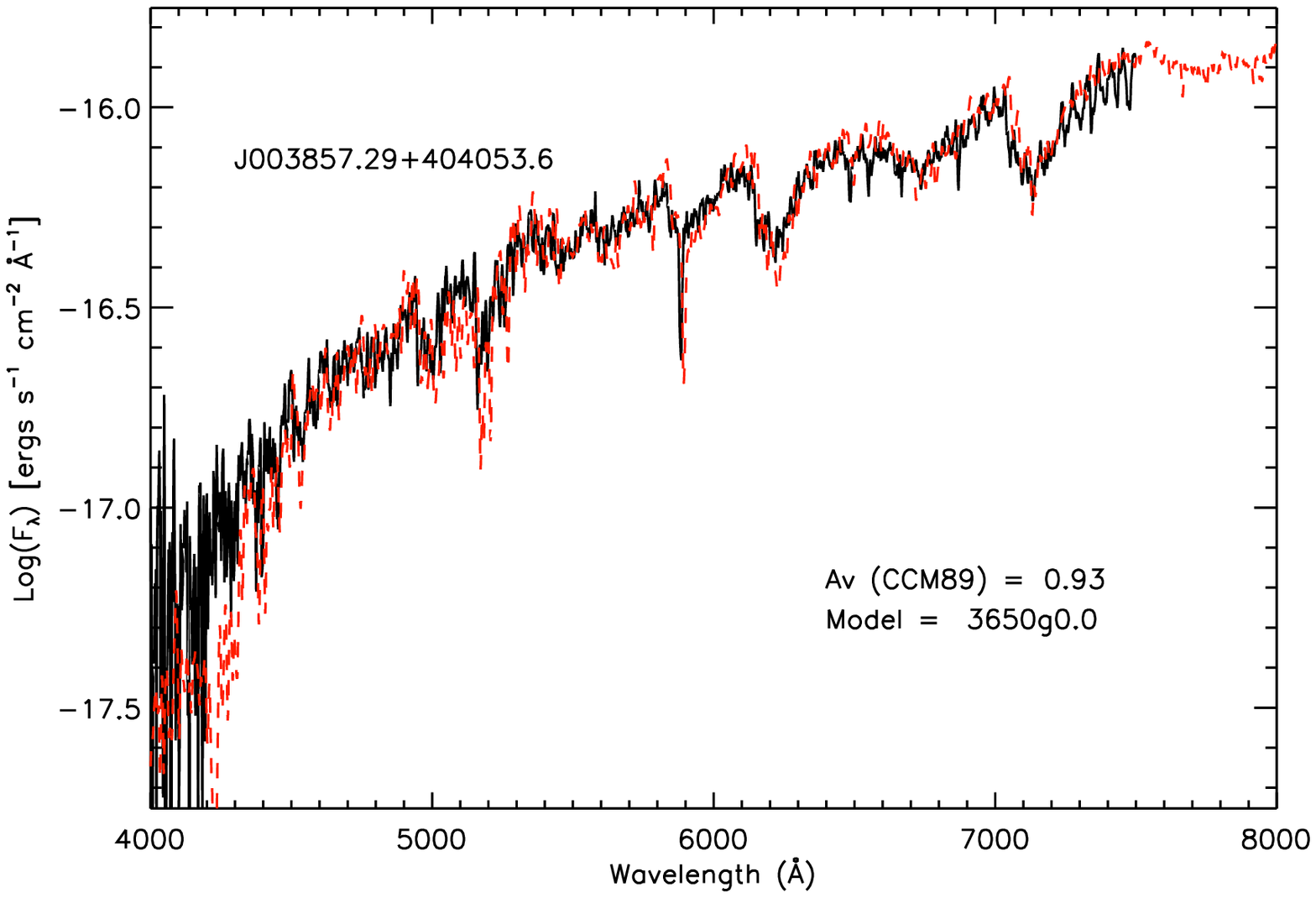}
\plotone{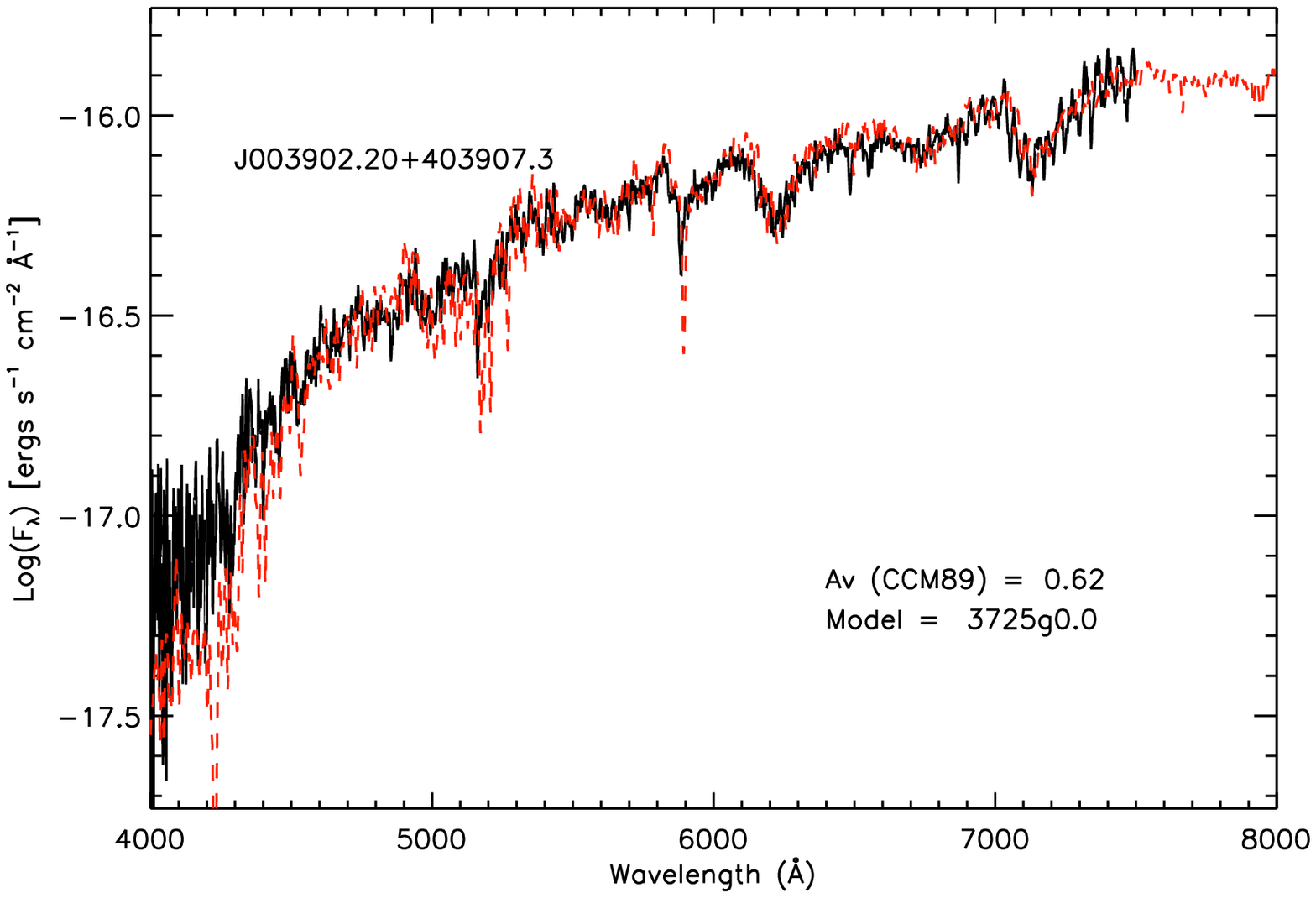}
\plotone{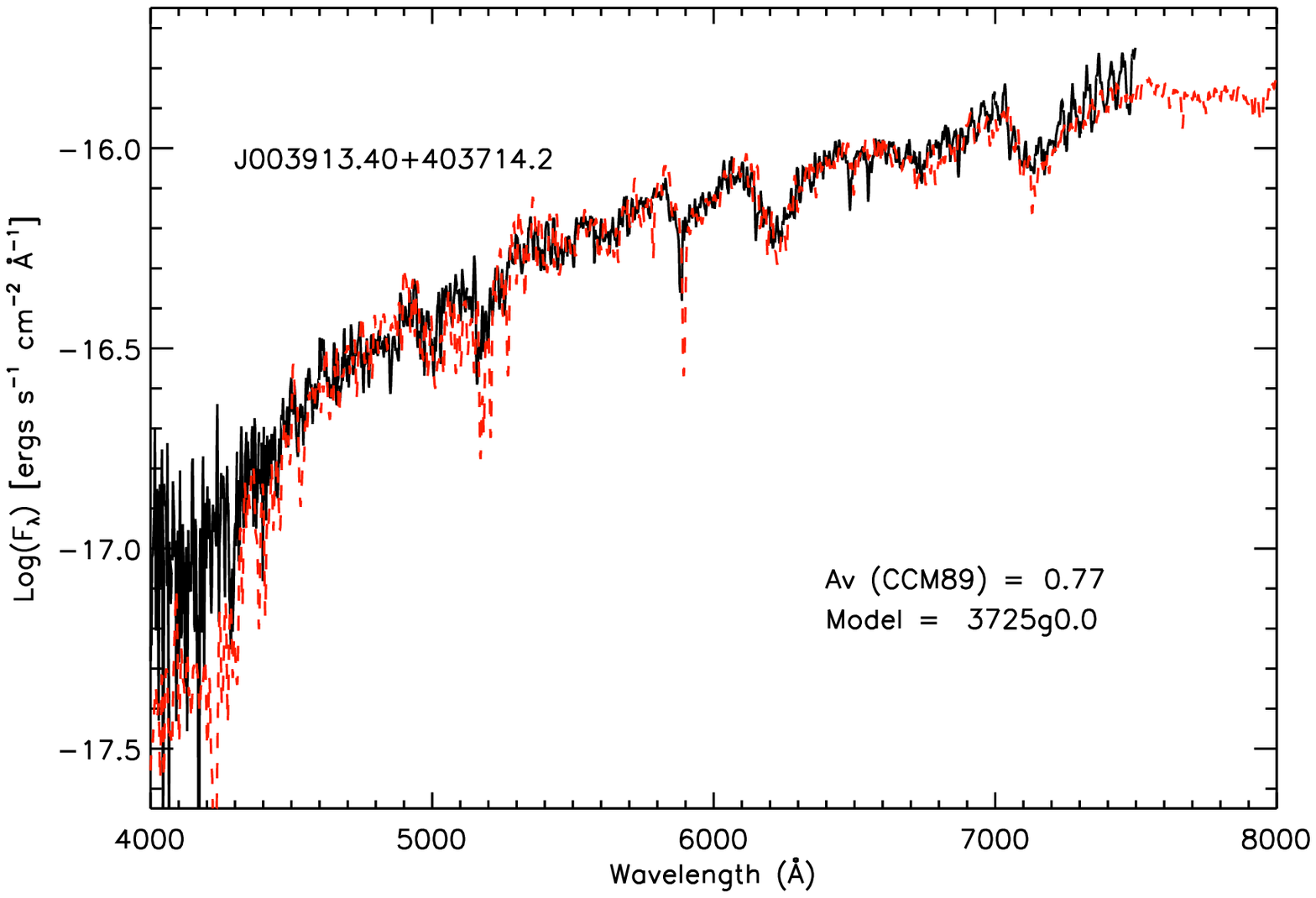}
\plotone{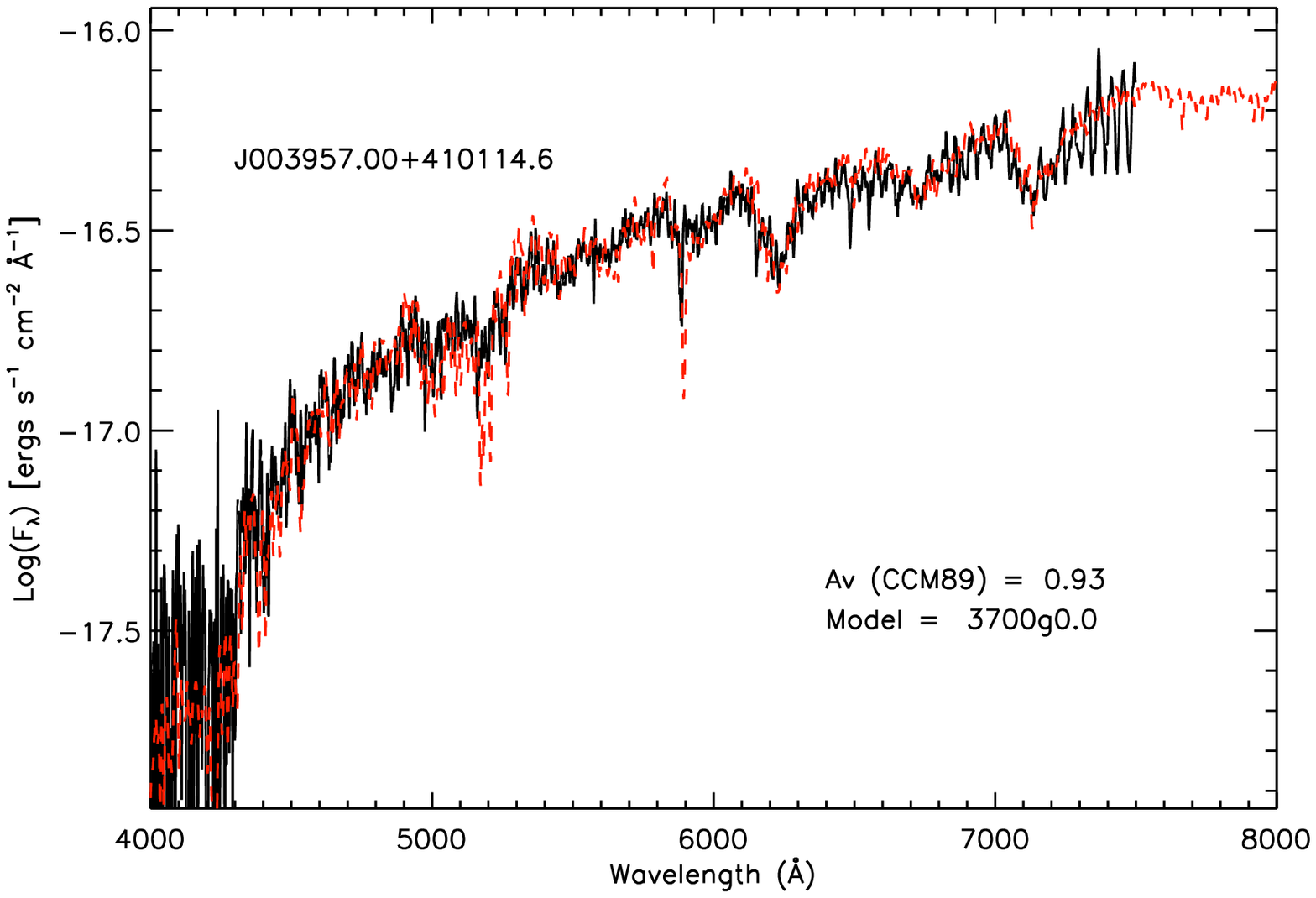}
\plotone{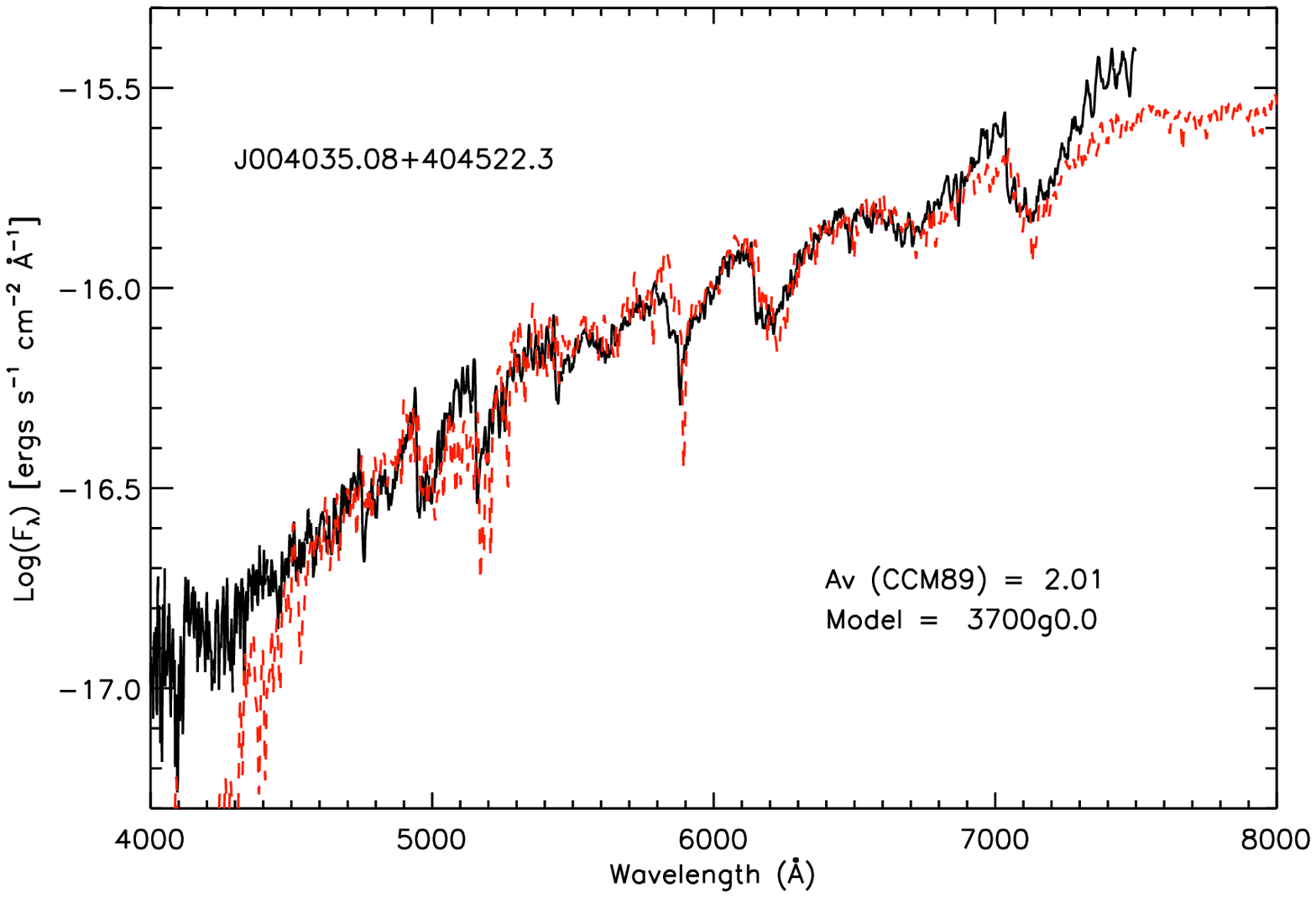}
\plotone{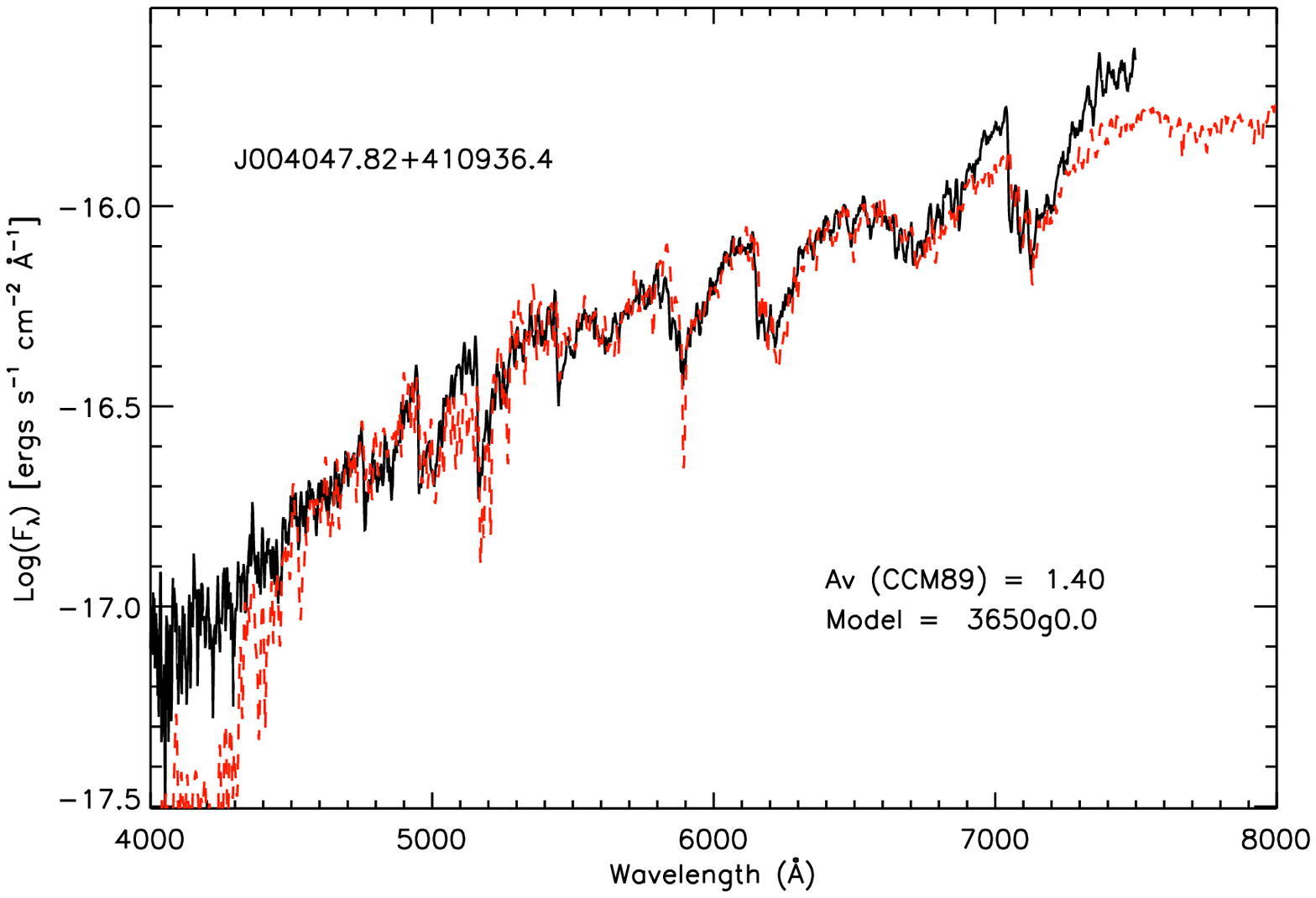}
\plotone{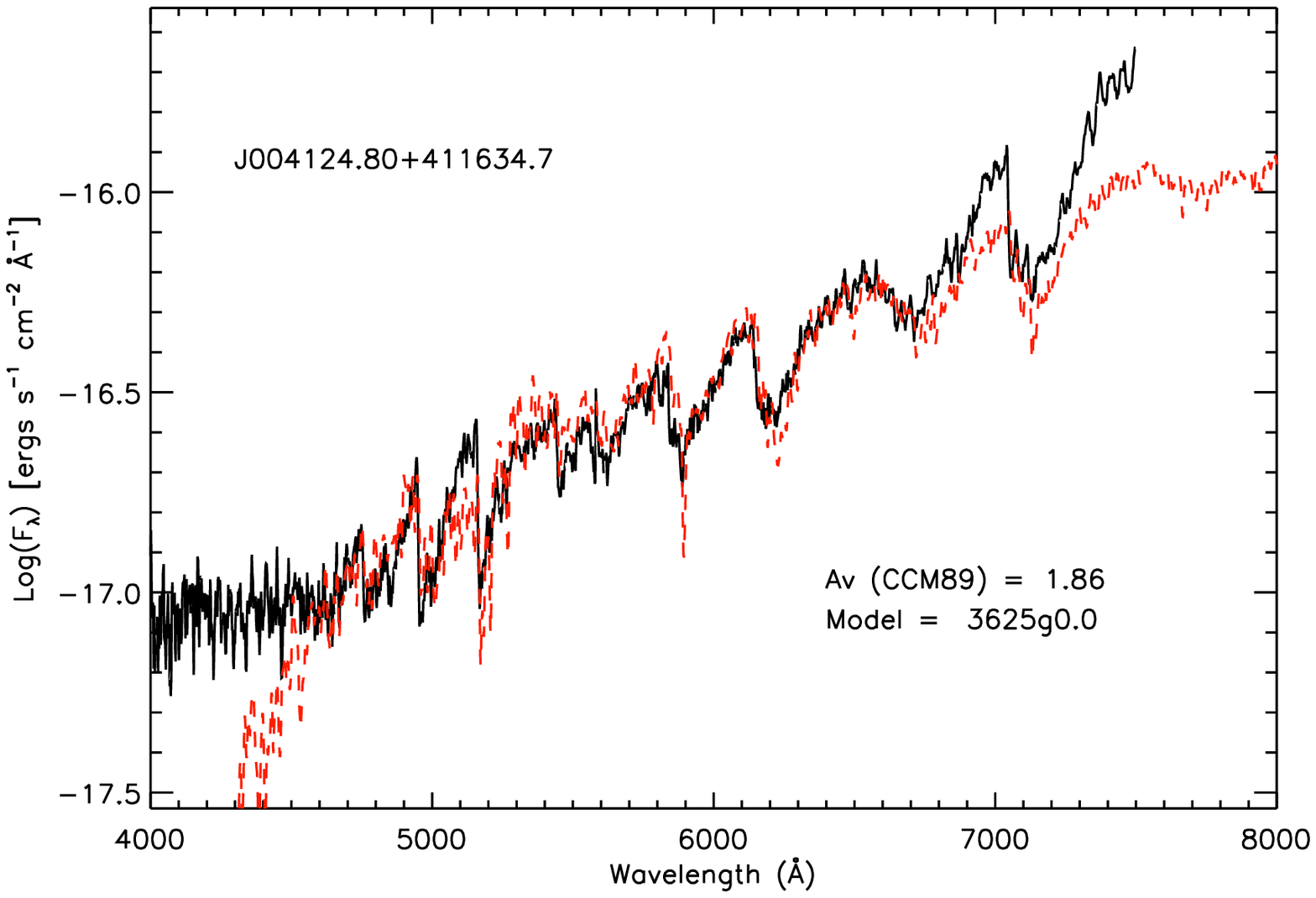}
\plotone{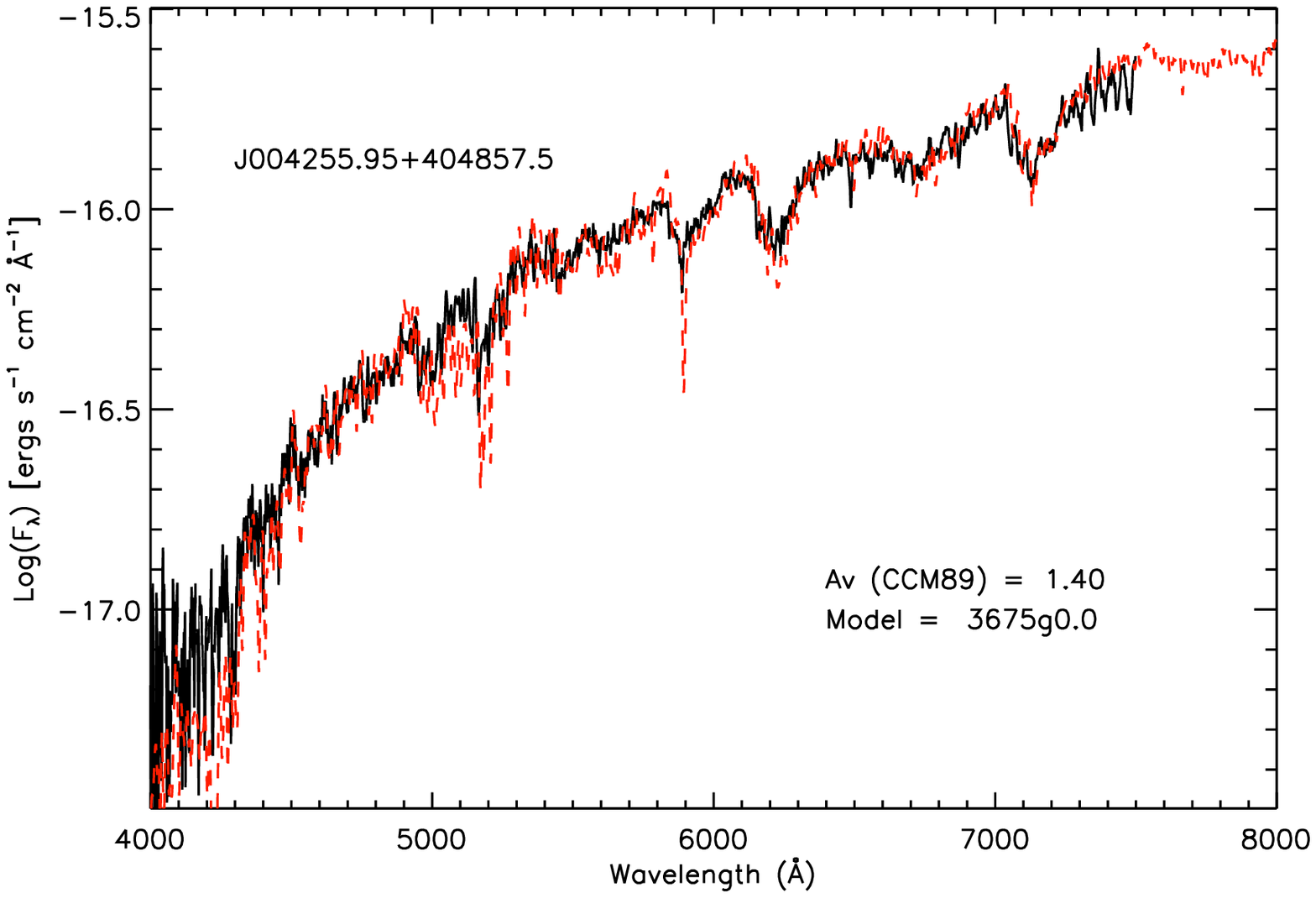}
\plotone{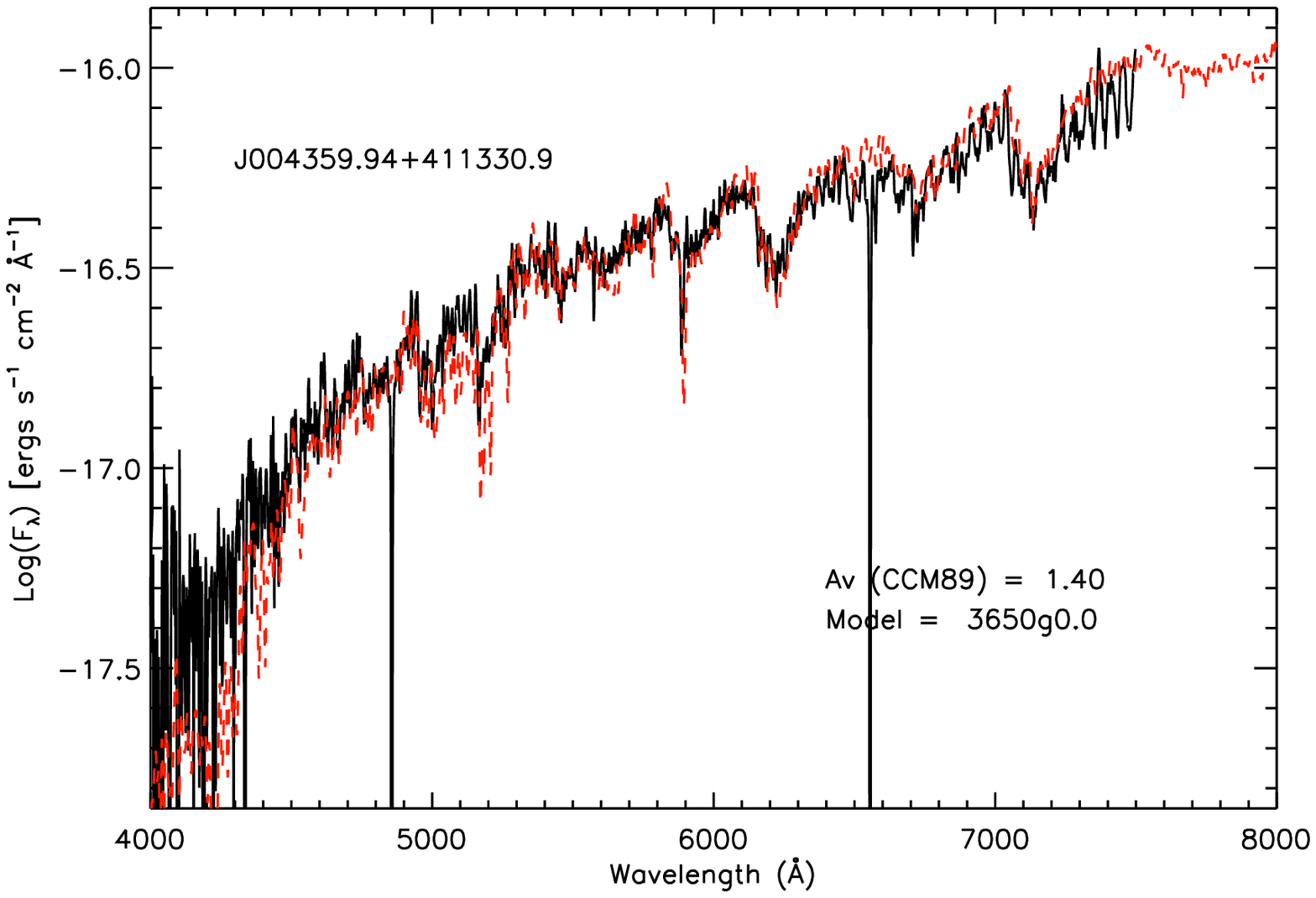}
\plotone{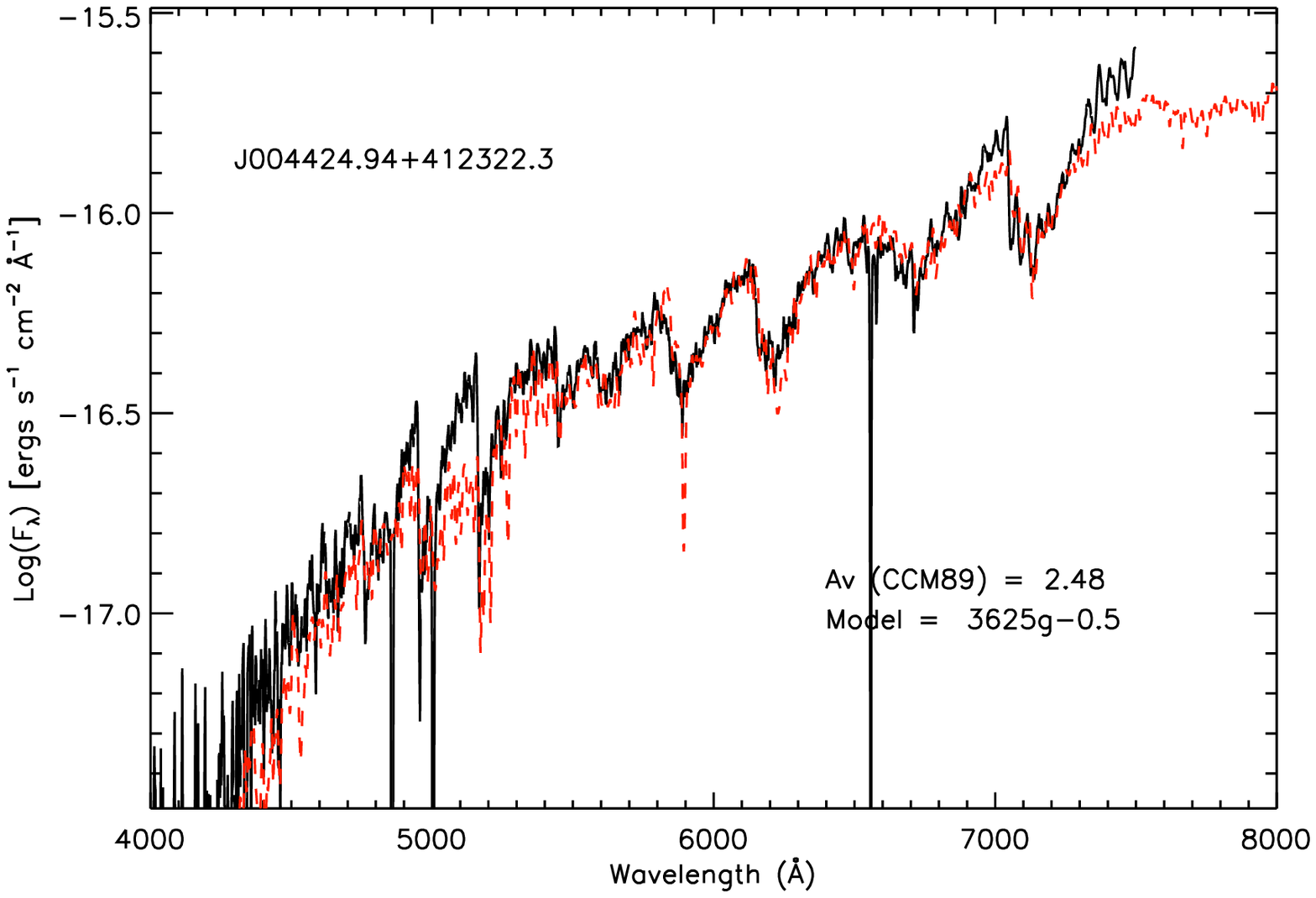}
\plotone{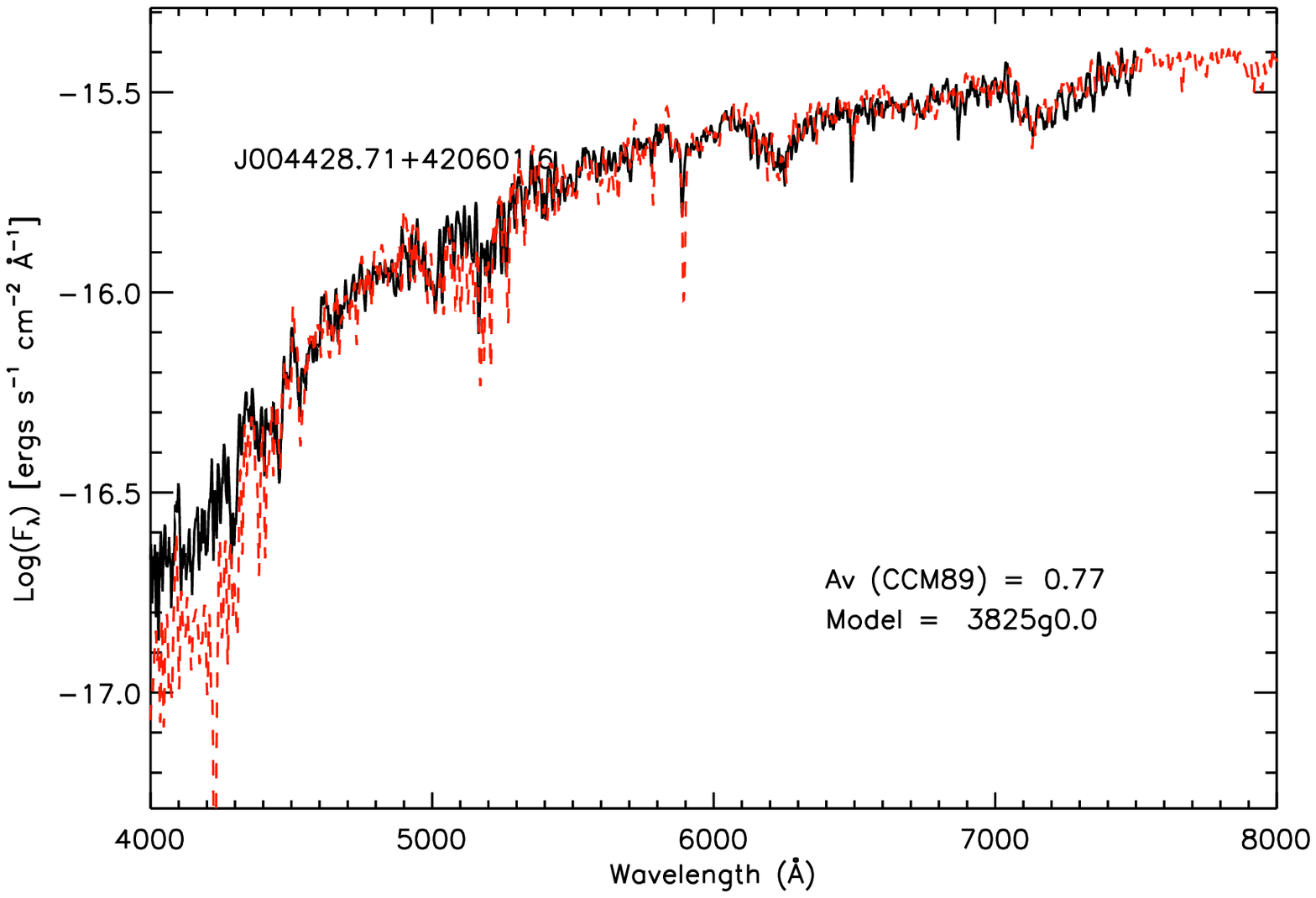}
\plotone{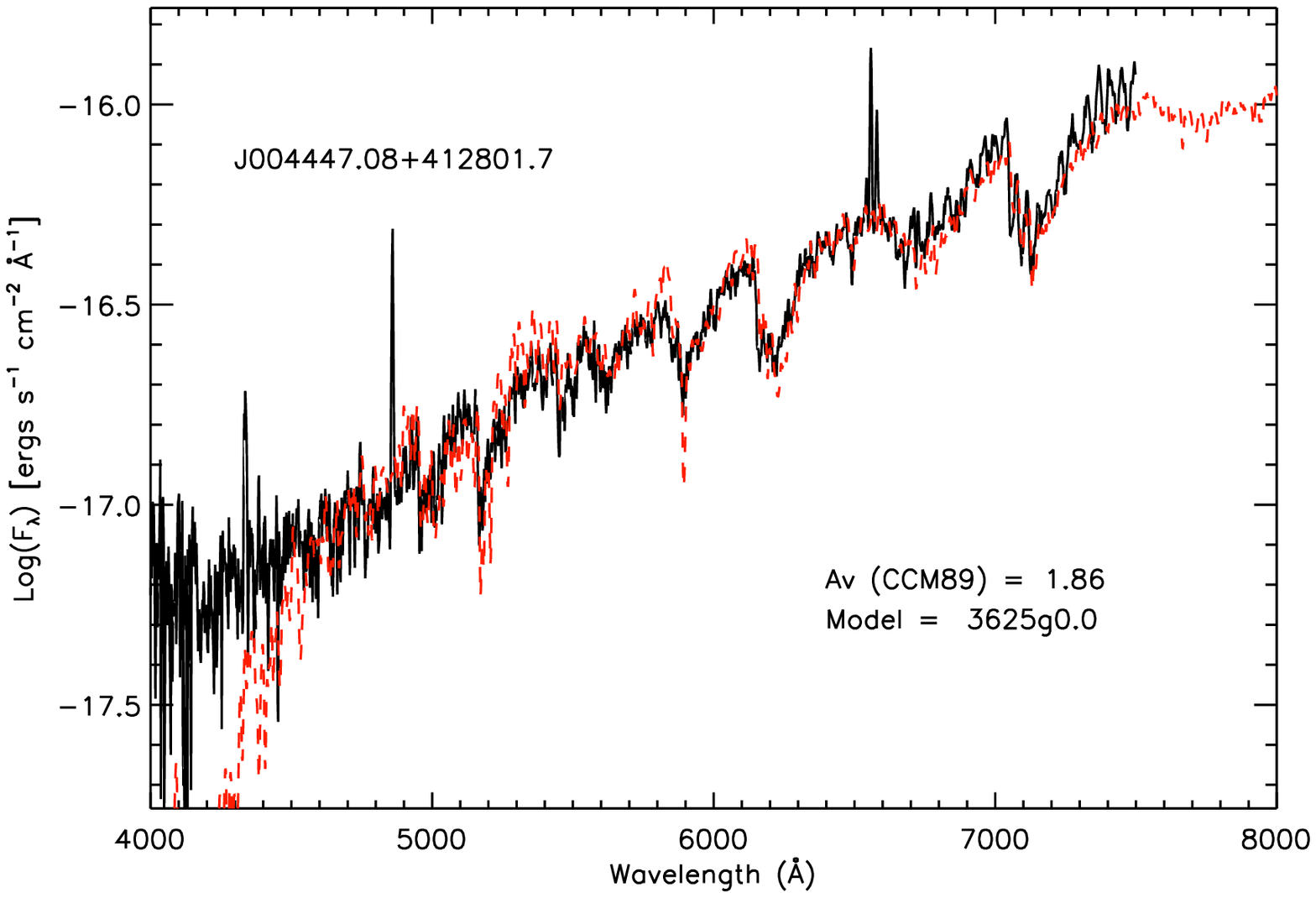}
\plotone{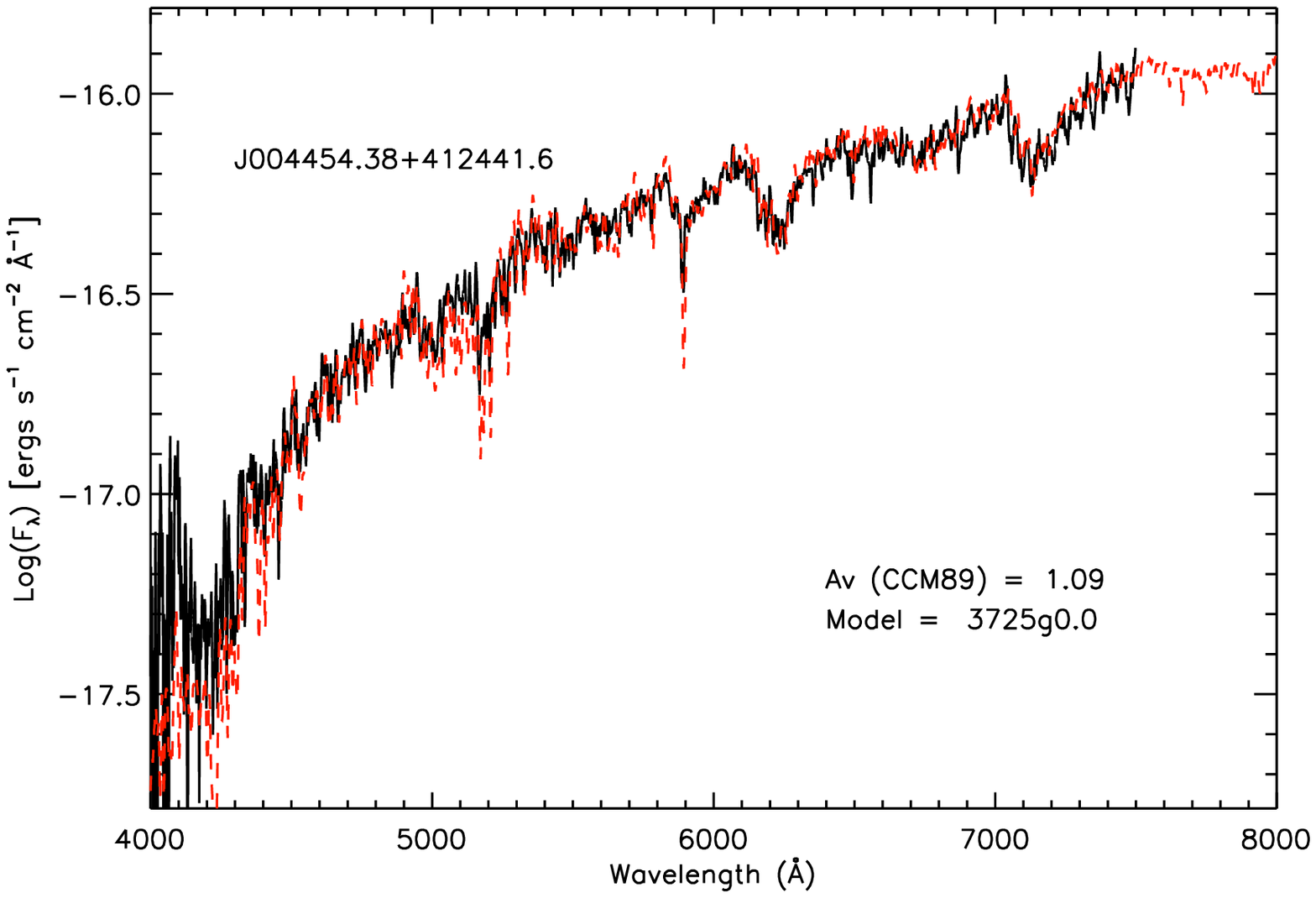}
\plotone{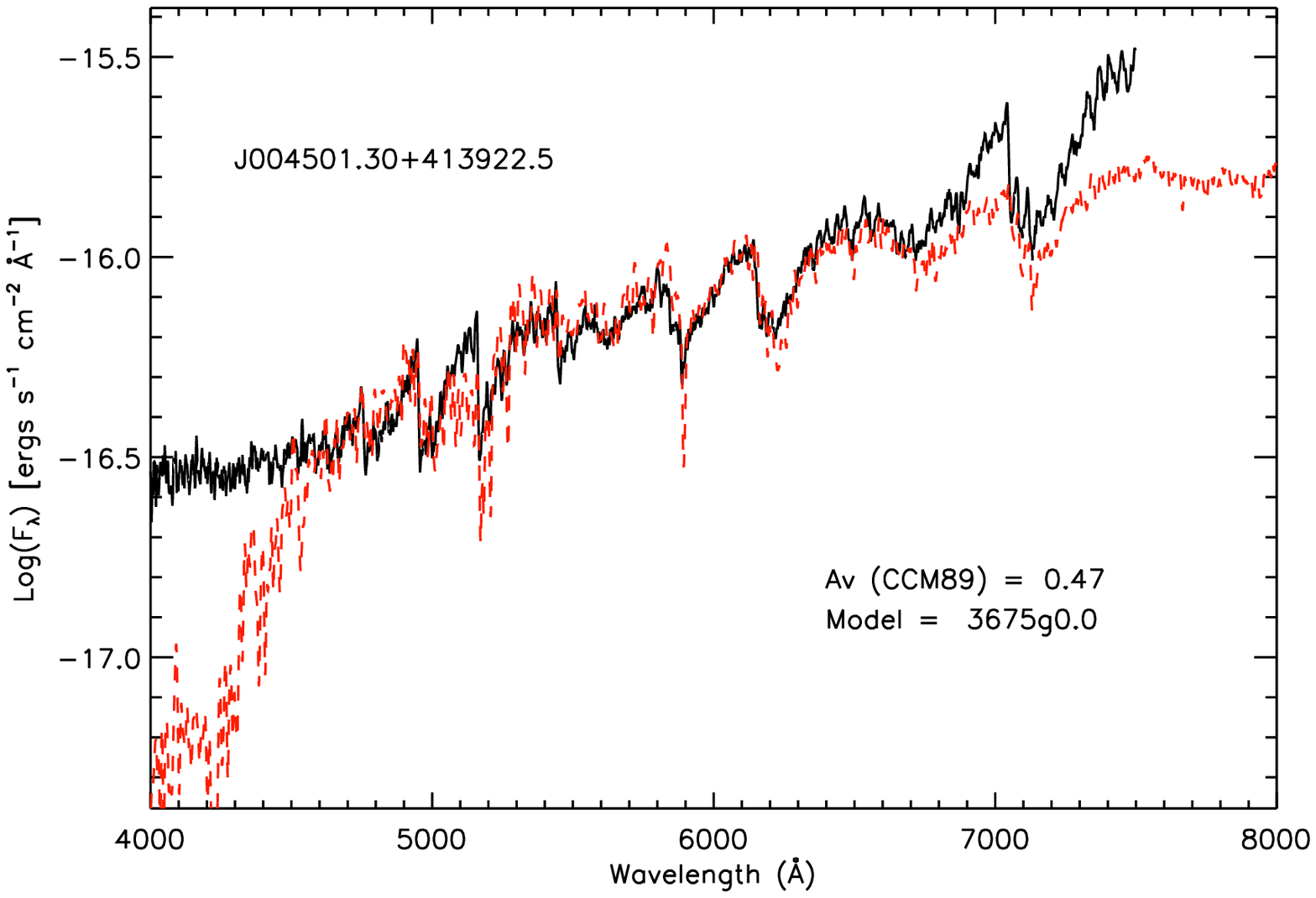}
\plotone{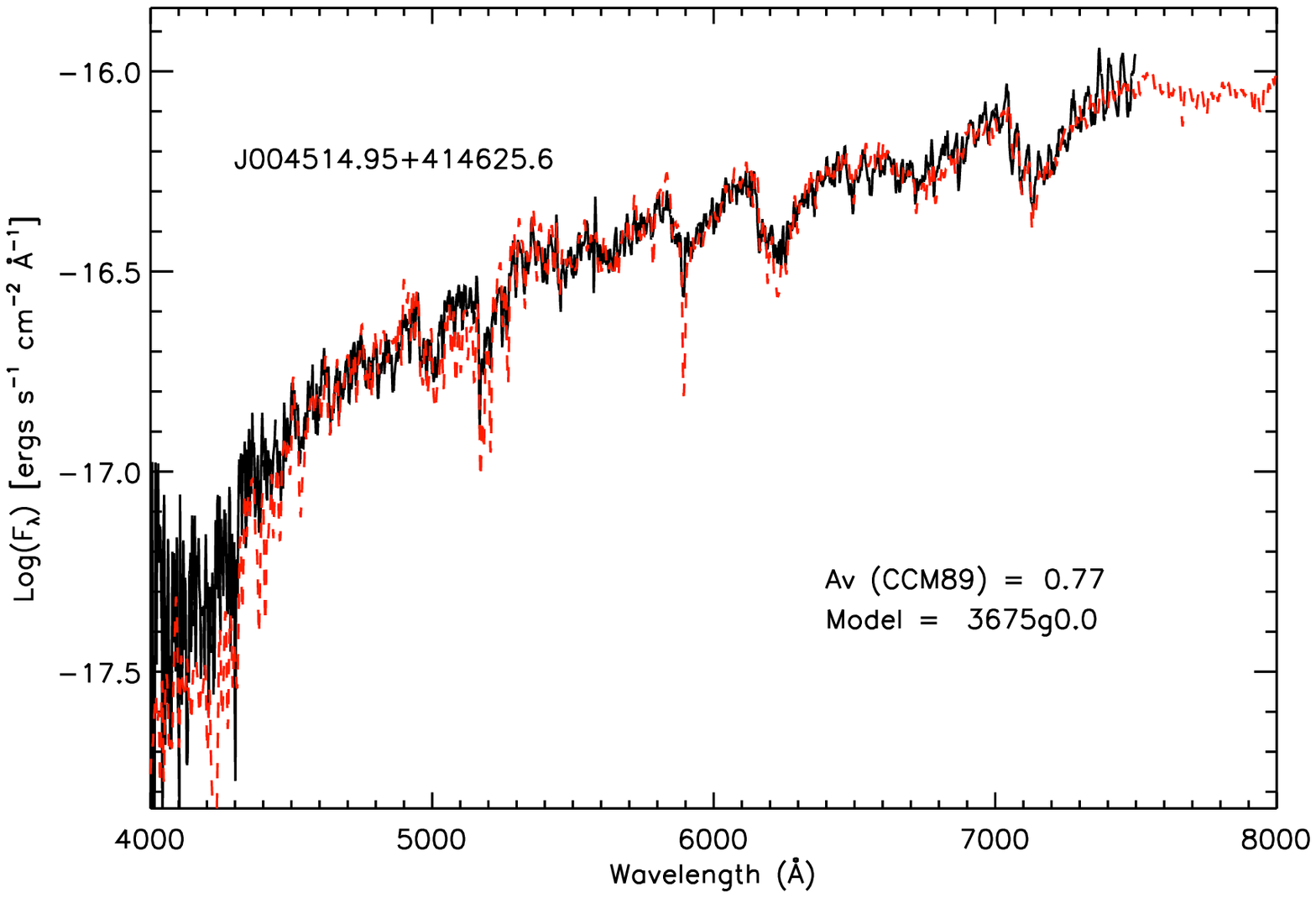}
\plotone{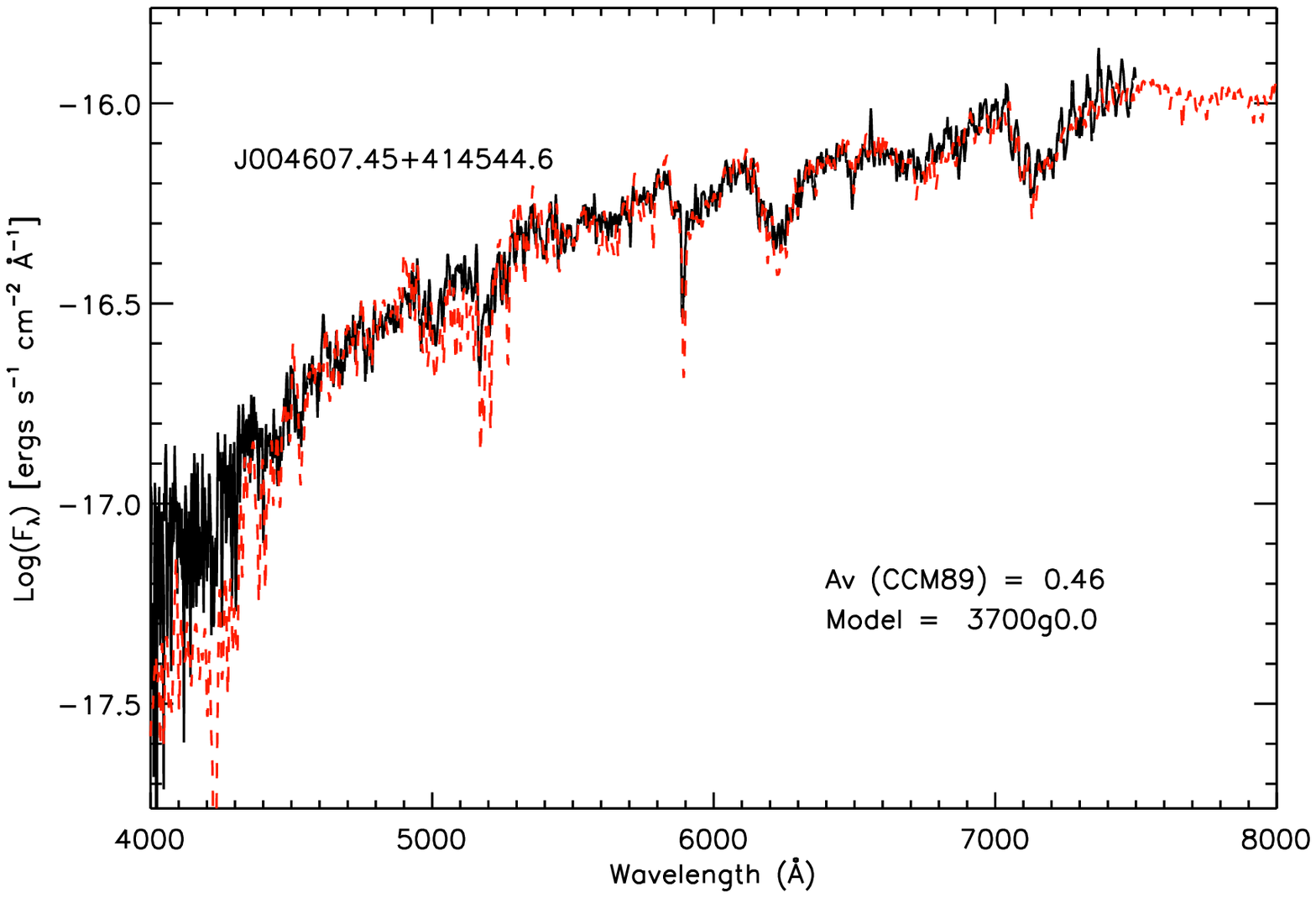}
\caption{\label{fig:fits} Fits of the MARCS stellar atmosphere models to our optical spectrophotometry. The observed spectral energy distributions are shown in black, and the adopted model fits are shown in red. The models have been reddened by the indicated amount using the standard $R_V = 3.1$ reddening law of Cardelli et al. (1989). Included are our relatively poor fits to the observed spectra of J004124.80+411634.7 and J004501.30+413922.5, which we conclude are likely binaries with  hot companions (\S~\ref{Sec-props}).
}
\end{figure}

We list the derived physical properties in Table~\ref{tab:spect}.   These stars are typically less luminous and smaller in radii
than our Galactic sample, which is in accordance with stellar evolutionary theory, as we discuss in the next section.
We show the fits in Fig.~\ref{fig:fits}.  In general, they are very good.  Our estimated uncertainty in fitting these spectra
are 25~K for $T_{\rm eff}$ and $0.10$ in $A_V$.   (The reader is referred to Fig.~3 of Levesque et al.\ 2009 for an example of
the sensitivity of the TiO bands to small changes in effective temperature.) 

However, there are two stars for which we obtained very poor fits, as shown in Fig.~\ref{fig:fits}:
J004124.80+411634.7 and J004501.30+413922.5, both with assigned spectral types of
M3~I.  A number of stars show a near-UV excess
compared to the models (as described below) but these two stars have a much worse problem in the NUV than the others, and,
more revealingly, there is considerably more flux in the far red than the models predict when fit to the center.  We will find in the
next section that these stars have considerably cooler temperatures according to their $V-K$ colors than the spectral fitting
allows.  This is consistent with these stars having hot companions, as the continuum of the hot star would tend to fill in
the TiO bands resulting in a  relatively poor fits, as shown in Fig.~\ref{fig:fits}.    Although we estimate the uncertainties in
the fitting parameters based upon our fitting, we consider these physical properties to be undermined.   In our Galactic and
Magellanic Cloud samples, our resolution and signal-to-noise in the NUV was good enough to be able to detect the upper
Balmer lines in stars we judged to have hot companions; here neither is sufficient for this direct determination. 

Note that these poor fits cannot simply be due to an $R_V$ value that differs significantly from the average 3.1 Galactic
value.  To produce a reasonable fit to the blue of 4000 A,  one would need
unrealistically large values ($\sim 7$).  In addition,  increasing the value for $R)V$ would lower the continuum at the far red end of the spectrum;
i.e., one can raise either the blue end of the fit, or the red end of the fit (but not both) by varying $R_V$.

Two other stars, J004035.08+404522.3 (M2.5 I) and J004047.82+410936.4 (M3 I) have
fits that are pretty good, but not great; in both cases the TiO band at $\lambda 7055$
is too weak in the model spectrum (Fig.~\ref{fig:fits}).  Going to a cooler model, though, would create worse
mismatches to other TiO bands.  Possibly these two stars are also affected by binary
companions but here we treat them as singles.

Many of the stars with good fits (longwards of 4200-4500 \AA) do show show the ``near-UV problem" alluded
to in Paper I, and discussed in detail in Massey et al.\ (2005),
These stars show more flux in the near-UV than the models
predict.  In discussing this previously we eliminated a number of possibilities, including the presence of a hot companion, as the
Balmer lines are not evident in the NUV/blue as they would be otherwise. Here we cannot use that test as our spectral resolution
is significantly lower. Instead, we see the same effect here as discussed
for the Galactic sample: the stars with the greatest NUV problem are also the ones with the largest values $A_V$ in Table
~\ref{tab:spect}.  Although there are certainly dusty locales in M31, in general the early-type stars have $A_V=0.4$, corresponding
to $E(B-V)=0.13$. Many
of the stars in Table~\ref{tab:spect} have values far in excess of this.  The median $A_V$ value is 1.0~mag (ignoring the two stars we suggest are binaries) ; the highest
value is 2.5~mag. For the Milky Way 
and Magellanic Clouds we found a similar enhancement in $A_V$ towards RSGs, and attributed
this to circumstellar dust, which we expect to be formed about these stars (Papers I, II, and Massey et al.\ 2005).  In fact, given the observed dust production rates
in Galactic RSGs, we expect that there should be a magnitude or more of extra extinction (Massey et al.\ 2005).  
The NUV excess is then easily 
explained as a scattering phenomenon
(Massey et al.\ 2005).

How much difference is there between stars of a given subtype in M31 and lower-metallicity systems?  There are six stars in M31 that we classified as M2~I;
the average effective temperature of these six is 3680~K ($\sigma_\mu=12$~K), and the median is 3675~K.  
This can be compared to 3660~K in the Milky Way, 3625~K in the LMC, and
3475~K in the SMC (Papers I and II).    On average we see that M31 M2~I stars are 20$\pm12$~K warmer than their Milky Way counterparts.   None of the other spectral
types contain more than 2 members.  The range in  effective temperatures for the M2~I stars in M31 is reasonable: 3650 to 3725~K, which makes sense if one considers
that the uncertainty in classifying a star is one spectral subtype, and that the average difference (in the Milky Way) in the effective temperature between M2 and the
adjacent types M1.5 and M2.5 is -50~K and +45~K, respectively, and that our temperature determinations are good to 25~K.  

Although this result is marginal, no larger difference is expected based upon the models.  Between the LMC and the Milky Way (a factor of 2 in metallicity)
we see a -50~K change for RSGs.  Here we measure a $20\pm12$~K change between the Milky Way and M31, another factor of 2 in metallicity.  We can see visually from the
models themselves that there is a saturation effect in the TiO band depths with effective temperatures, and that we would not expect as large a difference 
between the Milky Way and M31 as between the LMC and the Milky Way. In Fig~\ref{fig:models} we show a comparison between the MARCS models of varying metallicities for an effective temperature of 3700~K and a $\log g=0.0$.
Where there is a significant difference between the $0.5 Z_\odot$ (LMC-like) and $1.0 Z_\odot$ (Milky Way-like) models, the difference between
the $2.0 Z_\odot$ (M31-like) and $1.0 Z_\odot$ models is smaller.  We show below (\S~\ref{Sec-hrd}) that were we to have fit our spectra
with a solar abundance (rather than $2\times$ solar) model, the differences we derive would be -30~K to those we derive here, again consistent with
there being only a modest abundance effect in the effective temperature between solar and twice solar.

\subsection{Physical Properties from the Photometry}
\label{Sec-vmk}

The optical spectra offer us the advantage of fitting the effective temperature from the depth of the TiO bands, which then permit
an accurate measurement of using the continuum shape to determine $A_V$ by reddening the models, at least to within the
uncertainties of our assumptions (i.e., $R_V=3.1$, metallicity, etc.).  However, in Papers I and II we found there was also
an advantage to use the broad-band photometry to derive the physical properties, with $V-K$ determining the effective temperatures,
couple with the K-band luminosity to derive the bolometric luminosity. This serves as a useful complement
on the optical spectroscopy.  Although it is not completely independent (we adopt the value of $A_V$), the extinction in $A_K$ is
small (about 10\% that of $A_V$), and the bolometric correction at $K$ small and fairly insensitive to the temperature.  Furthermore,
this allows us to test the self-consistency of the MARCS models---do we get the same answers using the optical spectrometry as
with the broad-band colors?   In this regard, we note that we found a systematic problem in Paper II, 
that the effective temperatures derived from $(V-K)_0$ colors were on average
a bit higher than those derived from the optical spectrophotometry; the differences amount to
105 K for the Milky Way and LMC, and 170 K for the SMC.   We attributed this to an intrinsic limitation of  1-dimension model
atmospheres.  
Since the problem was larger for the lowest metallicity
galaxy, the SMC, we were curious to see what it was for M31, where the metallicity is thought to be higher than solar.

We begin by converting $K_s$ to $K$, using the relationship given by Carpenter (2001): $$K=K_s+0.04.$$  Next, we dereddened
the photometry using
$$(V-K)_0=V-K-0.88A_V, $$where the $A_V$ values were taken from the spectral fitting.  We note that this makes the results of our two methods
slightly intertwined.   We then determined
a theoretical relationship between $T_{\rm eff}$ and $(V-K)_0$  using the MARCS models and the filter bandpass and photometric
zero-points laid out by Bessell et al.\ (1998).  The results were similar to those found at lower metallicities in Papers I and II:
\begin{equation} T_{\rm eff}=8130.9 - 2113.22 (V-K)_0 + 327.883(V-K)_0^2 - 17.7886(V-K)_0^3. \label{eq:vmk} \end{equation}

Before comparing our effective temperatures to that from our spectrophotometry, let us briefly consider the uncertainties involved in these
quantities.  The photometric errors will result in an uncertainty in $(V-K)$ of 0.02-0.10~mag.  The typical uncertainty in
$A_V$ is 0.10~mag, resulting in a total uncertainty in $(V-K)_0$ of (roughly) 0.10-0.20. 
The uncertainty in $T_{\rm eff}$ derived from $V-K$ is 
$$\Delta T_{\rm eff}=-2113.22 +655.766 (V-K)_0-53.366 (V-K)_0^2) \Delta (V-K)_0.$$  So, for the warmest stars, with
$(V-K)_0=3.0$ the typical uncertainty in $T_{\rm eff}$ is about 100 K, while for the coolest stars, with $(V-K)_0=6.0$, it will
be about 20~K.  These values are remarkably similar to our uncertainties in fitting the spectra, which we estimate above
to be about 25~K for stars with strong TiO (i.e, the coolest stars).  In Papers I and II we conclude our uncertainty is about
100~K for the early K-types, and so the errors we obtain from $(V-K)_0$ are comparable to 
what we obtain from
the spectrophotometry.   We note that if we
lacked contemporaneous photometry, the errors associated with these $T_{\rm eff}$ would be 2 to 3 times larger, given
that the median variability in $V$ is about 0.2~mag, as judged from Table~\ref{tab:Vphot}.  Furthermore, if we lacked
spectrophotometry---so that we were forced to adopt some average value for the extinction $A_V$---then the uncertainty
in effective temperature would be much, {\it much} greater, given the range in $A_V$ we see in Table~\ref{fig:fits}.  We will return to this point
in \S~\ref{Sec-hrd}.

In Fig.~\ref{fig:irtemps} we compare the effective temperatures derived from $(V-K)_0$ with those from our spectral fitting.
The solid line shows the 1:1 relationship.  We see that in general the agreement is quite good.  Ignoring the two alleged binaries
(at the bottom) and the two stars with large uncertainties at the top, the median difference is +20~K (in the sense of spectral
fitting effective temperature minus $V-K$ effective temperature).  But clearly the dispersion is large.  In Papers I and II we
were able to do a far better comparison as the sample size was much larger, and here we restricted the sample to stars
that (we thought) would have similar effective temperatures.

There are two stars with small $V-K$ values (implying warm effective temperatures).
For J004428.71+420601.6 (the warmest star based on $V-K$) we have a good fit and the depth of the TiO bands in this star
precludes the warmer temperature implied by the $V-K$ colors.  It is true that the  $V$-band photometry that we adopt from
this star is  0.76~mag brighter than the LGGS value, leading to a significantly warmer
temperature: had we adopted the LGGS value, we would have derived an effective temperature of 3800 K from the $V-K$
photometry, in excellent agreement with the 3825 K value we derive from the spectrum.  We used the brighter value as it was
obtained closer in time than that of the LGGS, but this variability underscores the sort of uncertainty we are dealing with.
 Finally, the most discrepant result is for J004359.94+411330.9, an M2 I.  The fit for this star is also quite good. 
 We speculate that for the two unexplained outliers perhaps our ``contemporaneous" photometry wasn't quite
 contemporaneous enough.
 
We can perform a similar comparison for the bolometric luminosities.  
The bolometric magnitude derived from the K-band photometry is then found by: $$K_0=K-0.12A_V$$
$$M_K=K_0-24.40$$
$$M_{\rm bol}=M_K+BC_K,$$
analogous to the set of equations we used in the previous section.  Here 
the bolometric correction for $K$ will
be a positive number (see discussion in Bessell et al.\ 1989 and Paper I).  We derive the following from the MARCS stellar
atmospheres:
$${\rm BC}_K=7.149-1.5924(\frac{T_{\rm eff}}{1000\ {\rm K}})+0.10956(\frac{T_{\rm eff}}{1000\ {\rm K}})^2. $$

The uncertainty in the bolometric luminosity is a combination of the error in the $K$ photometry ($\sim 0.05$ mag),
the uncertainty in the reddening correction ($0.12\Delta A_V$, so a negligible 0.01~mag), and the error in the bolometric
correction due to the uncertainty in the effective temperature: 
$$\Delta BC_K=(-1.59 +0.218 \frac{T_{\rm eff}}{1000\ K}) (\frac{\Delta T_{\rm eff}}{1000\ {\rm K}}).$$  For
a typical M-type RSG we saw above that $\Delta T_{\rm eff} \sim 25$~K, so for a typical effective temperature of 3600~K this is
also a negligible 0.02~mag.  For a warmer star ($T_{\rm eff}\sim 4100$~K with $\Delta T_{\rm eff}\sim 80$~K), the error will
be 0.05~mag, about the same as our photometric error.  By contrast, the uncertainties in the bolometric luminosities derived
from the optical photometry will be several tenths of a magnitude due to the 0.1~mag uncertainty in $A_V$ and the stronger
dependence of the bolometric correction on effective temperature in the optical.  The latter can be derived from differentiating
Equation~\ref{equ-BCV}.  For $T_{\rm eff}=3600$~K and $\Delta T_{\rm eff}\sim 25$~K, the uncertainty in BC$_V$ is about 0.1~mag.
For a warmer star ($T_{\rm eff}=4100$~K with $\Delta T_{\rm eff}$=100~K), the uncertainty in BC$_V$ is the same.

We show the comparison in Fig.~\ref{fig:irmbols}.  The outliers from the effective temperature plots are again labeled.
The median difference of all of the data is +0.04~mag; that of the restricted sample (minus the two suspected binaries and
the two stars with discrepant effective temperatures) is also +0.04.  So, despite the difficulties in determining the effective
temperatures, we believe that the bolometric luminosities are well determined.

\subsection{Bolometric Luminosities Revisited}

The argument above motivated us to reexamine how we have been determining the bolometric
luminosities of RSGs throughout this series of papers.  Although we used the best available $V$-band
photometry for the Milky Way (Paper I) and Magellanic Cloud (Paper II) samples, these data were
by no means contemporaneous with our spectrophotometry.  Since the $V$-band
variability may be linked to variations in $A_V$, at least for some stars (see discussion in 
Massey et al.\ 2007a and Levesque et al.\ 2007), this introduces an additional source of error.
Finally, we note that in our analysis of VY CMa (Massey et al.\  2006a), a RSG enshrouded by circumstellar dust, our technique  derived a luminosity that was a factor of several too low
compared to the integrated photometry (Humphreys et al.\  2007), a fact which we take as
evidence of extensive grey extinction due to the dust  for this particular, peculiar
star (Massey et al.\ 2008). Since the
extinction at $K$ is only 12\% of that assumed at $V$, $M_K$ is much less sensitive to the
extinction, another advantage.

In Paper I we compared the methods for the Galactic sample, where the extinction can be
particularly large and troublesome.  There we found generally excellent agreement (see
Fig.~5c of Paper I), although five stars showed serious discrepancies, all in the sense that the
K-band luminosity was smaller.  

In a similar vein, we include in Table~\ref{tab:spect} the radii and bolometric luminosities we would
derive using the K-band data {\it with our spectral fitting effective temperatures.}   On average,
the agreement is excellent (+0.04 in the mean, $-0.05$ in the median, in the sense of K-band
bolometric luminosity {\it minus} the V-band bolometric luminosity), but with a large dispersion,
$\sim$0.6~mag.

Massey (1998) argued that there was a progression in the uppermost luminosities of RSGs
with metallicity, as evidenced by stars in NGC 6822, M33, and M31, with RSGs being more
luminous at lower metallicity than at higher.   That is in accord with the simple view that a massive
star becomes either a RSG at one mass or a WR at a higher, with the dividing line being
dependent upon metallicity.  At higher metallicity, mass-loss rates will be higher, and hence
it will be easier for a star to become a WR star; i.e., the mass limit should be lower at higher metallicity.

In Table~\ref{tab:alllum} we rederive  the luminosities for all of the stars in our series (Papers I, II,
and the present work).  For the SMC and LMC the K-band data comes from 2MASS, while
for the Galactic stars we used the K-band photometry of Josselin et al.\ (2000), as the stars
are too bright to have good photometry in 2MASS. (The only exception is the star CPD$-53^\circ$3621,
which was not observed by Josselin et al. 2000, but is faint enough to have good 2MASS data.)
Surprisingly, we find that the luminosities derived from the K-band and V-band differ
by $\sim$ 0.12~dex (0.3~mag) in all three galaxies, in the sense that the luminosities from the
$K$-band are slightly less luminous.   
We do not have a good explanation for this, particularly in light of the fact that
we find excellent agreement for the M31 data presented here.  The dispersions are 0.12 (0.3~mag) for
the SMC and LMC data, and 0.25~dex (0.6~mag) for the Milky Way.  Possibly this is one more
indication that the surface of  RSGs look different in the optical and the NIR, due to the
presence of large cool and warm patches on the surface (Freytag et al.\ 2002).  Young et al.\ (2000)
demonstrated that Betelgeuse has a slightly different effective radii and temperatures at different wavelengths due to such
spots.

We have excluded from this compilation VY CMa; as noted above, our optical analysis underestimated the luminosity of the star, although provided a much better effective temperature estimate.  However, the
revised distance to the star found by Choi et al.\ (2008) places the star at a luminosity comparable
to other Galactic RSGs rather than the extreme properties claimed by Humphreys et al.\ (2007).
We also exclude the highly unstable RSGs found in the Magellanic Clouds by Massey et al.\ (2007a)
and Levesque et al.\ (2007).

When all is said and done, we do not confirm the trend described by Massey (1998).  Regardless
of whether the K-band luminosities or the V-band luminosities are used, the medians of the
most luminous 5 stars in each sample are remarkably similar, as shown in Table~\ref{tab:toplums}.
Milky Way stars might show, if anything, a slightly higher luminosity than do the Magellanic Cloud
sample, i.e., in the opposite trend, but the difference is comparable to the uncertainties in the
luminosities.  It could also be that the result is affected by the small size of the M31 sample, but
we will see in the next section that including all of the stars with K-band photometry (regardless
of whether or not they were included in our spectral analysis) shows a similar distribution
in luminosities.  The most luminous RSGs have $\log L/L_\odot \sim$5.2-5.3.

Is this finding, then, a challenge to stellar evolutionary theory?   We think not.
An examination of the evolutionary tracks show that the situation is more complicated than we
described above.
For instance, the 25$M_\odot Z=0.004$ (SMC-like metallicity) evolutionary tracks terminate
at 4500~K (Meynet \& Maeder 2001) or even warmer (5600~K according to the
older, non-rotating models of Charbonnel et al.\ 1993), effective temperatures that 
correspond to those of a G-type supergiant,  not K- or M-type.
This suggests that a proper accounting
of ``red supergiants" and their luminosities must include yellow supergiants as well.
There is also a
complication relating the bolometric luminosity directly to mass, as the evolutionary tracks become
nearly vertical in the RSG phase, with considerable overlap between the luminosities of stars of
differing masses.   Given this, we instead must ask whether the distribution of RSGs in the H-R diagram is consistent with the evolutionary tracks, which we address in the next section.

 \section{H-R Diagram}
 \label{Sec-hrd}
 
 One of the primary motivations of this project was to compare the distribution of stars in the HRD to evolutionary tracks.
 We show the HRD in Fig.~\ref{fig:hrd}, upper left.
 The M31 RSGs for which we have spectrophotometry are
 shown by the circles.   The tracks reproduce the location of the stars in the HRD
 very well, with the highest luminosity RSGs consistent with the predictions of the
 tracks\footnote{Note that the $z=0.040$ rotating models for 20$M_\odot$ and 25$M_\odot$
 differ from those published by Meynet \& Maeder (2005), as those were known not to extend
 to realistically cool enough temperatures as a result of a numerical simplification made in computing the
 high metallicity models.  In general, the models consist of an envelope (containing at most 2-3\% of the mass of
 the star) and an interior region, which are treated separately.
 Partial ionization is accounted for only in the envelope, while the interior is assumed
 to be fully ionized.  The two treatments for the envelope and the interior are made consistent by choosing a mass
 coordinate, called the fitting point, where both treatments give the same solution for the pressure, temperature, radius,
 and luminosity.  Usually in the course of evolution evolution the position of the fitting point is changed.  Typically when
 the star evolves to the red part of the HR diagram, the fitting point should be lowered, since complete ionization will occur only
 in the deeper layers of the star. For the 25 $M_\odot$ and 20 $M_\odot z=0.040$ models of Meynet \& Maeder (2005) the fitting
 point was held constant at a constant value to avoid some numerical difficulties, as this had little effect on the
 Wolf-Rayet lifetimes and evolution which was the aim of those calculations.  But, for the present work a more exact treatment
 was needed.  We also include here the newly computed $z=0.040$ 15$M_\odot$ rotating track computed for the study
 of yellow supergiants by Drout et al.\ (2009).}, as
 well as the effective temperatures compared to the redwards extension of the tracks.
 
 We were curious what we would learn by increasing the sample size using the
 stars in Table~\ref{tab:all} for which there was only photometry.  We used the
 $V$ and $K$ data to derive the physical properties for these stars following
 the formulae given in \S~\ref{Sec-vmk}.  We see that with the exception of a few
 outliers the distribution of these additional stars (shown by crosses in 
 Fig.~\ref{fig:hrd}, upper right) match those derived from the spectrophotometry.
 We emphasize that the locations of the stars determined this way are less
 certain.  We have restricted the sample to only stars with the coolest temperatures
 ($T_{\rm eff}\le 3800$~K) as the errors quickly grow larger for warmer stars.
 The largest uncertainty in this process is the adoption of a value for the reddening. We
 have used the median value ($A_V=1.0$) from our spectroscopy, but recall that
 the values ranged from $A_V=0.46$ to 2.46; the standard deviation is 0.6~mag.
 If $A_V$ were uncertain by 0.6~mag, then $(V-K)_0$ is uncertain by 0.5~mag, and propagation of errors through the equations of \S~\ref{Sec-vmk} imply that the effective temperature would be uncertain by about 140~K (0.016~dex) at 3700~K,
 corresponding to a $(V-K)_0\sim 4.3$.  Combined with the uncertainty in correcting $K$ to $K_0$, and
the uncertainty in the bolometric correction, the luminosity would be uncertain by
about 0.11~mag, or 0.04 in $\log L/L_\odot$. figure.  The error bars are comparable
to the point sizes. We note although
we can determine the effective temperature much more precisely with fitting
the spectrophotometry, the actual error in the luminosity is comparable, or even less, by
using $K$.

As mentioned above (\S~\ref{Sec-props}),
 in recent years, the metallicity of the young population of M31 has become slightly controversial, with the analysis of four A-F supergiants by
 Venn et al.\ (2000) and a B-type supergiant  (Smartt 2001) which seemed to suggest that the oxygen abundance in M31 was more like solar than
 the $2\times$ solar implied by the analysis of HII regions by Zaritsky et al.\ (1994) and Blair et al.\ (1981).   However, N. Przybilla (2007, private communication)
 finds that when he adopts the new LGGS photometry for the A supergiants used in the Venn et al.\ (2000) analysis, he now obtains significantly higher
 oxygen abundances as well as more reasonable stellar parameters.     Here we use our M31 RSG to see if we can weigh in on this
 issue indirectly by seeing if we have significantly better agreement with the evolutionary tracks if we assume one metallicity rather than
 the other.  We have refit the spectra of our 16 M31 RSGs using the MARCS models computed at solar metallicity.  We find very little difference in
 the physical parameters, as suggested by the discussion above  in \S~\ref{Sec-props}.   On average, the difference is in the effective temperature is -32~K,
 in the sense expected; i.e., at lower metallicity a cooler temperature is needed to produce TiO as strong as what can be achieved by a slightly warmer temperature
 at higher metallicity.  Also as expected the bolometric luminosity is slightly higher, by 0.01~dex (0.03~mag).
 In the lower two panels of Fig~\ref{fig:hrd} we compare the  location of the M31 RSGs with the solar metallicity tracks, where we have made slight adjustments We see here that in fact the agreement is also good; i.e., the analysis here does not favor 2$\times$ solar metallicity over 1$\times$ solar metallicity, nor visa versa.   Note that our conclusion
 here differs from what we presented at recent conferences (e.g.,  Massey et al.\ 2008).  
 
 We agree with the conclusions of Crockett et al.\ (2006) that the HII regions
 of M31 require a reexamination with better data.  
 However, we can also make some progress on this front
 using observations of a much larger sample of RSGs: we know that the average spectral subtype depends upon metallicity of
 the parent galaxy, being K3~I in the SMC and M2~I in the Milky Way; see discussion in Elias et al.\ (1985), Massey \& Olsen (2003), and
 Paper II. Is the average spectral type
 of RSG M2~I or later in M31?  Our study was directed at the M2~I stars, but guided by our new $V-K$
 photometry future studies can sample the entire range of RSGs in this neighbor galaxy.
  
\section{The Curious Case of J004047.84+405602.6}
\label{Sec-wacky}

We had included the star J004047.84+405602.6 in our spectroscopic sample as the $V-K_S$ color (5.15) was quite red, and by all indications it was expected to be an M star with
similar properties to the others in Table~\ref{tab:spect}.  With reddening typical of the other RSGs we would expect $(V-K)_0\sim 4.3$, and thus $T_{\rm eff}=3700$~K from
Eq.~\ref{eq:vmk}.  We were therefore taken aback to find no TiO lines in the spectrum we obtained with the MMT (Fig.~\ref{fig:wacky}).   Our initial reaction to the spectrum
was that it was a heavily reddened late F-type or early K-type star, but closer examination revealed that there were no stellar lines visible in the optical, despite the good
signal to noise
(80 per spectral resolution element at 6000 \AA).     The Na D 
interstellar line is present; the only other strong feature at 6483 \AA\ turns out to be a CCD artifact.  Even the Balmer lines are not present.  In the red WIYN spectrum,
from which we measured the radial velocity, the Ca II triplet lines are strongly present, but no other features are detected.

There are several STIS images 
available in the {\it HST}  archive, and we have examined these carefully.  These establish that, at least at {\it HST} resolution, the object is single.

We conclude that this is potentially a very interesting object.  The radial velocity of the star ($-533$ km s$^{-1}$), in relatively good agreement with that expected for
its location ($-486$ km s$^{-1}$), makes it indisputably an M31 object.  Its very red colors ($V-K_s=5.15$, $B-V=2.65$) combined with the lack of molecular bands suggest
it is extremely highly reddened.  Given that, it could simply be an early-type star located in a particularly dusty patch of M31, but the lack of spectral features, even the Balmer
lines, argues that instead the spectrum is heavily ``veiled"\footnote{Or, as we concluded in haiku form: \begin{verse} Our favorite lines \\ Are just not there, anywhere! \\The spectrum is veiled.\end{verse}}.  Veiling is usually
symptomatic of a continuum source diluting the spectral features, e.g.,
from shock-heated gas in classical T Tauri stars (but see  Gahm et al.\ 2008),
or from the ionized envelope in Be stars.  However, the absorption spectra can
also be smeared out due to multiple scatterings in an expanding circumstellar dust shell
(Romanik \& Leung 1981), and that is what we suspect is happening here, given the
coincidence of extremely high reddening and a washed-out spectrum. We would expect such scattering to affect the blue region of the spectrum more strongly than the red, possibly explaining why the Ca II triplet lines are visible in the
far red.

We note that some RSGs exhibit very high dust production rates, and become dust-enshrouded: examples include VY CMa and NML Cyg
in the Milky Way (see, for example, Smith et al.\ 2001 and Schuster et al.\ 2006) and WOH G64 in the LMC (van Loon et al.\ 2005; 
Ohnaka et al.\ 2008; Levesque et al.\ 2009).  
No one understand physically what triggers such events.  Possibly this star servers as another example.  

\section{Summary and Future Work}
\label{Sec-summary}

Using the LGGS we were able to select RSG candidates from a two-color $B-V$, $V-R$ diagram, following Massey (1997).  Radial velocities obtained with WIYN established
that these stars are bona-fide RSGs rather than foreground Milky Way objects.  In order to determine the physical properties, we then obtained MMT optical spectra of 17 of these
stars, 16 of which were M-type supergiants.  In addition, we obtained contemporaneous $V-K$ values for these stars plus
a much larger sample.  The latter will be used to select stars for additional spectroscopy.

Comparing our new $V$-band photometry to the LGGS photometry, we find that on average
the M31 RSGs are variable in $V$ by 0.5~mag, smaller but comparable to
what was found for the normal RSGs in the Magellanic Clouds by Levesque et al.\ (2007).
 In contrast, our new $K_s$ photometry shows
no such variations when compared to the 2MASS values. This is consistent with what we know of
photometric variability in the Milky Way and Magellanic Clouds; see discussion in Levesque et al.\ (2007) and Josselin et al.\ (2000).  

Our analysis of the optical spectra with the MARCS stellar atmospheres reveals that the typical reddening ($A_V=0.5-2.6$~mag, with a 
median of 1.0~mag) is much higher than that of our early-type stars in M31 ($A_V=0.4$, according to Massey et al.\ 2007b).
This is similar to what we found in Massey et al.\ (2005) and Paper II for the Milky Way and Magellanic Clouds: on average, RSGs are
significantly more red.  Our interpretation is the same as in previous papers, namely that circumstellar dust is responsible.  At the same time
we see that there is a significant NUV excess compared to the models, which we expect to be a consequence of scattering by dust near the star.

\begin{figure}
\epsscale{0.6}
\plotone{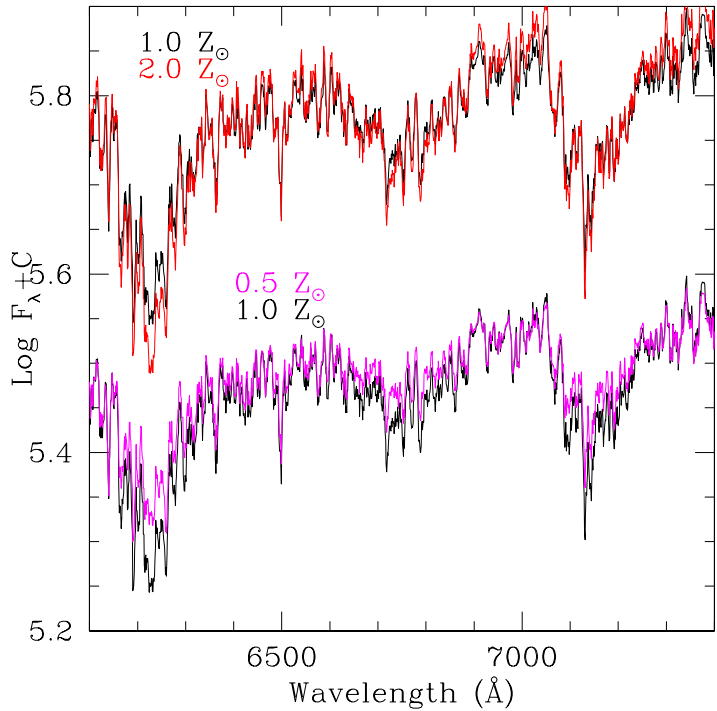}
\caption{\label{fig:models}Comparison of three 3700~K, $\log g=0.0$ models.  We show a small region of the spectrum containing three strong
TiO bands ($\lambda 6159$, $\lambda 6651$, and
$\lambda 7054$) for an effective temperature roughly corresponding to that of an M2 type.  In the upper comparison we superimpose a $2\times$ solar
metallicity model (M31-like, shown in 
red) on a solar metallicity model (black).  In the bottom comparison we superimpose a $0.5\times$ solar metallicity model (LMC-like, shown in magenta)
on a solar metallicity model (black).  There is a larger difference between the bottom two than the top, consistent with our finding of a smaller effective
temperature difference between the M2~I stars in the Milky Way and M31.}
\end{figure}

\begin{figure}
\epsscale{0.6}
\plotone{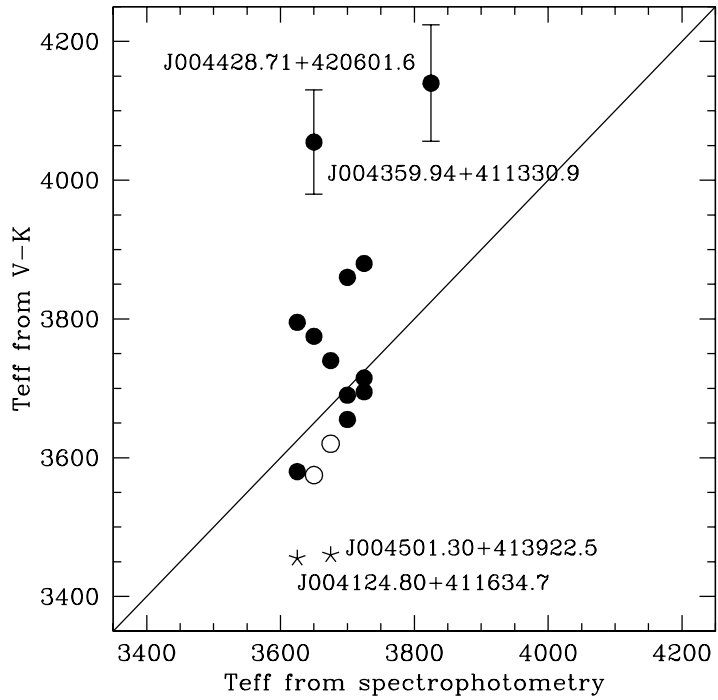}
\caption{\label{fig:irtemps} Comparison of effective temperatures derived from $V-K$ colors with those from
spectrophotometry.
The open circles denote the two stars with non-contemporaneous $V$ and/or $K$ measurements.  The labeled points 
(outliers) are discussed in the text, including the two stars denoted by astericks (which we believe have hot companions),
and the two stars with large error bars at the top.  Where error bars are not shown, they are similar in size to the point size.}
\end{figure}

\begin{figure}
\epsscale{0.6}
\plotone{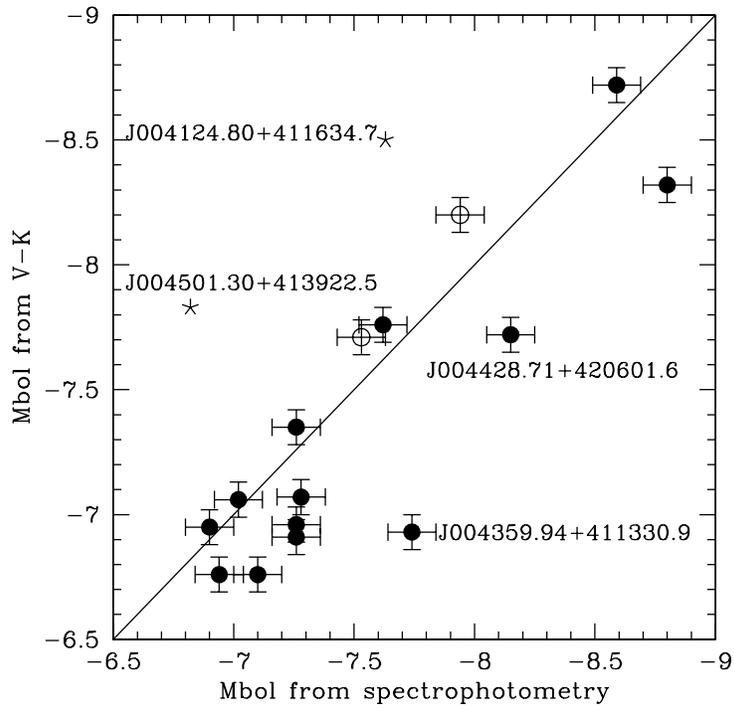}
\caption{\label{fig:irmbols} Comparison of the bolometric luminosities derived from $V-K$ colors with those from
spectrophotometry.  The symbols are the same as in Fig.~\ref{fig:irtemps}.}
\end{figure}

\begin{figure}
\epsscale{0.45}
\plotone{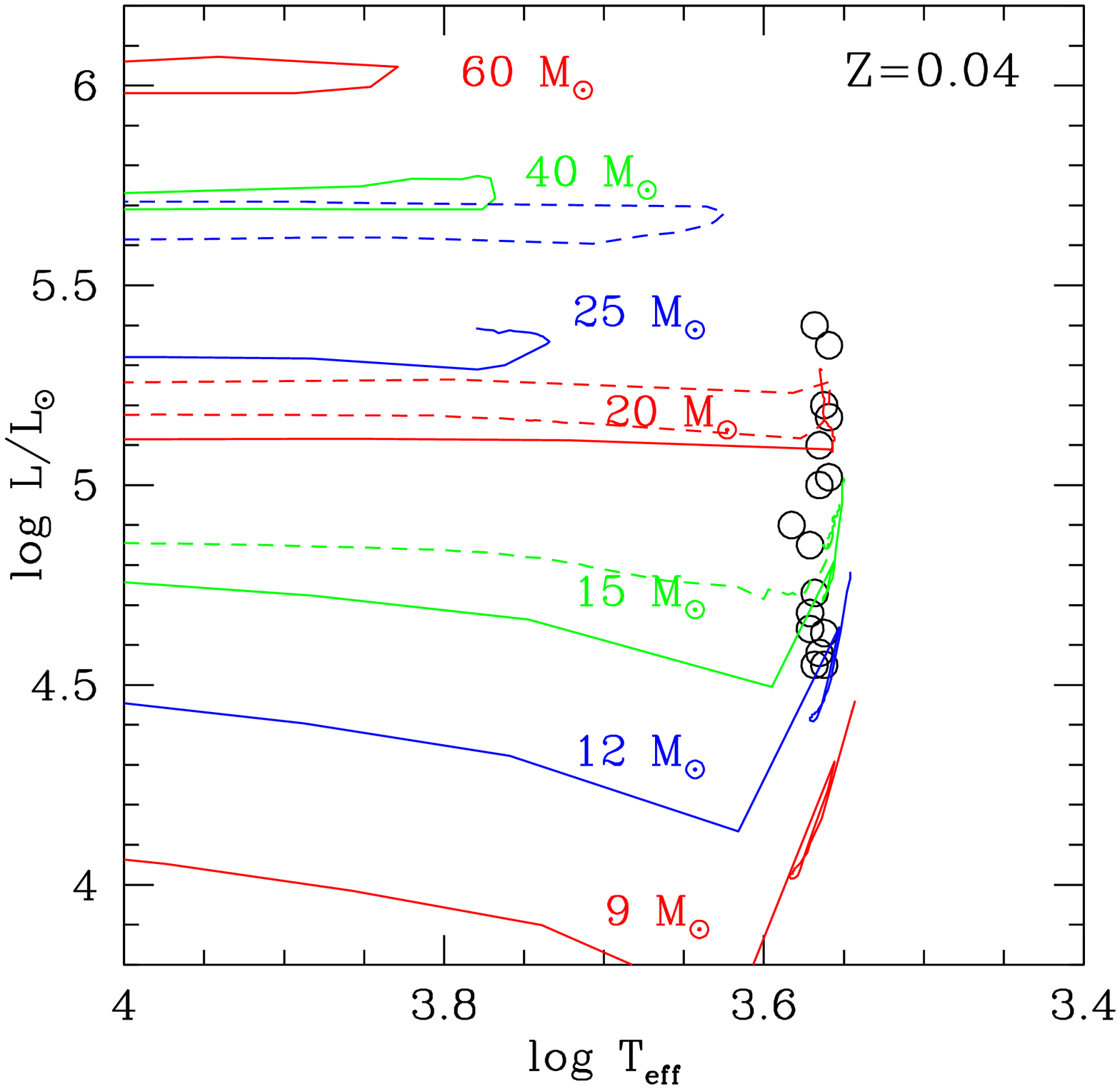}
\plotone{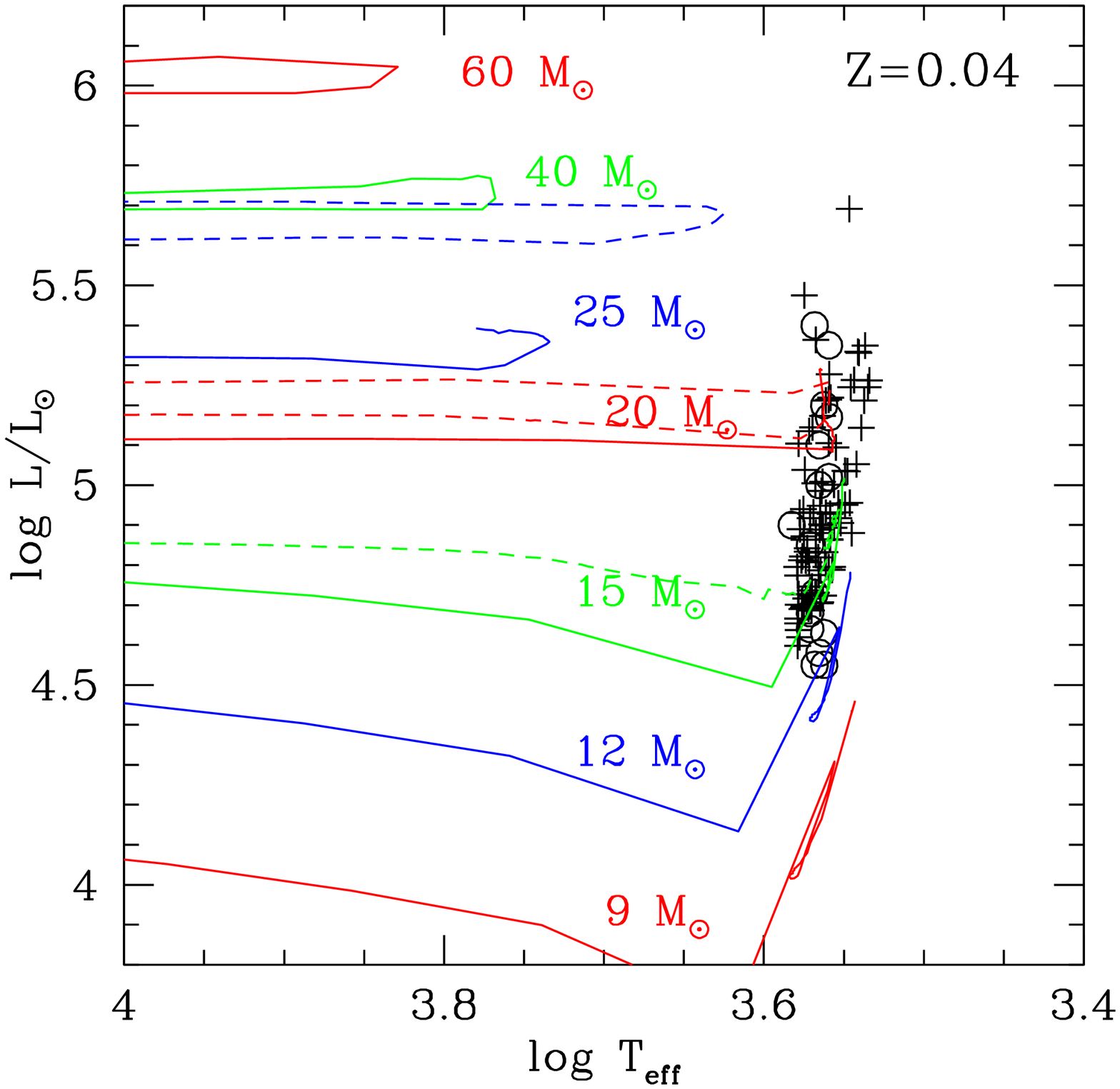}
\plotone{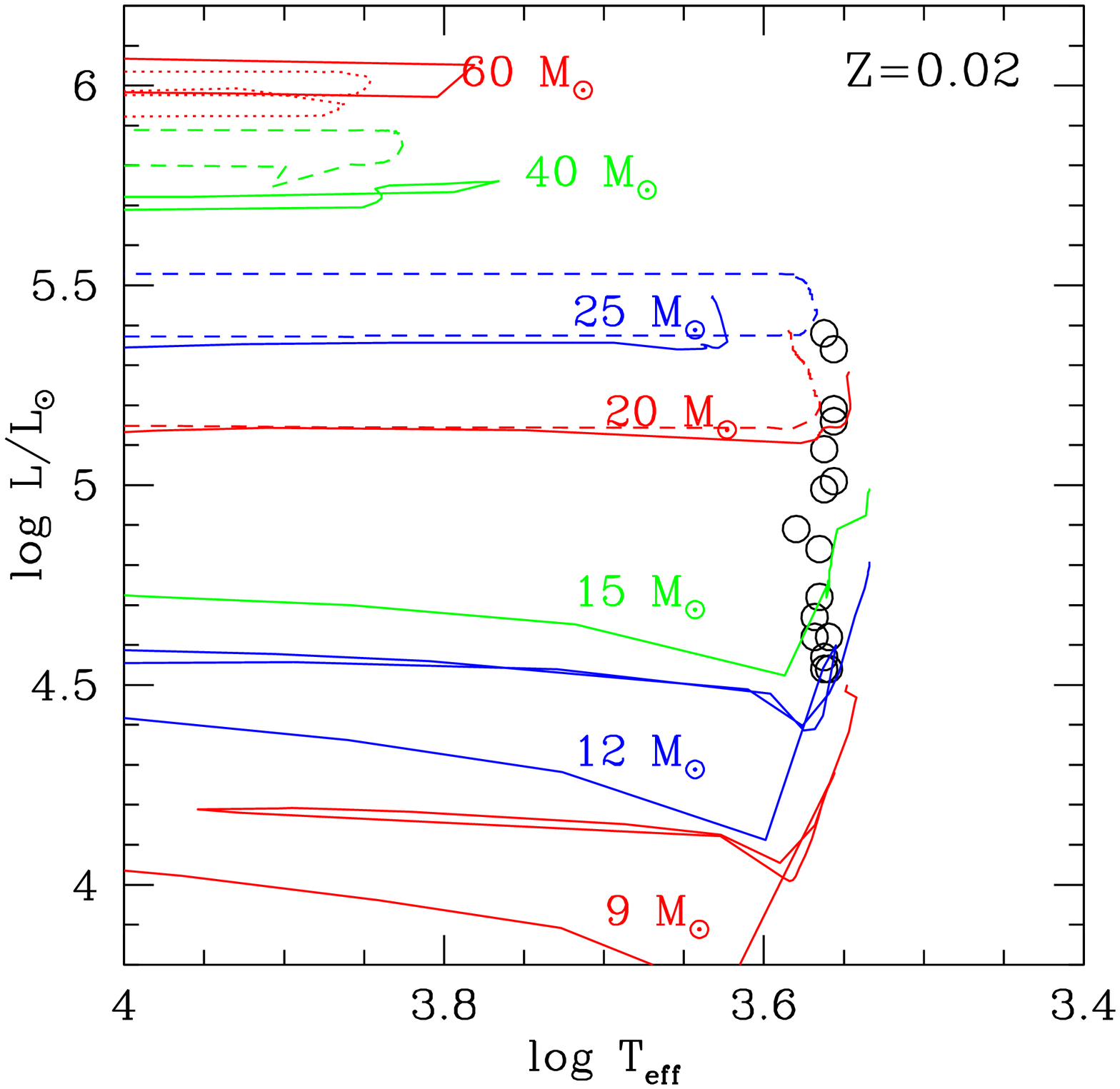}
\plotone{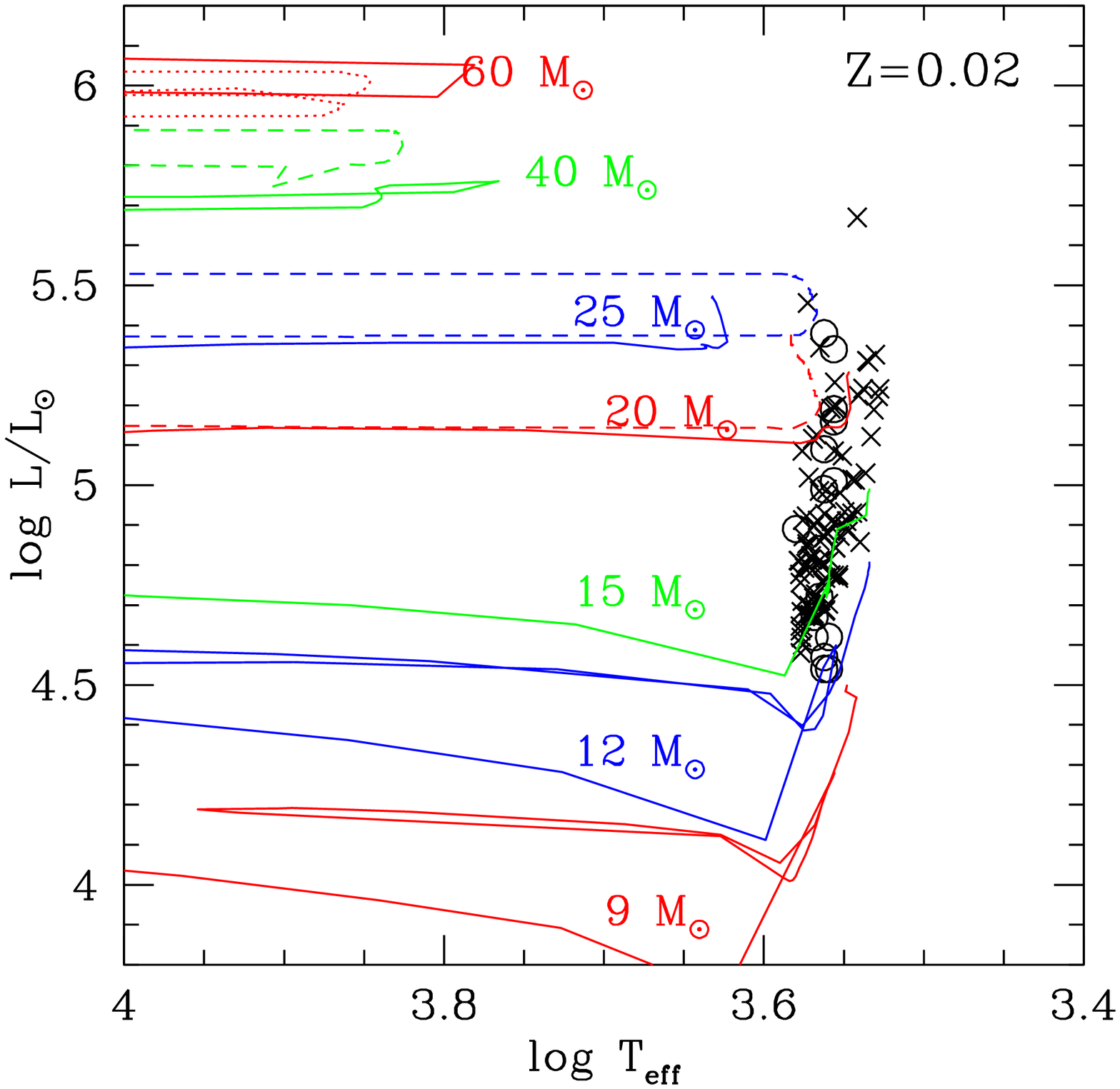}
\caption{\label{fig:hrd} Comparison with evolutionary tracks.  We show the location of the M31 RSGs in the HRD computed both with $Z=0.04$ metallicity (upper two panels)
and $Z=0.02$ (lower two panels).  The open circles are the
the physical parameters derived from the spectral types in Table~\ref{tab:spect}, minus the two suspected binaries.  For the two figures on the right, the crosses represent
the properties derived from the other stars in Table~\ref{tab:all} for which we have
only broad-band $V$ and $K$ photometry, where we restrict the sample to the uncrowded stars, and to stars with implied effective temperatures less than 3800~K,
as the errors become much larger at warmer temperatures.   For these additional
stars we have adopted $A_V=1.0$~mag.   For the evolutionary
tracks, the solid tracks are the ``old" Geneva tracks from Schaerer
et al.\ (1993) for $Z=0.04$, and from Schaller et al.\ (1992) for $Z=0.02$.  The dashed tracks are the ``new" Geneva tracks  which include the effects of rotation, with an initial
rotation speed of 300 km s$^{-1}$.  The rotating models come from Meynet \& Maeder (2003) for $Z=0.02$ and from Meynet \& Maeder (2005) for $Z=0.04$.  The latter have been supplemented by newly computed versions for the 25$M_\odot$, 20$M_\odot$, and 15$M_\odot$, as
explained in the text.} 
\end{figure}

\begin{figure}
\epsscale{0.6}
\plotone{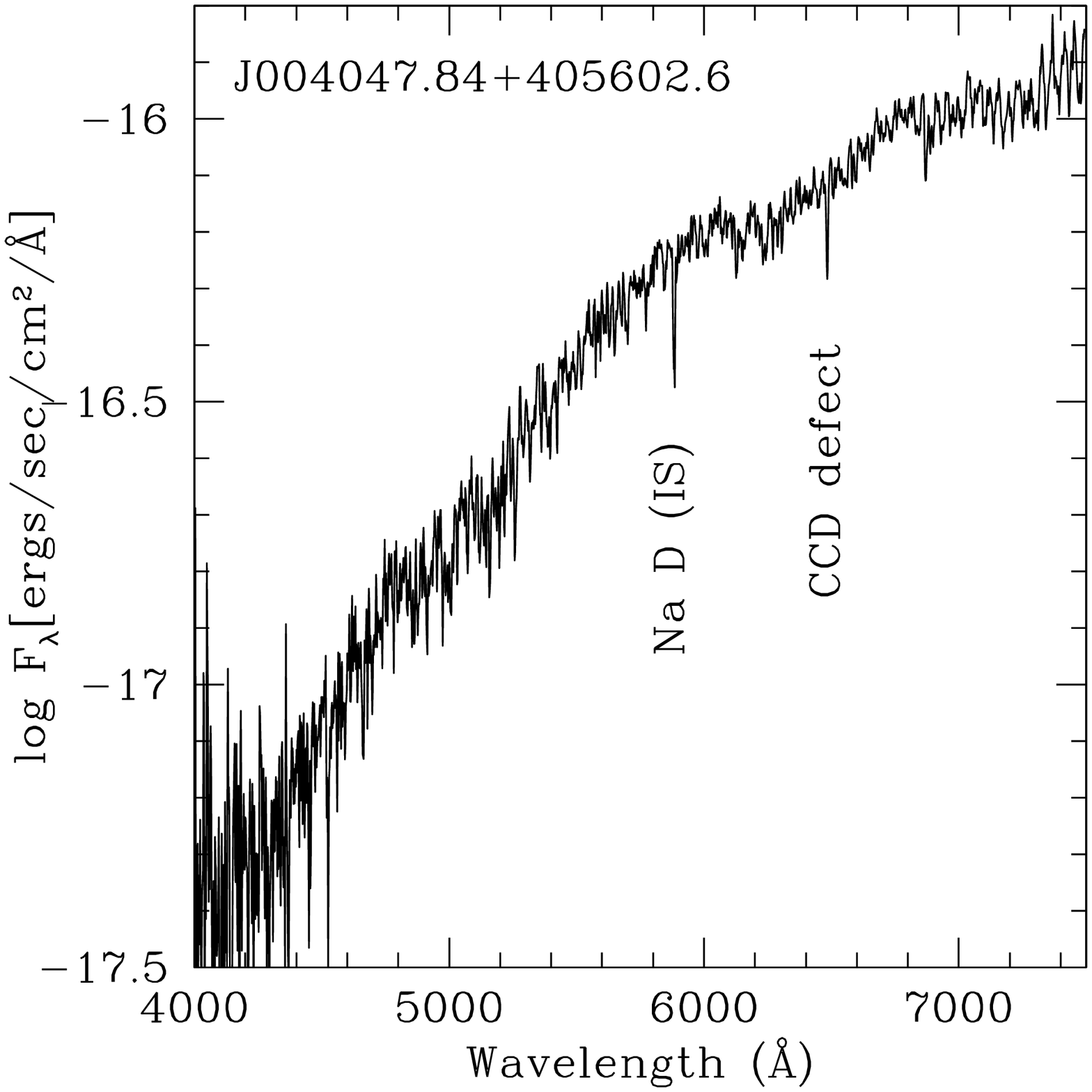}
\plotone{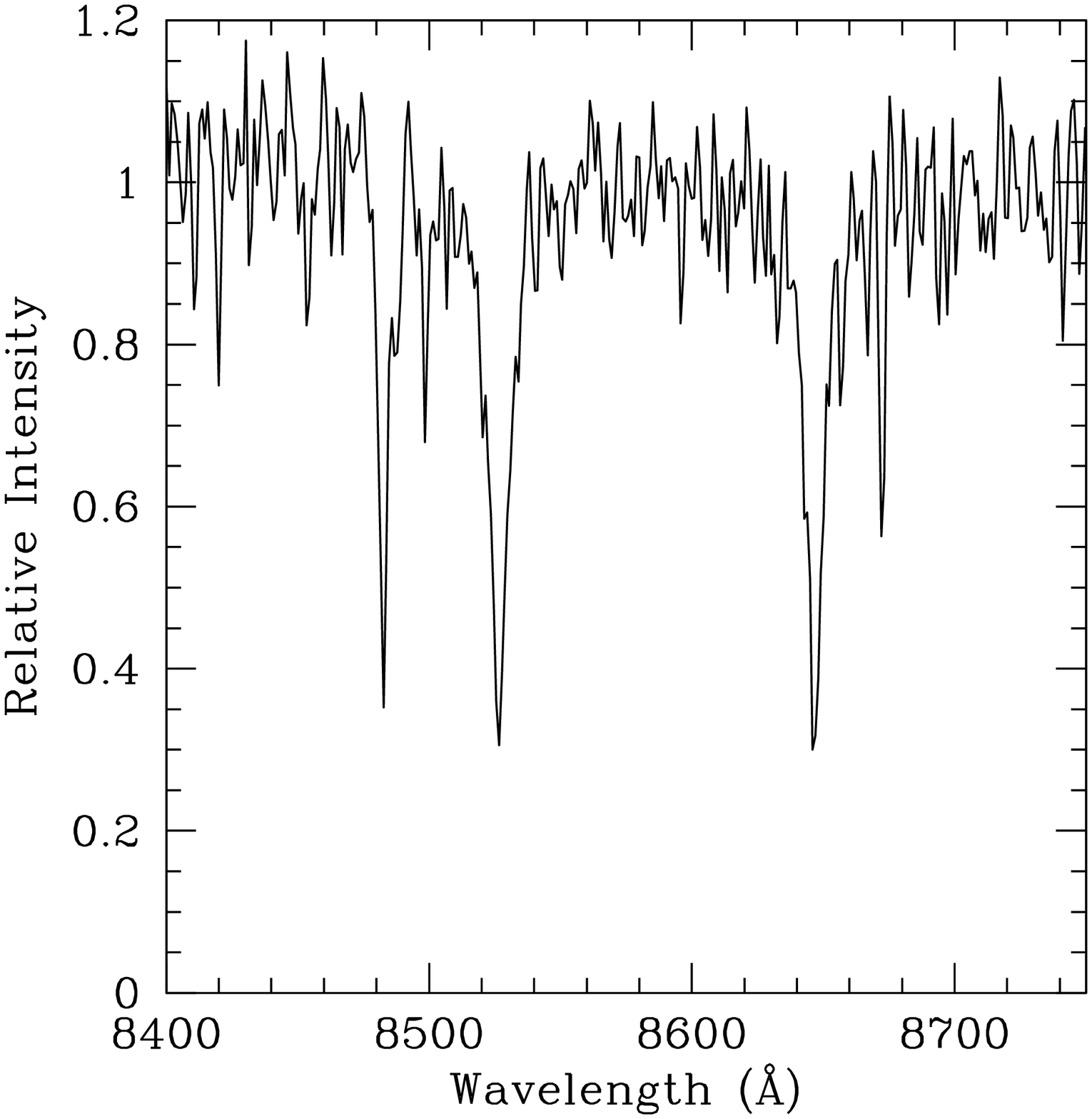}
\caption{\label{fig:wacky} The peculiar star J004047.84+405602.6.  No stellar features are visible in the blue-red region of the spectrum;
the Ca II triplet is visible in the far red.}
\end{figure}

For the six M31 stars we classify as M2~I, we find an effective temperature (on average) that is marginally warmer than their Galactic counterparts by 
20$\pm 12$~K.   This is in the sense expected, i.e., by adopting a 2$\times$ solar metallicity the TiO bands will become strong at slightly
warmer temperatures.  This difference is smaller than the typical 50~K difference seen between LMC and Milky Way RSGs (Paper II),  despite the fact that we assumed the LMC metallicity to be 0.5$\times$ solar,  suggesting that some mild
saturation is occurring in the formation of the molecular bands; we can see this directly from comparing the models (i.e., Figure~\ref{fig:models}).
 A comparison of the physical properties derived by $V-K$ and spectrophotometry shows good agreement, with the exception of four stars, two
of which we consider to be binaries.  

We use the K-band luminosities of our stars (derived using theeffective temperatures and reddenings from our spectral fitting)
to re-investigate the finding of Massey (1998) that the bolometricluminosities of RSGs depend upon the metallicites of the host galaxy.
We fail to confirm this trend; instead, the most luminous RSGs have
$\log L/L_\odot$ of about 5.2-5.3.  Note that this value is (broadly)
consistent with the $\log L/L_\odot=5.1$ upper limit of the most
luminous progenitor of a Type II-P SN (Smartt et al.\ 2009), which
are believed to have come from RSGs.  (We are grateful to theanonymous referee for noting that tackling this question from two
different directions leads to very similar answers.)  Smartt etal.\ (2009) dub the small gap the ``red supergiant problem", and
argue that the fate of the higher luminous RSGs is not yet understood,although in the case of M31 (with its higher metallicity) 
the  Geneva tracks with rotation show the most luminous RSGs
evolving back to the blue side of the HRD and becoming Wolf-Rayet
stars.

There appears to be a systematic offset of $\sim$0.3~mag in the luminosities derived from K-band
and V-band photometry, where we have adopted the same effective temperatures and reddenings 
as determined from the optical spectrophotometry.  Why this is remains unclear, but it could be
another indicator of the explanation we offered in Paper II that the surfaces of RSGs simply
look different in the optical and the NIR owing the large cool and warm patches on the surface
(Freytag et al.\ 2002).

A comparison of the location of these stars in the HRD with the  stellar evolutionary tracks indicates excellent agreement.  The effective temperatures
agree well with the redward most extension of the tracks, and the upper luminosities also agree well with what the tracks predicts.
We do find that we could not use the current data to distinguish between the 2$\times$ solar metallicity indicated from the HII regions
(Zaritzky et al.\ 1994) and a lower value.

Finally, we identify one peculiar object in M31. Its broad band colors ($B-V$ and $V-K_s$) suggest it is very red, but the spectrum shows none of the molecular bands
we associate with cool stars.  This is therefore a heavily reddened object.  Its lack of spectral features (other than Ca II triplet lines in the far red) suggests the spectrum is
heavily ``veiled"; given its extremely red colors (and lack of TiO features) we suggest that the
absorption lines are filled in  by scattering from within an expanding dust shell.  

As for the future: the distance of M31 results in its stars being about 200$\times$ fainter than
in the LMC.  The problem this introduces to such studies is that of more limited statistics. Using our
new $V-K$ photometry we plan to expand the current study, not only to a larger sample size, but
to later spectral types.  Earlier studies (Elias et al.\ 1985; Paper II) have shown that the average
spectral type of RSG varies from galaxy to galaxy based on metallicity (K3~I in the SMC, and M2~I
in the Milky Way); it would be of interest to determine this in M31.  Photometry alone is not sufficient,
given the large and variable reddening, which we largely attribute to circumstellar material.  Additional
spectroscopy would determine not only the spectral type distribution, but allow spectral fitting to determine
$A_V$.  Combined with $K$ band measurements this would provide for very accurate determinations of
the luminosities of these stars.
Determining the mass-dust loss rates for these stars would also be of interest, and we are in the
process of using archival Spitzer data to do so.  Finally, it is of paramount importance to be able
to place the population of M31's RSG in context---how do their numbers compare to the number
of OB star progenitors of similar mass?  The LGGS photometry provides a starting point for such
spectroscopic observations.

\acknowledgments
We thank the University of Florida Department of Astronomy
FLAMINGOS development team for
providing this instrument to the community but wish the original PI,
our friend Richard Elston, were still here to work with us.  Funding for Flamingos was provided
by NSF grant AST97-31180 and Kitt Peak National Observatory.
We also
thank Dick Joyce and Hal Halbedel for their usual level of cheerful
and competent assistance during our KPNO 4m observing run, John McAfee for making our MMT run as pleasant and productive as usual, and George Will for his able
assistance and good humor during the WIYN observations. This
publication makes use of data products from the Two Micron All Sky
Survey, which is a joint project of the University of Massachusetts
and the Infrared Processing and Analysis Center/California Institute
of Technology, funded by the National Aeronautics and Space
Administration and the National Science Foundation (NSF).  Support to PM
was provided by the NSF through AST-0604569.  We are indebted to Nelson Caldwell for thoughtful comments
on the spectrum of the peculiar J004047.84+405602.6, and to an anonymous referee for useful suggestions which
improved the paper.

\clearpage

\clearpage

\clearpage

\clearpage
\clearpage

\clearpage

\clearpage

\clearpage

\begin{deluxetable}{l c c c c c c c c c c c c c c l l c}
\tabletypesize{\tiny}
\rotate
\tablecaption{\label{tab:all} Photometrically Selected RSG Candidates in M31\tablenotemark{*}}
\tablewidth{0pt}
\tablecolumns{18}
\tablehead{
&\multicolumn{4}{c}{LGGS Photometry\tablenotemark{a}}
&&\multicolumn{2}{c}{$K_s$}
&&\multicolumn{2}{c}{$V-K_s$}
&&\multicolumn{2}{c}{Radial Vel.}
&
&\colhead{Other}
&\multicolumn{2}{c}{Sp.\ Type}  \\ \cline{2-5} \cline{7-8} \cline{10-11} \cline{13-14} \cline{17-18}
\colhead{Star}
&\colhead{$V$}
&\colhead{$B-V$}
&\colhead{$V-R$}
&\colhead{$R-I$}
&&\colhead{Value}
&\colhead{Ref.\tablenotemark{b}}
&&\colhead{Value}
&\colhead{Ref.\tablenotemark{c}}
&&\colhead{Value\tablenotemark{d}}
&\colhead{Ref.\tablenotemark{e}}
&\colhead{Crowding\tablenotemark{f}}
&\colhead{ID\tablenotemark{g}}
&\colhead{Value}
&\colhead{Ref.\tablenotemark{h}}
}
\startdata
J003703.64+402014.9  & 19.92 &  1.72 &  0.95 &  1.26 &&  15.72 & 1 &&   4.20 & 1 && \nodata& 0 & 0 & \nodata   & \nodata& 0 \\
J003721.98+401543.3  & 19.99 &  1.83 &  0.93 &  0.88 && \nodata& 0 && \nodata& 0 && \nodata& 0 & 0 & \nodata   & \nodata& 0 \\
J003722.34+400012.1  & 18.93 &  1.87 &  0.89 &  0.84 &&  15.02 & 2 &&   3.97 & 3 && -531.8 & 1 & 0 & \nodata   & \nodata& 0 \\
J003723.56+401715.5  & 18.86 &  1.90 &  0.98 &  1.02 &&  14.83 & 1 &&   4.03 & 1 && \nodata& 0 & 0 & \nodata   & \nodata& 0 \\
J003724.48+401823.3  & 19.62 &  2.08 &  1.08 &  1.06 &&  15.11 & 1 &&   4.51 & 1 && \nodata& 0 & 0 & \nodata   & \nodata& 0 \\
\enddata
\tablenotetext{*}{The full version of this table is available in the on-line edition.}
\tablenotetext{a}{LGGS photometry from Massey et al.\ 2006.}
\tablenotetext{b}{Reference for $K_s$: 0=no value; 1=2MASS; 2=Present work.}
\tablenotetext{c}{Reference for $V-K_s$: 0=no value; 1=$V$ from LGGS and $K_s$ from 2MASS;
2=$V$ from LGGS and $K_s$ frm present work; 3=$V$ and $K_s$ both from present work.}
\tablenotetext{d}{Radial velocity in units of km s$^{-1}$.}
\tablenotetext{e}{Reference for radial velocity: 0=no value; 1=present work; 2=Massey 1998.}
\tablenotetext{f}{Crowding was determined by computing the contamination from
neighbors under 1.2\arcsec seeing with a  1.2 \arcsec slit; the value reported is the worst fraction in $B$, $V$, or $R$:  0=$<$1\%; 1=1-5\%; 2=5-10\%; 3=$>$10\%}
\tablenotetext{g}{From Massey 1998 for OB cross-IDs, and from Humphreys et al.\ 1988 for R95.}
\tablenotetext{h}{Reference for spectral type: 0=no value; 1=present work; 2=Massey 1998; 3=Humphreys et al.\ 1988.}
\end{deluxetable}

\begin{deluxetable}{c r r r c r l r l l}
\tabletypesize{\tiny}
\rotate
\tablecaption{\label{tab:rvs} New Radial Velocities (RV) of Photometrically Selected RSG Candidates}
\tablewidth{0pt}
\tablecolumns{10}
\tablehead{
\colhead{Star}
&\colhead{RV$_{\rm obs}$}
&\colhead{$N_{\rm obs}$}
&\colhead{$\sigma_\mu$}
&\colhead{$X/R$}
&\colhead{RV$_{rot}$}
&\multicolumn{3}{c}{Previous\tablenotemark{a}}     
&\colhead{New} \\ \cline{7-9}
&km s$^{-1}$&&km s$^{-1}$&&&\colhead{ID}&{RV$_{\rm obs}$}&\colhead{Type} & \colhead{Type}
}
\startdata
J003722.34+400012.1  &    -531.8 &1&  0.6 &  -0.994 & -535 &\nodata & \nodata & \nodata & \nodata              \\
J003739.41+395835.0 &    -523.5 &1&  0.4 &  -0.999 & -536 &\nodata & \nodata & \nodata & \nodata            \\
J003829.23+403100.2\tablenotemark{b} &    -413.9 &1&  0.8 &  -0.805 & -490 &\nodata & \nodata & \nodata & \nodata             \\
J003857.29+404053.6 &    -543.0 &1&  0.3 &  -0.697 & -463 &\nodata & \nodata & \nodata & M2 I             \\
J003902.20+403907.3  &    -484.7 &1&  0.4 &  -0.756 & -477 &\nodata & \nodata & \nodata & M1 I             \\
J003903.28+403042.7  &    -543.8 &1&  0.5 &  -0.930 & -520 &\nodata & \nodata & \nodata  & \nodata           \\
J003912.77+404412.1  &    -448.7 &1&  0.4 &  -0.676 & -458 &\nodata & \nodata & \nodata  & \nodata           \\
J003913.40+403714.2 &    -522.1 &1&  0.3 &  -0.844 & -499 &\nodata & \nodata & \nodata &M1 I             \\
J003915.67+403559.0 &    -502.3 &1&  0.4 &  -0.880 & -508 &\nodata & \nodata & \nodata   & \nodata           \\
J003921.18+402611.9 &    -548.1 &1&  0.3 &  -1.000 & -536 &\nodata & \nodata & \nodata   & \nodata          \\
J003923.76+401852.6 &    -530.8 &1&  0.5 &  -0.955 & -526 &\nodata & \nodata & \nodata  & \nodata           \\
J003935.86+402705.7 &    -542.5 &1&  0.5 &  -0.993 & -535 &\nodata & \nodata & \nodata  & \nodata           \\
J003936.90+405120.5 &    -441.1 &1&  0.6 &  -0.584 & -436 &\nodata & \nodata & \nodata   & \nodata          \\
J003940.67+405125.4  &    -432.3 &1&  0.3 &  -0.595 & -439 &\nodata & \nodata & \nodata  & \nodata           \\
J003944.05+403234.5  &    -476.4 &1&  0.6 &  -0.999 & -536 &\nodata & \nodata & \nodata & \nodata            \\
J003947.20+403854.1 &    -525.7 &1&  0.3 &  -0.948 & -524 &\nodata & \nodata & \nodata & \nodata            \\
J003954.26+404004.4 &    -514.3 &2&  0.5 &  -0.953 & -525 &\nodata & \nodata & \nodata & \nodata             \\
J003957.00+410114.6 &    -442.4 &1&  0.3 &  -0.403 & -392 &\nodata & \nodata & \nodata  & M0 I            \\
J004013.47+400524.9 &    -385.8 &1&  0.6 &  -0.627 & -446 &\nodata & \nodata & \nodata  & \nodata           \\
J004015.18+405947.7 &    -406.8 &1&  0.3 &  -0.483 & -412 &\nodata & \nodata & \nodata  & \nodata           \\
J004019.15+404150.8  &    -551.9 &1&  0.8 &  -0.997 & -536 &\nodata & \nodata & \nodata  & \nodata           \\
J004023.81+403351.8 &    -575.5 &1&  0.4 &  -0.919 & -517 &\nodata & \nodata & \nodata  & \nodata            \\
J004023.89+403554.7 &    -559.2 &1&  0.6 &  -0.950 & -524 &\nodata & \nodata & \nodata    & \nodata          \\
J004024.52+404444.8 &    -550.5 &1&  0.3 &  -0.983 & -532 &\nodata & \nodata & \nodata  & \nodata            \\
J004025.75+404254.8 &    -572.8 &1&  0.5 &  -0.999 & -536 &\nodata & \nodata & \nodata   & \nodata           \\
J004026.66+405019.7 &    -519.4 &1&  0.3 &  -0.862 & -503 &\nodata & \nodata & \nodata  & \nodata            \\
J004027.36+410444.9  &    -374.3 &1&  0.3 &  -0.383 & -387 &\nodata & \nodata & \nodata  & \nodata            \\
J004029.07+404210.0  &    -572.6 &1&  0.5 &  -0.998 & -536 &\nodata & \nodata & \nodata   & \nodata           \\
J004029.08+404003.6 &    -564.7 &1&  0.6 &  -0.984 & -533 &\nodata & \nodata & 
\nodata & \nodata             \\
J004030.43+410244.6 &    -406.4 &1&  0.4 &  -0.445 & -402 &\nodata & \nodata & \nodata  & \nodata             \\
J004031.18+403659.5  &    -557.5 &1&  0.5 &  -0.934 & -520 &\nodata & \nodata & \nodata     & \nodata         \\
J004031.92+404313.0 &    -559.5 &1&  0.3 &  -0.999 & -536 &\nodata & \nodata & \nodata   & \nodata           \\
J004032.89+410155.1 &    -399.3 &1&  0.3 &  -0.478 & -410 &\nodata & \nodata & \nodata & \nodata             \\
J004033.84+404658.2 &    -505.\tablenotemark{c}  &2& $>$40 &  -0.982 & -532 &\nodata & \nodata & \nodata  & \nodata             \\
J004035.08+404522.3  &    -560.7 &1&  0.3 &  -0.998 & -536  &ob78-300 & -550 & M2.5 I  &M2.5 I                 \\
J004036.08+403823.1  &    -552.2 &1&  0.4 &  -0.933 & -520 &\nodata & \nodata & \nodata  & \nodata           \\
J004044.41+404402.6 &    -554.2 &1&  0.3 &  -0.982 & -532 &\nodata & \nodata & \nodata   & \nodata            \\
J004047.82+410936.4  &    -361.5 &1&  0.4 &  -0.299 & -367 &\nodata & \nodata & \nodata  &M3 I             \\
J004047.84+405602.6 &    -533.2 &1&  0.4 &  -0.792 & -486 &\nodata & \nodata & \nodata& Peculiar             \\
J004052.31+404356.1 &    -479.5 &1&  2.3 &  -0.946 & -524 &\nodata & \nodata & \nodata & \nodata              \\
J004103.55+410750.0 & -404.7 &1&  0.4 &  -0.381 & -387 &\nodata & \nodata & \nodata  & \nodata             \\
J004109.61+404920.4\tablenotemark{b} &    -527.6 &1&  0.3 &  -0.955 & -526 &\nodata & \nodata & \nodata & \nodata              \\
J004112.38+410918.5 &    -455.1 &1&  0.4 &  -0.354 & -380 &\nodata & \nodata & \nodata & \nodata              \\
J004114.18+403759.8 &    -464.5 &1&  0.3 &  -0.680 & -459 &\nodata & \nodata & \nodata  & \nodata             \\
J004114.22+411732.7 &    -337.5 &1&  0.3 &  -0.145 & -330 &\nodata & \nodata & \nodata  & \nodata             \\
J004118.29+404940.3 &    -519.4 &1&  0.2 &  -0.901 & -513 &\nodata & \nodata & \nodata  & \nodata             \\
J004120.25+403838.1  &    -473.6 &1&  0.3 &  -0.649 & -452 &\nodata & \nodata & \nodata  & \nodata             \\
J004120.96+404125.3  &    -476.7 &1&  0.2 &  -0.689 & -461 &\nodata & \nodata & \nodata  & \nodata             \\
J004120.99+404902.5 &    -524.7 &1&  0.4 &  -0.860 & -503 &\nodata & \nodata & \nodata  & \nodata             \\
J004121.42+403412.7 &    -438.4 &1&  0.3 &  -0.587 & -437 &\nodata & \nodata & \nodata & \nodata              \\
J004122.48+411312.9 &    -380.6 &1&  0.6 &  -0.246 & -355 &\nodata & \nodata & \nodata & \nodata             \\
J004124.80+411634.7 &    -355.8 &1&  0.2 &  -0.162 & -334 &\nodata & \nodata & \nodata  &M3+? I            \\
J004133.42+403721.1  &    -444.7 &1&  0.2 &  -0.554 & -429 &\nodata & \nodata & \nodata  & \nodata            \\
J004138.35+412320.7  &    -302.1 &1&  0.3 &  -0.030 & -302 &\nodata & \nodata & \nodata  & \nodata            \\
J004143.80+412134.8 &    -295.1 &1&  0.2 &  -0.049 & -307 &\nodata & \nodata & \nodata  & \nodata            \\
J004144.65+405446.8 &    -507.0 &1&  0.3 &  -0.788 & -485 &\nodata & \nodata & \nodata & \nodata              \\
J004149.19+411024.8 &    -393.5 &1&  0.3 &  -0.462 & -407 &\nodata & \nodata & \nodata  & \nodata            \\
J004158.68+410833.7 &    -550.4 &1&  0.3 &  -0.762 & -479 &\nodata & \nodata & \nodata & \nodata             \\
J004158.78+413057.0  &    -272.7 &1&  0.7 &   0.091 & -273 &\nodata & \nodata & \nodata  & \nodata            \\
J004210.50+405527.1 &    -446.4 &1&  0.4 &  -0.513 & -419 &\nodata & \nodata & \nodata & \nodata             \\
J004213.75+412524.7 &   -326.2 &1&  0.8 &   0.083 & -275 &\nodata & \nodata & \nodata  & \nodata            \\
J004217.56+413504.1  &    -276.4 &1&  0.3 &   0.169 & -254 &\nodata & \nodata & \nodata  & \nodata            \\
J004217.99+410912.7 &    -440.1 &1&  0.9 &  -0.986 & -533 &\nodata & \nodata & \nodata   & \nodata            \\
J004219.25+405116.4  &    -401.1 &1&  0.3 &  -0.404 & -393 &\nodata & \nodata & \nodata  & \nodata            \\
J004221.33+404735.0 &    -414.5 &1&  0.3 &  -0.374 & -385 &\nodata & \nodata & \nodata  & \nodata            \\
J004235.88+405442.2  &    -352.9 &1&  0.2 &  -0.321 & -373 &\nodata & \nodata & \nodata & \nodata             \\
J004240.58+413603.5 &    -249.4 &1&  0.4 &   0.260 & -232 &\nodata & \nodata & \nodata  & \nodata            \\
J004249.19+415244.3  &    -211.6 &1&  0.4 &   0.291 & -225 &\nodata & \nodata & \nodata& \nodata              \\
J004252.10+414516.4 &    -256.5 &1&  0.3 &   0.306 & -221 &\nodata & \nodata & \nodata  & \nodata            \\
J004252.37+412727.7  &   -275.0 &1&  0.3 &   0.358 & -208 &\nodata & \nodata & \nodata  & \nodata            \\
J004255.75+412300.3&    -140.4 &1&  0.5 &   0.512 & -171 &\nodata & \nodata & \nodata& \nodata               \\
J004255.95+404857.5  &    -389.6 &1&  0.4 &  -0.239 & -353 &\nodata & \nodata & \nodata &M2 I             \\
J004258.62+414446.0  &    -243.0 &1&  0.3 &   0.332 & -215 &\nodata & \nodata & \nodata & \nodata             \\
J004307.51+414548.7  &    -221.5 &1&  0.4 &   0.368 & -206 &\nodata & \nodata & \nodata & \nodata             \\
J004313.45+415044.7 &    -232.7 &1&  0.5 &   0.376 & -204 &\nodata & \nodata & \nodata  & \nodata            \\
J004319.91+411759.8 &    -200.0 &1&  0.4 &   0.290 & -225 &\nodata & \nodata & \nodata & \nodata              \\
J004323.39+411751.6  &    -223.6 &1&  0.4 &   0.268 & -230 &\nodata & \nodata & \nodata & \nodata              \\
J004325.42+415043.1  &    -197.7 &1&  0.5 &   0.428 & -192 &\nodata & \nodata & \nodata & \nodata              \\
J004329.85+415203.0  &    -231.7 &1&  0.3 &   0.441 & -188 &\nodata & \nodata & \nodata & \nodata              \\
J004330.37+412813.9  &     -99.1 &1&  0.2 &   0.983 &  -58 &\nodata & \nodata & \nodata   & \nodata            \\
J004338.12+415351.5 &    -203.8 &1&  0.3 &   0.468 & -182 &\nodata & \nodata & \nodata  & \nodata             \\
J004341.77+412550.1 &     -88.4 &1&  0.6 &   0.785 & -105 &\nodata & \nodata & \nodata & \nodata              \\
J004342.75+411442.8 &    -289.3 &1&  0.4 &   0.131 & -263 &\nodata & \nodata & \nodata  & \nodata             \\
J004342.98+412850.0\tablenotemark{b} &     -96.6 &1&  5.8 &   0.975 &  -60 &\nodata & \nodata & \nodata  & \nodata             \\
J004344.78+415727.8 &    -174.2 &1&  0.3 &   0.474 & -180 &\nodata & \nodata & \nodata  & \nodata             \\
J004354.03+411742.0  &    -268.3 &1&  0.3 &   0.217 & -242 &\nodata & \nodata & \nodata  & \nodata             \\
J004354.50+411108.3  &    -281.6 &1&  0.3 &   0.069 & -278 &\nodata & \nodata & \nodata  & \nodata             \\
J004356.22+413717.6  &    -112.3 &1&  0.3 &   0.920 &  -73 &\nodata & \nodata & \nodata  & \nodata             \\
J004358.00+412114.1 &    -227.3 &1&  0.4 &   0.341 & -213 &\nodata & \nodata & \nodata & \nodata              \\
J004359.94+411330.9 &    -250.2 &1&  0.2 &   0.116 & -267 &\nodata & \nodata & \nodata  & M2 I             \\
J004402.86+414921.9  &    -195.5 &1&  0.3 &   0.667 & -134 &\nodata & \nodata & \nodata & \nodata              \\
J004405.72+411551.7 &    -269.7 &1&  0.3 &   0.166 & -255 &\nodata & \nodata & \nodata  & \nodata             \\
J004415.76+411750.7  &    -246.4 &1&  0.2 &   0.210 & -244 &\nodata & \nodata & \nodata & \nodata              \\
J004420.52+414650.5  &    -125.6 &1&  0.5 &   0.860 &  -87 &\nodata & \nodata & \nodata  & \nodata             \\
J004423.17+412107.9 &    -223.4 &1&  0.3 &   0.286 & -226 &\nodata & \nodata & \nodata & \nodata              \\
J004423.64+412217.0  &    -219.9 &1&  0.5 &   0.319 & -218 &\nodata & \nodata & \nodata  & \nodata             \\
J004423.83+413749.8 &       1.3 &1&  0.5 &   0.979 &  -59 &\nodata & \nodata & \nodata    & \nodata           \\
J004424.94+412322.3 &    -222.6 &1&  0.4 &   0.350 & -211 &\nodata & \nodata & \nodata  & M3 I            \\
J004425.23+413211.1 &     -99.9 &1&  0.3 &   0.736 & -117 &\nodata & \nodata & \nodata   & \nodata            \\
J004425.29+415516.4 &    -146.9 &1&  0.7 &   0.719 & -121 &\nodata & \nodata & \nodata   & \nodata            \\
J004425.77+413659.6 &     -51.6 &1&  0.7 &   0.947 &  -66 &\nodata & \nodata & \nodata  & \nodata             \\
J004428.71+420601.6 &    -188.9 &1&  1.0 &   0.601 & -150 &\nodata & \nodata & \nodata & M0 I              \\
J004430.47+412435.8 &    -235.4 &1&  0.5 &   0.376 & -204 &\nodata & \nodata & \nodata & \nodata              \\
J004432.48+415033.8  &    -104.7 &1&  0.3 &   0.861 &  -87 &\nodata & \nodata & \nodata  & \nodata             \\
J004438.65+412934.1\tablenotemark{b}  &    -164.6 &1&  0.7 &   0.527 & -168 &R95 & \nodata & M1-2 I& \nodata  \\
J004442.26+415122.9  &     -96.9 &1&  0.3 &   0.907 &  -76 &\nodata & \nodata & \nodata   & \nodata            \\
J004444.91+412530.3  &    -195.5 &1&  0.4 &   0.369 & -206 &\nodata & \nodata & \nodata  & \nodata             \\
J004447.08+412801.7 &    -181.2 &1&  0.2 &   0.440 & -189 &\nodata & \nodata & \nodata & M2.5 I              \\
J004451.13+415808.4 &     -96.9 &1&  0.4 &   0.830 &  -95 &\nodata & \nodata & \nodata   & \nodata            \\
J004451.76+420006.0  &    -141.0 &1&  0.5 &   0.799 & -102 &\nodata & \nodata & \nodata  & \nodata             \\
J004454.38+412441.6 &    -219.3 &1&  0.5 &   0.331 & -215 &\nodata & \nodata & \nodata  &M2 I             \\
J004457.32+415144.9\tablenotemark{b} &     -75.4 &1&  0.4 &   0.972 &  -60 &\nodata & \nodata & \nodata  & \nodata             \\
J004501.30+413922.5 &    -144.6 &1&  0.2 &   0.793 & -103 &\nodata & \nodata & \nodata  & M3+? I             \\
J004502.52+412926.7 &    -169.3 &1&  0.3 &   0.438 & -189 &\nodata & \nodata & \nodata  & \nodata             \\
J004507.90+413427.4 &    -141.8 &1&  0.5 &   0.576 & -156 &\nodata & \nodata & \nodata  & \nodata             \\
J004514.95+414625.6 &     -96.3 &1&  0.4 &   0.935 &  -69 &\nodata & \nodata & \nodata   &M2 I            \\
J004517.25+413948.2 &    -107.8 &1&  0.3 &   0.716 & -122 &ob48-120 & -113 & M2.5 I  & \nodata\\
J004531.41+413400.6  &     -94.5 &1&  0.4 &   0.483 & -178 &\nodata & \nodata & \nodata & \nodata              \\
J004607.45+414544.6 &    -122.3 &1&  0.4 &   0.671 & -133 &\nodata & \nodata & \nodata &M2 I              \\
J004614.57+421117.4 &     -48.2 &1&  1.0 &   0.978 &  -59 &\nodata & \nodata & \nodata& \nodata               \\
J004627.53+420950.9 &     -42.4 &1&  0.8 &   1.000 &  -54 &ob102-573 & -74 & \nodata& \nodata \\
J004638.81+415456.1 &    -100.5 &1&  1.2 &   0.777 & -107 &\nodata & \nodata & \nodata& \nodata               \\
J004655.39+415827.0  &    -120.6 &1&  0.7 &   0.794 & -103 &\nodata & \nodata & \nodata  & \nodata             \\
J004748.71+420423.8 &    -157.6 &1&  0.6 &   0.745 & -115 &\nodata & \nodata & \nodata& \nodata              \\
\enddata
\tablenotetext{a}{From Massey 1998, except for the star R95, classified as M1-2 I by Humphreys et al.\ 1988.}
\tablenotetext{b}{Crowded.}
\tablenotetext{c}{Observations made on two different nights of this star differed by 85 km s$^{-1}$, although each measurement
was well determined.}
\end{deluxetable}

\begin{deluxetable}{l c c r c c r r c r }
\pagestyle{empty}
\tabletypesize{\small}
\tablecaption{\label{tab:dave} New $K_s$ Photometry and Comparison with 2MASS PSC}
\tablewidth{0pt}
\tablecolumns{10}
\tablehead{
\colhead{Star}
&\multicolumn{3}{c}{New}
&
&\multicolumn{2}{c}{2MASS} \\  \cline{2-4} \cline{6-7}
&\colhead{$K_s$}
&\colhead{$\sigma_\mu$}
&\colhead{$N$}
&&\colhead{$K_s$}
&\colhead{$\sigma_\mu$}
&\colhead{$\Delta K_s$\tablenotemark{a}} 
&\colhead{$\sigma_{\Delta K_s}$}
&\colhead{N$\sigma$} 
}
\startdata
J003722.34+400012.1&  15.02&   0.04&   47&  &15.03&   0.13&   0.00&   0.13&  0.0\\
J003739.41+395835.0&  15.19&   0.03&   48&  &15.13&   0.15&   0.06&   0.15&  0.4\\
J003857.29+404053.6&  14.86&   0.02&   90&  &14.96&   0.12&  -0.10&   0.12& -0.8\\
J003902.20+403907.3&  14.74&   0.02&   69&  &14.76&   0.11&  -0.02&   0.11& -0.2\\
J003912.77+404412.1&  14.30&   0.01&   46&  &14.34&   0.08&  -0.04&   0.08& -0.5\\
J003913.40+403714.2&  14.34&   0.02&   47&  &14.49&   0.09&  -0.15&   0.09& -1.6\\
J003915.67+403559.0&  14.65&   0.02&   45&  &14.80&   0.12&  -0.16&   0.12& -1.3\\
J003944.05+403234.5&  15.66&   0.04&   33&  &15.74&   0.22&  -0.08&   0.22& -0.4\\
J003954.26+404004.4&  14.15&   0.01&   50&  &14.20&   0.08&  -0.05&   0.08& -0.6\\
J003957.00+410114.6&  15.08&   0.03&   49&  &14.95&   0.14&   0.13&   0.14&  1.0\\
J004023.81+403351.8&  14.79&   0.02&   49&  &14.79&   0.10&   0.00&   0.10&  0.0\\
J004023.89+403554.7&  15.22&   0.03&   48&  &15.13&   0.12&   0.09&   0.12&  0.8\\
J004026.66+405019.7&  14.43&   0.02&   45&  &14.31&   0.07&   0.12&   0.07&  1.7\\
J004027.36+410444.9&  13.38&   0.01&   40&  &13.34&   0.03&   0.04&   0.03&  1.2\\
J004030.43+410244.6&  15.19&   0.02&   38&  &15.27&   0.15&  -0.07&   0.15& -0.5\\
J004032.89+410155.1&  14.91&   0.02&   37&  &15.13&   0.13&  -0.23&   0.13& -1.7\\
J004035.08+404522.3&  13.09&   0.01&   47&  &13.19&   0.04&  -0.10&   0.04& -2.5\\
J004044.41+404402.6&  14.42&   0.01&   46&  &14.40&   0.09&   0.01&   0.09&  0.2\\
J004047.84+405602.6&  14.25&   0.01&   47&  &14.21&   0.06&   0.04&   0.06&  0.6\\
J004052.31+404356.1&  14.98&   0.03&   45&  &15.44&   0.19&  -0.46&   0.19& -2.4\\
J004114.18+403759.8&  13.24&   0.01&   50&  &13.36&   0.04&  -0.11&   0.04& -3.2\\
J004114.22+411732.7&  15.20&   0.02&  117&  &15.20&   0.15&   0.00&   0.15&  0.0\\
J004120.25+403838.1&  13.66&   0.01&   47& & 13.66&   0.04&   0.00&   0.04&  0.0\\
J004120.96+404125.3&  14.12&   0.01&   50&  &14.16&   0.06&  -0.04&   0.06& -0.7\\
J004122.48+411312.9&  14.48&   0.01&   50& & 14.40&   0.07&   0.07&   0.07&  1.0\\
J004124.80+411634.7&  13.13&   0.00&   98&  &13.13&   0.04&   0.00&   0.04& -0.1\\
J004133.42+403721.1&  14.68&   0.02&   49& & 14.46&   0.08&   0.23&   0.08&  2.9\\
J004138.35+412320.7&  13.36&   0.00&   99& & 13.33&   0.03&   0.03&   0.03&  0.8\\
J004143.80+412134.8&  15.03&   0.02&   94& & 15.23&   0.15&  -0.20&   0.15& -1.4\\
J004252.10+414516.4&  15.51&   0.04&   43& & 15.60&   0.21&  -0.09&   0.21& -0.4\\
J004258.62+414446.0&  15.54&   0.04&   44& & \nodata&  \nodata&  \nodata&  \nodata&  \nodata\\ 
J004307.51+414548.7&  15.66&   0.04&   47&&  15.46&   0.19&   0.20&   0.19&  1.1\\
J004313.45+415044.7&  14.83&   0.03&   76& & 14.81&   0.11&   0.02&   0.11&  0.2\\
J004325.42+415043.1&  15.86&   0.08&   40&&   \nodata&  \nodata&  \nodata&  \nodata&  \nodata\\
J004329.85+415203.0&  14.91&   0.04&   46&&  14.77&   0.11&   0.14&   0.11&  1.3\\
J004338.12+415351.5&  15.42&   0.05&   47& & 15.30&   0.17&   0.12&   0.17&  0.7\\
J004344.78+415727.8&  15.08&   0.05&   33& & 15.34&   0.18&  -0.26&   0.18& -1.4\\
J004356.22+413717.6&  14.39&   0.02&   46&&  14.43&   0.07&  -0.04&   0.07& -0.6\\
J004358.00+412114.1&  14.16&   0.01&   46&&  14.12&   0.07&   0.04&   0.07&  0.5\\
J004359.94+411330.9&  15.10&   0.04&   48& & 15.08&   0.12&   0.01&   0.12&  0.1\\
J004415.42+413409.5&  15.52&   0.04&   47& & 15.58&   0.20&  -0.05&   0.20& -0.3\\
J004420.52+414650.5&  15.08&   0.03&   41&&  15.10&   0.12&  -0.02&   0.12& -0.2\\
J004423.17+412107.9&  15.48&   0.05&   42& & 15.60&   0.19&  -0.12&   0.19& -0.6\\
J004423.64+412217.0&  14.35&   0.16&   70&  &14.45&   0.18&  -0.10&   0.18& -0.6\\
J004423.83+413749.8&  15.84&   0.03&   43& & 15.21&   1.00&   0.63&   1.00&  0.6\\
J004424.94+412322.3&  13.66&   0.01&   61& & 13.77&   0.05&  -0.11&   0.05& -2.3\\
J004425.29+415516.4&  14.78&   0.02&   48& & 15.06&   0.11&  -0.28&   0.11& -2.6\\
J004428.71+420601.6&  14.30&   0.01&   46& & 14.09&   0.06&   0.21&   0.06&  3.8\\
J004432.48+415033.8&  15.35&   0.05&   43& & 15.39&   0.16&  -0.04&   0.16& -0.3\\
J004442.26+415122.9&  14.91&   0.02&   67& & 15.02&   0.10&  -0.11&   0.10& -1.1\\
J004444.91+412530.3&  15.18&   0.03&   48& & 15.20&   0.12&  -0.01&   0.12& -0.1\\
J004447.08+412801.7&  13.97&   0.01&   47&&  14.26&   0.05&  -0.29&   0.05& -5.4\\
J004451.13+415808.4&  14.69&   0.02&   47& & 14.71&   0.09&  -0.02&   0.09& -0.3\\
J004454.38+412441.6&  14.91&   0.03&   48& & 14.91&   0.10&   0.00&   0.10&  0.0\\
J004501.30+413922.5&  13.64&   0.01&   65& & 13.59&   0.03&   0.05&   0.03&  1.3\\
J004502.52+412926.7&  14.43&   0.02&   49& & 14.42&   0.11&   0.01&   0.11&  0.1\\
J004503.83+413737.0&  13.88&   0.01&   66& & 13.89&   0.05&  -0.01&   0.05& -0.2\\
J004507.90+413427.4&  14.62&   0.02&   65& & 14.63&   0.08&  -0.01&   0.08& -0.1\\
J004509.33+413633.1&  14.80&   0.02&   65& & 14.62&   0.08&   0.18&   0.08&  2.2\\
J004513.96+413858.3&  15.26&   0.03&   64& & 15.17&   0.13&   0.09&   0.13&  0.7\\
J004514.95+414625.6&  14.97&   0.03&   46& & 15.02&   0.11&  -0.05&   0.11& -0.5\\
J004519.91+413857.6&  14.77&   0.02&   64& & 14.63&   0.08&   0.14&   0.08&  1.8\\
J004528.15+413916.7&  14.86&   0.02&   63& & 14.95&   0.11&  -0.09&   0.11& -0.8\\
J004528.88+413827.8&  14.62&   0.02&   62& & 14.77&   0.09&  -0.15&   0.09& -1.7\\
J004531.41+413400.6&  15.38&   0.03&   64& & 15.02&   0.11&   0.35&   0.11&  3.1\\
J004607.45+414544.6&  14.59&   0.02&   45& & 14.55&   0.08&   0.05&   0.08&  0.6\\
J004614.57+421117.4&  14.09&   0.01&   46& & 14.19&   0.06&  -0.10&   0.06& -1.6\\
J004625.83+421011.5&  15.03&   0.03&   46& & 15.05&   0.12&  -0.02&   0.12& -0.2\\
J004628.17+421119.3&  15.54&   0.04&   42&&  15.71&   0.20&  -0.17&   0.20& -0.8\\
J004638.81+415456.1&  14.48&   0.01&   49&&  14.41&   0.07&   0.07&   0.07&  1.0\\
J004655.39+415827.0&  15.22&   0.02&   47&&  15.20&   0.14&   0.03&   0.14&  0.2\\
\enddata
\tablenotetext{a} {$\Delta K_s$ is our new $K_s$ {\it minus} the 2MASS $K_s$.}
\end{deluxetable}

\begin{deluxetable}{l c c r c c r r c r }
\pagestyle{empty}
\tabletypesize{\footnotesize}
\tablecaption{\label{tab:Vphot} New $V$ Photometry and Comparison with LGGS}
\tablewidth{0pt}
\tablecolumns{10}
\tablehead{
\colhead{Star}
&\multicolumn{3}{c}{New}
&
&\multicolumn{2}{c}{LGGS} \\  \cline{2-4} \cline{6-7}
&\colhead{$V$}
&\colhead{$\sigma_\mu$}
&\colhead{$N$}
&&\colhead{$V$}
&\colhead{$\sigma_\mu$}
&\colhead{$\Delta V$\tablenotemark{a}} 
&\colhead{$\sigma_{\Delta V}$}
&\colhead{N$\sigma$} 
}
\startdata
J003722.34+400012.1& 18.99&  0.00&  4&& 18.94&  0.01&  0.06&  0.01&    8.7\\
J003739.41+395835.0& 19.18\tablenotemark{b}&  0.02\tablenotemark{b}&  4&& 19.35&  0.01& -0.17\tablenotemark{b}&  0.03&  -5.6\tablenotemark{b}\\
J003857.29+404053.6& 19.75&  0.01&  6&& 19.57&  0.01&  0.18&  0.01&   15.9\\
J003902.20+403907.3& 19.55&  0.01&  6&& 19.58&  0.00& -0.03&  0.01&   -3.9\\
J003903.28+403042.7& 18.64&  0.01&  3&& 18.75&  0.00& -0.11&  0.01&  -12.2\\
J003912.77+404412.1& 19.58&  0.00&  6&& 19.77&  0.01& -0.19&  0.01&  -18.9\\
J003913.40+403714.2& 19.34&  0.01&  3&& 19.33&  0.00&  0.01&  0.01&    0.9\\
J003915.67+403559.0& 19.76&  0.01&  6&& 19.74&  0.01&  0.01&  0.01&    1.1\\
J003935.86+402705.7& 19.46&  0.01&  3&& 19.55&  0.00& -0.09&  0.01&   -9.4\\
J003936.90+405120.5& 19.85&  0.00&  1&& 19.72&  0.01&  0.13&  0.01&   19.1\\
J003954.26+404004.4& 19.32&  0.01&  3&& 19.54&  0.00& -0.22&  0.01&  -17.2\\
J003957.00+410114.6& 19.74&  0.02&  4&& 19.82&  0.01& -0.08&  0.02&   -4.4\\
J004023.81+403351.8& 18.71\tablenotemark{b}&  0.05\tablenotemark{b}&  3&& 18.86&  0.00& -0.15\tablenotemark{b}&  0.05\tablenotemark{b}&  -3.0\tablenotemark{b}\\
J004023.89+403554.7& 19.93\tablenotemark{b}&  0.05\tablenotemark{b}&  3&& 19.91&  0.01&  0.02\tablenotemark{b}&  0.05\tablenotemark{b}&    0.4\tablenotemark{b}\\
J004026.66+405019.7& 19.91&  0.02&  6&& 19.65&  0.00&  0.26&  0.02&   12.0\\
J004027.36+410444.9& 18.86&  0.01&  4&& 18.85&  0.00&  0.01&  0.01&    1.3\\
J004030.43+410244.6& 19.65&  0.01&  4&& 19.75&  0.01& -0.10&  0.01&  -10.0\\
J004032.89+410155.1& 19.68&  0.01&  4&& 19.62&  0.00&  0.06&  0.01&    4.6\\
J004035.08+404522.3& 19.33\tablenotemark{b}&  0.04\tablenotemark{b}&  3&& 19.38&  0.00& -0.05\tablenotemark{b}&  0.04\tablenotemark{b}&   -1.2\tablenotemark{b}\\
J004044.41+404402.6& 19.48&  0.00&  3&& 19.71&  0.00& -0.23&  0.01&  -41.7\\
J004047.82+410936.4& 19.54&  0.01&  7&& 19.69&  0.01& -0.15&  0.01&  -10.9\\
J004047.84+405602.6& 19.40&  0.02&  3&& 19.69&  0.01& -0.29&  0.02&  -12.6\\
J004050.87+410541.1& 19.67&  0.00&  1&& 19.72&  0.00& -0.05&  0.00&   -9.8\\
J004112.38+410918.5& 19.71&  0.00&  3&& 19.95&  0.01& -0.24&  0.01&  -28.7\\
J004114.18+403759.8& 19.83&  0.01&  3&& 19.80&  0.01&  0.03&  0.01&    2.4\\
J004114.22+411732.7& 19.60\tablenotemark{b}&  0.04\tablenotemark{b}&  6&& 19.74&  0.01& -0.14\tablenotemark{b}&  0.05\tablenotemark{b}&  -2.8\tablenotemark{b}\\
J004120.25+403838.1& 18.75&  0.00&  3&& 18.88&  0.00& -0.13&  0.00&  -29.3\\
J004120.96+404125.3& 20.23&  0.01&  3&& 19.97&  0.01&  0.26&  0.01&   24.4\\
J004122.48+411312.9& 19.79&  0.02&  3&& 20.00&  0.01& -0.21&  0.02&   -8.7\\
J004124.80+411634.7& 20.40&  0.01&  6&& 19.73&  0.01&  0.68&  0.01&   57.9\\
J004133.42+403721.1& 19.82\tablenotemark{b}&  0.02\tablenotemark{b}&  3&& 19.77&  0.00&  0.05\tablenotemark{b}&  0.03&    1.7\tablenotemark{b}\\
J004138.35+412320.7& 18.21&  0.01&  3&& 18.47&  0.01& -0.26&  0.02&  -17.1\\
J004143.80+412134.8& 19.45&  0.01&  6&& 19.63&  0.00& -0.18&  0.01&  -14.6\\
J004313.45+415044.7& 18.90&  0.00&  8&& 18.94&  0.00& -0.04&  0.01&   -6.6\\
J004329.85+415203.0& 19.15&  0.01&  6&& 19.01&  0.01&  0.14&  0.01&   14.4\\
J004356.22+413717.6& 19.60&  0.03&  3&& 19.42&  0.00&  0.18&  0.03&    6.2\\
J004358.00+412114.1& 19.76&  0.01&  4&& 19.55&  0.00&  0.22&  0.01&   19.2\\
J004359.94+411330.9& 19.74&  0.01&  3&& 19.82&  0.01& -0.08&  0.01&   -9.5\\
J004420.52+414650.5& 19.67&  0.01&  3&& 19.77&  0.01& -0.10&  0.01&   -9.9\\
J004423.64+412217.0& 19.74&  0.01&  8&& 19.63&  0.00&  0.12&  0.01&    7.8\\
J004424.94+412322.3& 19.85&  0.02&  8&& 19.98&  0.01& -0.12&  0.02&   -7.7\\
J004425.29+415516.4& 19.78&  0.01&  7&& 19.63&  0.01&  0.15&  0.02&    9.8\\
J004428.71+420601.6& 18.22&  0.00&  4&& 18.98&  0.00& -0.76&  0.00& -181.1\\
J004442.26+415122.9& 19.04&  0.01& 10&& 19.18&  0.00& -0.14&  0.01&  -14.1\\
J004444.91+412530.3& 19.90&  0.01&  8&& 19.93&  0.01& -0.03&  0.01&   -2.4\\
J004447.08+412801.7& 20.41&  0.02&  6&& 19.34&  0.00&  1.07&  0.02&   54.0\\
J004451.13+415808.4& 19.99&  0.01&  4&& 19.96&  0.01&  0.03&  0.02&    1.8\\
J004454.38+412441.6& 19.66&  0.02&  4&& 19.55&  0.01&  0.11&  0.02&    5.8\\
J004501.30+413922.5& 19.37&  0.01&  3&& 19.64&  0.01& -0.27&  0.01&  -27.4\\
J004502.52+412926.7& 18.63&  0.00&  4&& 18.57&  0.00&  0.06&  0.01&    9.6\\
J004503.83+413737.0& 19.57&  0.00&  3&& 19.89&  0.01& -0.32&  0.01&  -39.4\\
J004507.90+413427.4& 19.76&  0.01&  3&& 20.00&  0.01& -0.24&  0.01&  -18.9\\
J004509.33+413633.1& 19.76&  0.02&  3&& 19.43&  0.00&  0.33&  0.02&   14.5\\
J004513.96+413858.3& 19.29&  0.01&  3&& 19.63&  0.01& -0.34&  0.01&  -29.4\\
J004514.95+414625.6& 19.82&  0.00&  3&& 19.69&  0.01&  0.13&  0.01&   17.5\\
J004519.91+413857.6& 18.58&  0.00&  3&& 18.76&  0.00& -0.18&  0.01&  -30.3\\
J004528.15+413916.7& 19.77&  0.00&  3&& 19.89&  0.01& -0.12&  0.01&  -19.8\\
J004528.88+413827.8& 20.04&  0.01&  3&& 20.16&  0.01& -0.12&  0.01&  -11.4\\
J004531.41+413400.6& 19.85&  0.01&  3&& 19.67&  0.01&  0.18&  0.02&   11.3\\
J004607.45+414544.6& 19.35&  0.00&  4&& 19.25&  0.00&  0.10&  0.01&   16.3\\
J004614.57+421117.4& 19.60&  0.01&  4&& 19.61&  0.00& -0.01&  0.01&   -1.3\\
J004625.83+421011.5& 19.10&  0.01&  4&& 19.08&  0.00&  0.03&  0.01&    3.1\\
J004633.50+421109.4& 19.76\tablenotemark{b}&  0.03\tablenotemark{b}&  4&& 19.33&  0.00&  0.43\tablenotemark{b}&  0.03&   14.3\tablenotemark{b}\\
J004638.81+415456.1& 19.35&  0.01&  4&& 19.39&  0.00& -0.04&  0.01&   -5.5\\
J004655.39+415827.0& 19.32&  0.01&  4&& 19.36&  0.01& -0.04&  0.01&   -4.3\\
\enddata
\tablenotetext{a} {$\Delta V$ is our new $V$ {\it minus} the LGGS $V$.}
\tablenotetext{b} {Affected by crowding at the 0.02-0.05~mag level.}
\end{deluxetable}

\begin{deluxetable}{l l l l l l l r r r r r r r r r r r r }
\pagestyle{empty}
\tabletypesize{\tiny}
\rotate
\tablecaption{\label{tab:spect} Physical Properties of Stars with Spectrophotometry}
\tablewidth{0pt}
\tablecolumns{19}
\tablehead{
\colhead{Star}
&\colhead{Type}
&\colhead{$V$\tablenotemark{a}}
&\colhead{$K_S$\tablenotemark{a}}
&\multicolumn{10}{c}{Spectral Fitting}
&
&\multicolumn{4}{c}{$(V-K)_0$} \\ \cline{5-14} \cline{16-19}
&&&&&&&Model&Comp&
\multicolumn{2}{c}{$V$-band}&&
\multicolumn{2}{c}{$K$-band} \\ \cline{10-11} \cline{13-14}
&
&
&
&\colhead{$A_V$\tablenotemark{b}}
&\colhead{$M_V$}
&\colhead{$T_{\rm eff}$\tablenotemark{c}}
&\colhead{$\log g$}
&\colhead{$\log g$}
&\colhead{$R/R_\odot$}
&\colhead{$M_{\rm bol}$}
&
&\colhead{$R/R_\odot$}
&\colhead{$M_{\rm bol}$}
&
&\colhead{$M_K$}
&\colhead{$T{\rm eff}$}
&\colhead{$R/R_\odot$}
&\colhead{$M_{\rm bol}$}
}
\startdata
J003857.29+404053.6 &M2    &19.75 &14.86  &   0.93 &  -5.58 &  3650 &  0.0 &  0.1 &   630 &   -7.26&&  510 &   -6.82 &&   -9.61 &  3775 &   500 &   -6.91\\
J003902.20+403907.3 &M1    &19.55 &14.74  &   0.62 &  -5.47 &  3725 &  0.0 &  0.2 &   510 &   -6.90&&  525 &   -6.96 &&   -9.69 &  3715 &   525 &   -6.95\\
J003913.40+403714.2 &M1    &19.34 &14.34  &   0.77 &  -5.83 &  3725 &  0.0 &  0.1 &   605 &   -7.26&&  635 &   -7.37 &&  -10.11 &  3695 &   640 &   -7.35\\
J003957.00+410114.6 &M0    &19.74 &15.08  &   0.93 &  -5.59 &  3700 &  0.0 &  0.1 &   570 &   -7.10&&  460 &   -6.63 &&   -9.39 &  3860 &   445 &   -6.76\\
J004035.08+404522.3 &M2.5  &19.33 &13.09  &   2.01 &  -7.08 &  3700 &  0.0 & -0.3 &  1130 &   -8.59&& 1220 &   -8.75 &&  -11.51 &  3655 &  1230 &   -8.72\\
J004047.82+410936.4 &M3    &19.54 &13.47\tablenotemark{d} &   1.40: &  -6.26 &  3650: &  0.0 & -0.1 &   860 &   -7.94&& 1000 &   -8.26 &&  -11.06 &  3575 &  1010 &   -8.20\\
J004124.80+411634.7\tablenotemark{e} &M3+?  &20.40 &13.13  &   1.86: &  -5.86 &  3625: &  0.0 & -0.1 &   760 &   -7.63&& 1205 &   -8.64 &&  -11.45 &  3455 &  1240 &   -8.50\\
J004255.95+404857.5 &M2    &19.86\tablenotemark{f}&14.00\tablenotemark{d}&   1.40 &  -5.94 &  3675 &  0.0 &  0.0 &   700 &   -7.53&&  780 &   -7.75 &&  -10.53 &  3620 &   785 &   -7.71\\
J004359.94+411330.9 &M2    &19.74 &15.10  &   1.40 &  -6.06 &  3650 &  0.0 & -0.1 &   785 &   -7.74&&  470 &   -6.63 &&   -9.43 &  4055 &   440 &   -6.93\\
J004424.94+412322.3 &M3    &19.85 &13.66  &   2.48 &  -7.03 &  3625 & -0.5 & -0.4 &  1300 &   -8.80&&  975 &   -8.18 &&  -11.00 &  3795 &   945 &   -8.32\\
J004428.71+420601.6 &M0    &18.22 &14.30  &   0.77 &  -6.95 &  3825 &  0.0 & -0.1 &   860 &   -8.15&&  635 &   -7.49 &&  -10.15 &  4140 &   605 &   -7.72\\
J004447.08+412801.7 &M2.5  &20.41 &13.97  &   1.86 &  -5.85 &  3625 &  0.0 & -0.1 &   755 &   -7.62&&  815 &   -7.80 &&  -10.61 &  3580 &   825 &   -7.76\\
J004454.38+412441.6 &M2    &19.66 &14.91  &   1.09 &  -5.83 &  3725 &  0.0 &  0.1 &   605 &   -7.26&&  500 &   -6.84 &&   -9.58 &  3880 &   485 &   -6.96\\
J004501.30+413922.5\tablenotemark{e} &M3+?  &19.64 &13.64  &   0.47: &  -5.23 &  3675: &  0.0 &  0.2 &   505 &   -6.82&&  875 &   -8.00 &&  -10.78 &  3460 &   910 &   -7.83\\
J004514.95+414625.6 &M2    &19.82 &14.97  &   0.77 &  -5.35 &  3675 &  0.0 &  0.2 &   535 &   -6.94&&  480 &   -6.71 &&   -9.48 &  3740 &   475 &   -6.76\\
J004607.45+414544.6 &M2    &19.35 &14.59  &   0.47 &  -5.51 &  3700 &  0.0 &  0.2 &   545 &   -7.02&&  560 &   -7.07 &&   -9.83 &  3690 &   560 &   -7.06\\
\enddata
\tablenotetext{a}{Contemporaneous with the spectrophotometry, except as noted.}
\tablenotetext{b}{Uncertainties are $\pm$0.10~mag unless marked ``:" ($\pm$0.15~mag).}
\tablenotetext{c}{Uncertainties are $\pm$25~K unless marked ``:" ($\pm$50~K). }
\tablenotetext{d}{From 2MASS.}
\tablenotetext{e}{Hot companion?}
\tablenotetext{f}{From LGGS.}
\end{deluxetable}

\begin{deluxetable}{l l l r r r r r r r r }
\pagestyle{empty}
\tabletypesize{\tiny}
\tablecaption{\label{tab:alllum} Derived Bolometric Luminosities for RSGs\tablenotemark{a}}
\tablewidth{0pt}
\tablecolumns{11}
\tablehead{
&&&&&&\multicolumn{2}{c}{Using $V$}
&
&\multicolumn{2}{c}{Using $K$} \\  \cline{7-8} \cline{10-11}
\colhead{Gal}
&\colhead{Star}
&\colhead{Type}
&\colhead{$A_V$}
&\colhead{$T_{\rm eff}$}
&\colhead{$M_V$}
&\colhead{$R/R_\odot$}
&\colhead{$\log L/L_\odot$}
&\colhead{$M_K$}
&\colhead{$R/R_\odot$}
&\colhead{$L/L_\odot$}
}
\startdata
SMC& [M2002] SMC005092      &  M2 I    & 0.40 &3475& -6.40 &1220 &5.29 &-10.84 & 935 &5.06\\
SMC& [M2002] SMC008930      &  M0 I     &0.56 &3625& -6.78 &1070 &5.25 &-10.61&  815 &5.01\\
SMC &[M2002] SMC010889      &  M0 I     &0.09 &3600& -6.79 &1130 &5.29 &-11.10& 1025 &5.20\\
SMC &[M2002] SMC011101      &  K2.5 I   &1.43&4200 &-6.79 & 540 &4.92 & -9.79  &500 &4.85\\
SMC &[M2002] SMC011709       & K5-M0 I  &0.09 &3725 &-6.56 & 820 &5.07 &-10.34  &705& 4.94\\
\enddata
\tablenotetext{a}{The complete version of this table is made available in machine readable form in the
on-line edition.}
\end{deluxetable}

\begin{deluxetable}{l c c}
\tablecaption{\label{tab:toplums}Most Luminous RSGs}
\tablecolumns{3}
\tablehead{
\colhead{Galaxy}
&\colhead{K-band $\log L/L_\odot$}
&\colhead{V-band $\log L/L_\odot$}
}
\startdata
SMC & 5.40 & 5.25 \\
LMC & 5.33 & 5.21 \\
MW  & 5.63 & 5.43 \\
M31 & 5.16 & 5.20 \\
\enddata
\end{deluxetable}
\end{document}